\documentclass[
 reprint,
 superscriptaddress,
showpacs,
 amsmath,amssymb,
 aps,
 prd,
floatfix,
]{revtex4-1}

\usepackage{bm}

\usepackage{latexsym}
\usepackage{amssymb}
\usepackage{epsfig}

\newcommand{\vect}[1]{\boldsymbol{#1}}

\begin{document}


\title{Comparison with experimental data of different theoretical
  approaches to high-energy electron bremsstrahlung including quantum
  coherence effects }

\author{A.~Mangiarotti}
\affiliation{Instituto de F\'{\i}sica da Universidade de S\~{a}o
  Paulo, Rua do Mat\~{a}o 1371, 05508-090 S\~{a}o Paulo, Brazil}
\author{P. Sona}
\affiliation{INFN, Sezione di Firenze, Polo Scientifico,
 Via G. Sansone 1, 50019 Sesto Fiorentino, Italy\\}
\author{U. I. Uggerh{\o}j}
\affiliation{Department of Physics and Astronomy, University
 of Aarhus, Ny Munkegade 120, 8000 Aarhus, Denmark}

\date{\today}

\begin{abstract}
  The basic expressions for the differential nuclear bremsstrahlung
  cross section at high electron energy, as derived under different
  theoretical approaches and approximations to quantum coherence
  effects, are compared. The Baier-Katkov treatment is reformulated to
  allow introduction of the same value of the radiation length in all
  calculations.  A dedicated Monte Carlo code is employed for
  obtaining photon energy spectra in the framework of the Baier-Katkov
  approach taking into account multiphoton emission, attenuation by
  pair production, and pile-up with photons from the background.  The
  results of Monte Carlo simulations for both the Migdal and
  Baier-Katkov descriptions are compared to all available data that
  show the Landau-Pomeranchuk-Migdal suppression.  The issue of the
  sensitivity of the experiments to the difference of the two
  approaches is investigated.
\end{abstract}

\maketitle

\section{\label{sec:intro}Introduction}

Ever since the pioneering papers by Sauter~\cite{Sauter:1934}, Bethe
and Heitler~\cite{Bethe:1934a,Bethe:1934b}, and
Racah~\cite{Racah:1934a}, the bremsstrahlung process undergone by
electrons and positrons in crossing matter has attracted the interest
of theorists and experimentalists. As a matter of fact, a huge range
of disciplines directly or indirectly benefits from any progress of
the research in this field, from elementary particle physics, to
cosmic ray studies, which need accurate simulations of shower
development. At the highest energies, one of the most important steps
forward in the theory was the recognition by Landau and
Pomeranchuk~\cite{Landau:1953a,Landau:1953b} of the possible
suppression of the intensity of the bremsstrahlung radiation due to a
reduction of the coherence length caused by the multiple scattering
suffered by the radiating particle. These authors used pure classical
arguments, while Migdal~\cite{Migdal:1956} developed a fully
quantum-mechanical theory of the bremsstrahlung process including the
mentioned suppression. This effect, named the LPM effect after the
discoverers (Landau-Pomeranchuk-Migdal), becomes progressively more
important as the energy of the electron/positron increases since the
suppression extends up to a larger fraction of the total radiated
spectrum~\cite{Klein:1999}.  Early review papers further clarifying
the LPM effect were written by Feinberg and
Pomeranchuk~\cite{Feinberg:1956} and Galitsky and
Gurevich~\cite{Galitsky:1964}, while more recent surveys are due to
Klein~\cite{Klein:1999} and Baier and
Katkov~\cite{Baier:2005}. Although the LPM effect was postulated for
QCD~\cite{Zakharov:1996a,Wiedemann:1999,Baier:2000b,Wiedemann:2000,Armesto:2004}
and weak interactions~\cite{Raffelt:1991}, direct experimental
evidence under controlled conditions was gathered only in the QED
case~\cite{Anthony:1995,Anthony:1996,Anthony:1997,Hansen:2003,Hansen:2004}.

Conceptually, the LPM effect is interesting because it goes beyond the
standard perturbative formulation of quantum field theory describing
the point interaction of free particles: the photon emission requires
a formation length, which can also be interpreted as a coherence
length, to be fully emitted and behave as a particle independent from
the radiating electron. Any disturbance during this phase, reduces the
coherence length thereby leading to a suppression. In the QED case,
the electromagnetic interaction has an infinite range and this gives
rise to a nonperturbative effect also lying outside the standard
perturbative formulation: the Coulomb correction. Bethe and Maximon
discovered that such nonradiative corrections can be treated, with
appropriate wave functions, at the leading order of $\alpha\,Z$ in the
high-energy limit~\cite{Bethe:1954a}. Data with systematic
uncertainties low enough to validate this conclusion were only
recently obtained~\cite{Mangiarotti:2021}. The original quantum
treatment of the LPM suppression by Migdal~\cite{Migdal:1956} did not
include the Coulomb correction, which has since been taken into
account semiempirically by expressing his cross section in terms of
the radiation length and then adding the Coulomb correction to the
latter~\cite{Anthony:1995}. Needless to say, without this rescaling,
Migdal theory cannot reproduce the data for high-$Z$ elements. A
complete treatment of the LPM effect directly embodying the Coulomb
correction was developed by Baier and Katkov~\cite{Baier:2005}. Recent
work was also published improving the original approach by Migdal and
essentially justifying the renormalization procedure to the radiation
length~\cite{Voskresenskaya:2014}. However, the comparison with data
of these approaches: the one by Migdal, with rescaling, and the one by
Baier and Katkov, directly dealing with the Coulomb correction, was
performed in a quite unsystematic way up to now. The Migdal treatment
was implemented in several Monte Carlo codes allowing to accurately
take into account the emission of multiple photons by the same
electron crossing the target and the absorption of the photons by pair
production. Moreover, in some experimental conditions, it is necessary
to consider the pile-up of photons from the target with photons from
the background, which reach simultaneously the calorimeter. In the
case of the Baier-Katkov approach, only multiphoton emission was taken
into account by an approximate analytic treatment~\cite{Baier:2005},
while the other effects were never accounted for and are, altogether,
actually larger than the difference between the theories. The first
Monte Carlo implementation of the Baier-Katkov description is
performed in the present work, allowing a comparison with measurements
on the same footing as was done for the Migdal one. Moreover, again
for the first time, the Baier-Katkov formulae are rewritten in a form
that allows to use the same value of the radiation length introduced
in the Migdal ones. Only then it is possible to find out whether the
data have enough sensitivity to show a clear preference. We cover all
available accelerator experiments for amorphous targets with good
quality: SLAC E-146~\cite{Anthony:1995,Anthony:1996,Anthony:1997},
CERN LPM~\cite{Hansen:2003,Hansen:2004}, and CERN
LOW-$Z$~\cite{Andersen:2013}.

From the point of view of applications, it should be stressed that, to
the best of our knowledge, all publicly available Monte Carlo codes
for particle physics (GEANT4~\cite{Schaelicke:2008},
EGS5~\cite{Kirihara:2010}, and EPICS~\cite{Kasahara:Epics}) or cosmic
rays (AIRES~\cite{Cillis:1999}, CORSIKA~\cite{Heck:1998}, and
COSMOS~\cite{Kasahara:Cosmos}) implement the Migdal cross
sections. The highest-energy single-photon spectra measured under
controlled laboratory conditions were collected by the LHCf
collaboration~\cite{Adriani:2011s} reaching up to $\approx 3$~TeV. The
authors found discrepancies in the shower profile between experiment
and EPICS simulations, especially in the initial part of the shower,
where the LPM suppression is strong. They attribute them to problems
in the channel-to-channel calibration or the description of the LPM
suppression (based on the Migdal approach, as mentioned). Important
motivations to improve as far as possible the basic electromagnetic
cross sections under strong LPM suppression, used in simulations,
derive also from the recent opening of PeV gamma ray
astronomy~\cite{Amenomori:2021s,Cao:2021s} and the present differences
between experiments in the measured electron and positron primary
cosmic ray fluxes in the TeV
region~\cite{Kobayashi:2012,Adriani:2018s}.

The paper is structured as follows. The available theoretical
descriptions of the LPM suppression are presented in
Sec.~\ref{sec:theo}, with particular emphasis on the Migdal (see
Sec.~\ref{sec:Migdal}) and Baier-Katkov (see Sec.~\ref{sec:BK})
ones. The reformulation of the Baier-Katkov approach to use a given
radiation length is described in Sec.~\ref{sec:BK_norm}. A first
simplistic comparison between the theories, without the Monte Carlo,
is discussed in Sec.~\ref{sec:th_comp}. Then the need for Monte Carlo
simulations is fully motivated in Sec.~\ref{sec:multiph} by showing
that the multiphoton effect (see Sec.~\ref{sec:dmultiph}) and the
pile-up with the background (see Sec.~\ref{sec:multiph_bksub}) are
more important than the difference between the approaches. The
implementation of the Monte Carlo code is briefly summarized in
Sec.~\ref{sec:imp}, while most of the technical details are reported
in Appendixes~\ref{sec:imp_D1_D2}, \ref{sec:BK_ldisc},
\ref{sec:BK_sigmatot}, and~\ref{sec:BK_samp}. The comparison with the
data is presented in Sec.~\ref{sec:exp} and, finally, the conclusion
is offered in Sec.~\ref{sec:con}.

\section{\label{sec:theo}The basic formulae for the different theoretical approaches}

The most important expressions that different authors proposed over
the years to account for bremsstrahlung at high energies are described
in the following with particular focus on those that take into account
quantum coherence effects. The basic quantity is the probability per
unit length $dp_{\gamma}/dx$ for an ultrarelativistic electron of
total energy $E$ to emit a photon of energy $k$, differential in the
fractional photon energy $x=k/E$, while crossing matter.  For a thin
amorphous and homogeneous target of number density $n$ this can be
related to the usual differential cross section per atom by the
expression
\begin{equation}
  \frac{dp_{\gamma}}{dx}=n\frac{d\sigma}{dx}
  \;\;.
  \label{eq:defp}
\end{equation}

The expressions may not coincide with those in the original papers and
are reported in the precise form utilized in our calculations since
they have been ``normalized'' (as far as possible) to a common
(modern) radiation length as given in Tsai~\cite{Tsai:1974,Tsai:1977},
in order to allow a meaningful detailed comparison.  The reader
interested in the historical development of the theory can find more
detailed information in the review papers quoted above.

\subsection {Bethe-Heitler}

As the simplest expression for the radiation probability in the
collision of a high-energy electron with a single atom we will adopt
the one appearing as Eq.~(3.84) in the review paper by
Tsai~\cite{Tsai:1974} (corresponding to Eq.~(11) by
Klein~\cite{Klein:1999}),
\begin{equation}
 \frac{dp^{\mathrm{BH}}_{\gamma}}{dx}=\frac{1} {3\,X_{0}\,x}
  \left\{x^{2}+2\,\left[1+(1-x)^{2}\right]\,\right\}
 \label{eq:BH}
\end{equation}
valid for the so-called ``complete screening limit''. In the following,
we refer to this result as the Bethe-Heitler (BH) expression.

For the radiation length $X_{0}$ we assume the numerical values
tabulated in Ref.~\cite{Tsai:1974} according to the definition
\begin{equation}
  \begin{split}
    X_{0}&=\left\{4\,n\,\alpha\,r_{\mathrm{e}}^{2}\,
    \left[{Z^{2}\left(\frac{\phi_1}{4}-\frac{1}{3}\ln(Z)
    -{f(\alpha\,Z)}\right)}+\right.\right.\\
    &\left.\left.
    {Z\left(\frac{\psi_1}{4}-\frac{2}{3}\ln(Z)\right)}\right]\right\}^{-1}
  \;\;,
  \end{split}
  \label{eq:x0}
\end{equation}
where $\alpha$ is the fine-structure constant, $r_\mathrm{e}$ the
classical electron radius and $Z$ the atomic number of the material.
The function $f(\alpha\,Z)$ describes the Coulomb correction which is
due to the distortion suffered by the impinging plane wave of the
electron as it approaches the atomic nucleus~\cite{Bethe:1954a}, while
$\phi_1$ and $\psi_1$ are $Z$-independent quantities characterizing
the amount of screening of the nuclear and electrons electric
field. Their detailed expressions, calculated using the Moli\'{e}re
representation of the Thomas-Fermi model, are also given in
Ref.~\cite{Tsai:1974}.  In particular, in case of complete screening,
$-\phi_1/4=5.216=\ln(184)\equiv \ln(B)$~\footnote{Slightly different
  numbers for $B$ appear in the literature, ranging from
  183~\cite{Baier:1998} to 190~\cite{Migdal:1956}; in the following we
  will assume B=$184$ even when quoting (or referring to) equations of
  the original papers.}. For later use we also define a ``simplified''
radiation length $X^{\mathrm{c}}_{0}$ which does not include the
Coulomb correction $f(Z\alpha)$ and the inelastic scattering term
(proportional to $Z$), namely
\begin{equation}
  X^{\mathrm{c}}_{0}=[4\,n\,\alpha\,r_\mathrm{e}^{2}\,Z^{2}\,
  \ln(B/Z^{\frac{1}{3}})]^{-1}
  \;\;.
  \label{eq:x0c}
\end{equation}
Equation~(\ref{eq:BH}) constitutes a very useful reference commonly
adopted to characterize quantitatively the amount of radiation
suppression occurring when the influence of the medium on the
bremsstrahlung process is taken into account~\cite{Baier:2000a}.

\begin{figure}[t!!!!]
  \centering
  \begin{tabular}{c}
    \includegraphics[width=.425\textwidth]{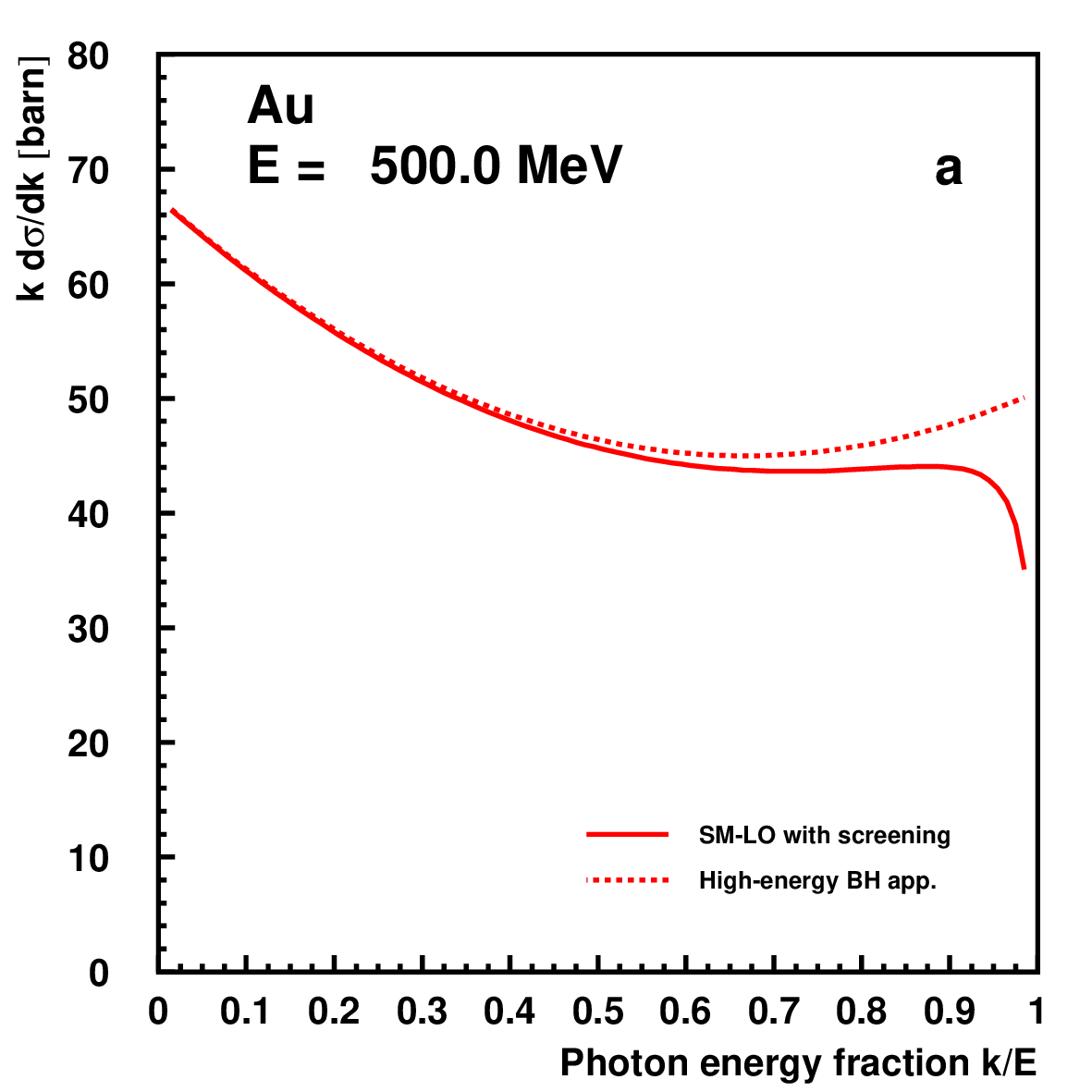}\\
    \includegraphics[width=.425\textwidth]{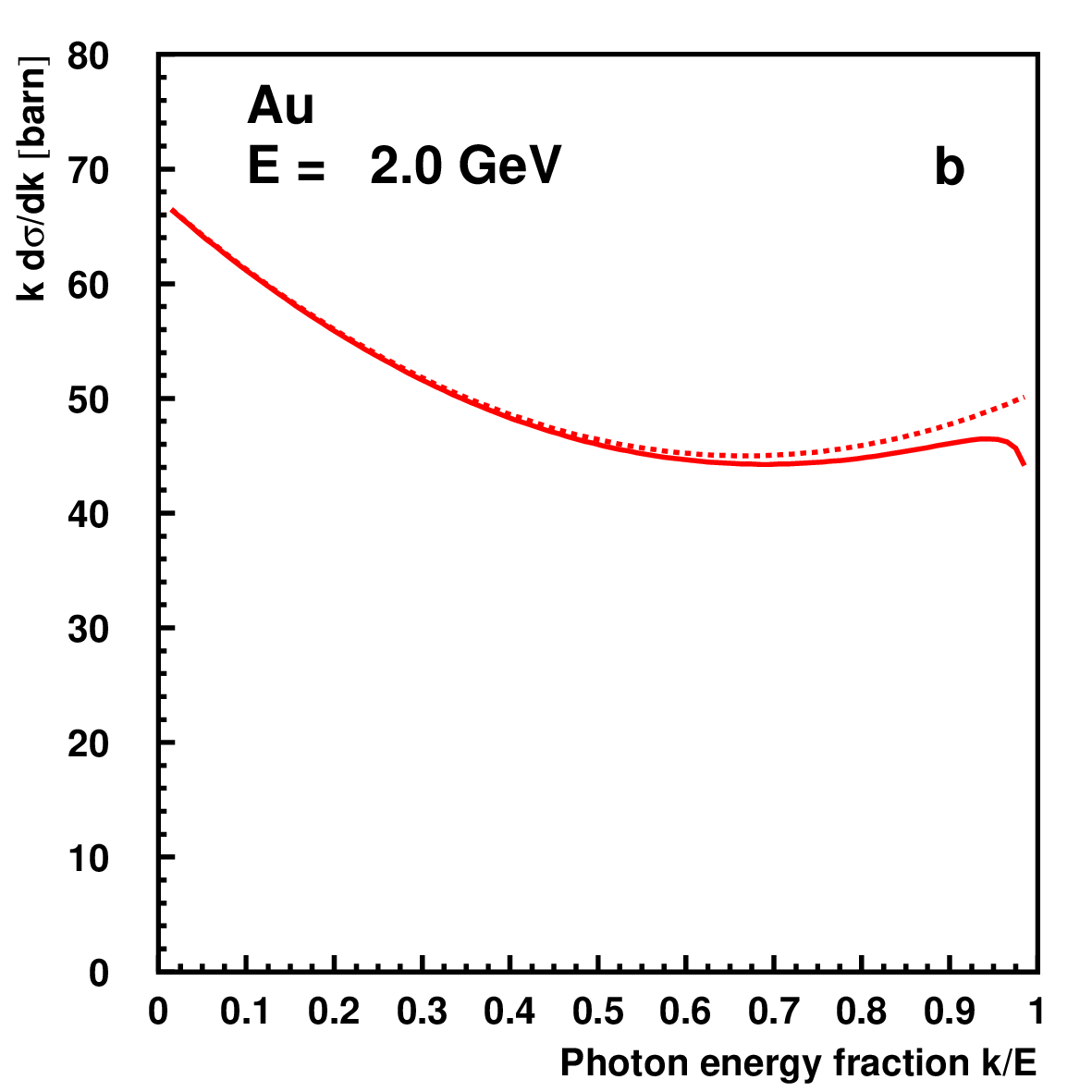}\\
  \end{tabular}
  \caption{\label{fig:BH_app}Comparison of the best available
    bremsstrahlung cross sections (SM-LO continuous line), obtained
    with the Furry-Sommerfeld-Maue wave functions at the leading
    order, with the BH high-energy approximation (dashed line) given
    in Eq.~(\ref{eq:BH}). The power spectrum $k\,d\sigma/dk$ is
    actually shown to cancel the BH divergence at low values of
    $k$. The SM-LO results take into account the screening correction
    with the Olsen-Maximon-Wergeland additivity rule and the atomic
    form factors by Hubbell et al.~\cite{Hubbell:1975,Hubbell:1977}.}
\end{figure}

To arrive at Eq.~(\ref{eq:BH}), it is necessary to neglect a term
which does not scale with $X_{0}$ [compare Eq.~(3.83) with Eq.~(3.84)
in Ref.~\cite{Tsai:1974}] amounting to a maximum reduction of cross
section for $x\approx 0$ between $\approx 1.6\%$ at low $Z$ and
$2.4\%$ at high $Z$. Moreover, the complete screening condition,
implying that the atomic elastic and inelastic form factors are
evaluated in the low momentum transfer limit, is applied for all
values of $x$. While this is certainly correct in the high-energy
limit for $E$ and at small $x$, it is bound to fail when $x\approx 1$
(i.e.~at the high-energy end of the spectrum, denoted as
short-wavelength limit (SWL) in the rest of this work), when the
momentum transfer is not small. It is then interesting to test
Eq.~(\ref{eq:BH}) against a more complete calculation, including the
Coulomb correction at the leading
order~\cite{Bethe:1954a,Elwert:1969,Roche:1972} with the
Furry-Sommerfeld-Maue wave
functions~\cite{Furry:1934,Sommerfeld:1935a} and screening with the
Olsen-Maximon-Wergeland~\cite{Olsen:1957} additivity rule. Such an
approach was recently validated against accurate experimental data
measured for $500$-MeV electrons at MAMI in
Ref.~\cite{Mangiarotti:2021} and using exact partial-wave calculations
at low energies~\cite{Jakubassa-Amundsen:2019a}. The comparison is
shown in Fig.~\ref{fig:BH_app}, adopting the realistic
Hartree-Fock-Slater atomic form factors by Hubbell et
al.~\cite{Hubbell:1975,Hubbell:1977}. When $x$ is small,
Eq.~(\ref{eq:BH}) is indeed accurate as expected and any limitation
due to the Moli\'{e}re approximation of the form factors or the
exclusion of the term not proportional to $X_{0}$ is hardly
visible. However, for $E=500$~MeV the discrepancy increases for
$x\gtrsim 0.5$, reaching $\approx 10\%$ for $x\approx 0.9$. It grows
further for $x>0.9$. For $E=2$~GeV, the situation is better and at
$x\approx 0.9$ the discrepancy is about half the value quoted for
$E=500$~MeV.

\subsection{\label{sec:fl}The formation  length}

The BH expression for the probability of emission of a quantum of
radiation of energy $k$ refers to the interaction of a high-energy
electron with a single atom and as such does not take into account the
effect of the environment (typically represented by the presence of
nearby atoms in a target bombarded by an electron beam) on the
radiative process.  The idea of a formation (or coherence) length of
the photon, which was first conceived by Ter-Mikaelyan in the early
fifties of the past century~\cite{TerMikaelian:1953b}, proved to be of
crucial importance in providing physical insight in the description of
high-energy electromagnetic processes and in estimating the effect of
the different influences of the environment or medium on the radiated
energy spectrum. The concept was described at length in the old and
recent literature (see
e.g. Refs.~\cite{Feinberg:1956,Galitsky:1964,TerMikaelian:1972,Klein:1999,Baier:2000a}).
In its simplest form, which considers an electron, with a high energy
$E$, colliding on a single atom and neglects the small scattering
angle of the order $mc^{2}/E=1/\gamma$, it is based on the recognition
that for $\gamma\gg1$, the minimum value of the uncertainty of the
component of the momentum delivered to the atom, parallel to the
momentum of the impinging electron, is given approximately by the
expression~\footnote{This relation (as several others in the
  following) is only valid up to photon energies $k$ such that
  $(E-k)\gg mc^2$ i.e.~not too close to the SWL.}
\begin{equation}
  q_{\parallel}=\frac{m^{2}c^{3}\,x}{2E\,(1-x)}
  \;\;.
  \label{eq:qpar}
\end{equation}
It follows, from the Heisenberg uncertainty relation, that the actual
longitudinal dimension of the region in which the photon is created is
of the order
\begin{equation}
  l_{\mathrm{f}0}(E,k)=\frac{\hbar}{q_{\parallel}}=
  \frac{2\hbar\,E\,(E-k)}{m^{2}c^{3}\,k}
  \;\;.
  \label{eq:lf}
\end{equation}

This is usually called the ``vacuum formation length'' and is thought
of as the distance where coherence of the emitted wavelets is
maintained. It can be shown (see Ref.~\cite{Galitsky:1964}) that the
radiation intensity (as well as the cross section per atom
differential in the photon energy) emitted in the whole solid angle,
for any given $k$ and $E$, is proportional to the formation length,
meaning that any influence of the medium in altering this length will
reflect as a factor on the radiated intensity. Limiting the analysis
to radiative electromagnetic processes occurring in amorphous media,
it turns out that the presence of the medium inevitably results in a
shortening of the formation length compared to
$l_{\mathrm{f}0}$. Strictly speaking, this only applies to the nuclear
bremsstrahlung process, namely the one having a $Z^{2}$ dependence,
while the possible suppression of the bremsstrahlung involving atomic
electrons is still an open question, perhaps to be investigated by
studying bremsstrahlung on low-$Z$ elements.  In turn, the relation
$l_{\mathrm{f}}\leq l_{\mathrm{f}0}$ suggests that the BH expression,
given above for the emission probability in a collision with a single
atom, is to be interpreted as a kind of maximum for the radiation
intensity (see e.g. Fig.~1 in Ref.~\cite{Feinberg:1956}). Such an
upper limit must be respected by any theory utilizing the same basic
physical ingredients and parameters (like electron screening and
radiation length) as Eq.~(\ref{eq:BH}) and aimed at quantitatively
describing the influence of the medium in reducing (by a factor
depending both on $E$ and $k$) the intensity of the associated
bremsstrahlung. It is also worth noting that, for a fixed energy $E$
of an electron impinging on a given medium, the radiation probability
given in Eq.~(\ref{eq:BH}) is proportional to the formation length
$l_{\mathrm{f}0}$ (up to first-order terms in $x=k/E$). This further
supports the idea of an upper limit for the radiation emission,
although no definite and rigorous argument valid for the whole
radiated spectrum is known to the authors.

Leaving aside the case in which an external field is present (typical
of the facilities for synchrotron radiation), the two most important
``influences'' of the environment in modifying the formation length
and the ones most relevant to the present work, are the multiple
scattering affecting an electron during the time it takes for the
photon to be formed~\cite{Landau:1953a,Landau:1953b} and the so-called
``polarization''
effect~\cite{TerMikaelian:1953a,TerMikaelian:1953b}. The main features
of the two processes are briefly discussed hereafter.  As is well
known, the average multiple scattering angle $\theta_{\mathrm{ms}}$
for an electron of energy $E$ crossing a distance $d$ of a medium
characterized by the radiation length $X_{0}$ is given by
$\theta_{\mathrm{ms}}=E_{\mathrm{ms}}/E\,\sqrt{d/X_{0}}$ where
$E_{\mathrm{ms}}=mc^{2}\,\sqrt{4\pi/\alpha}$.  This results in an
additional contribution to the longitudinal momentum transferred to
the nucleus (to be added to that given in Eq.~(\ref{eq:qpar})) which
is of the order $\simeq k\,\theta_{ms}^{2}/c$ and thereby to an
overall formation length $l_{\mathrm{f}}<l_{\mathrm{f}0}$. Obviously,
it is expected that the multiple scattering will be important when
this contribution to $q_{\parallel}$ exceeds the quantity in
Eq.~(\ref{eq:qpar}). In particular, when the latter is much smaller
than the former, the coherence length is found to be given by the
expression
$l_{\mathrm{f}}=mc^{2}/E_{\mathrm{s}}\,\sqrt{2\,X_{0}\,l_{\mathrm{f}0}}$
and a suppression factor due to the multiple scattering can be defined
as
\begin{equation}
  S_{\mathrm{ms}}(E,k)\equiv\frac{l_{\mathrm{f}}}{l_{\mathrm{f}0}}=
  \frac{m^{2}c^{4}}
  {E_{\mathrm{s}}\,\sqrt{\frac{k\,X_{0}}{E(E-k)\hbar c}}}
  \;\;,
  \label{eq:ms}
\end{equation}
which actually reduces the section of the spectrum below a photon
energy $k_{LPM}\simeq(E^{2}\,E_{s}^{2}\,\hbar
c)/(m^{4}c^{8}\,X_{0})$. For instance, $k_{LPM}\simeq 125$ MeV for
$25$-GeV electrons in lead.

The polarization effect (also called the ``longitudinal density
effect'') is caused by a phase change in the wave function of the
emitted photon due to a sequence of forward Compton scatterings on the
electrons of the medium. This can be macroscopically described to a
good approximation (see page~128 in Ref.~\cite{TerMikaelian:1972}) by
considering the modified phase velocity of the photon $v=c/{\cal N}(k)
$ where $ {\cal N}(k) $ is the refraction index of the medium for the
photon energy $k$. The relevant medium parameter in this case is the
well-known plasma frequency $\omega_p$ given by
$\omega_p=\sqrt{(4\pi\,n\,Z\,\alpha\,\hbar c^{3})/(mc^{2})}$.  For
photon frequencies far exceeding the atomic ones (as is here the case,
the minimum photon energy we are interested in being $\simeq 1$ MeV)
one can write ${\cal N}(k)=1-k_{\epsilon}^{2}/k^2$ where
$k_{\epsilon}=\hbar\,\omega_{\mathrm{p}}$.  Again the process implies
the addition of one more term to the longitudinal momentum imparted to
the atom leading to a reduced formation length and finally resulting
in an attenuation factor
\begin{equation}
  S_{d}(E,k)=\frac{k^{2}}{k^{2}+(\gamma\,k_{\epsilon})^{2}}
  \;\;.
\end{equation}
Clearly, this effect is particularly important for photon energies
$k\leq k_{\mathrm{p}}=\gamma\,k_{\epsilon}$.  As it turns out (see
page~118 in Ref.~\cite{TerMikaelian:1972}), the polarization effect
significantly reduces the intensity of the photon spectrum only for
$k\leq 10^{-4}$--$10^{-5}E$.

We only have given a short summary of the individual main effects
leading to a reduced intensity of the radiated photon spectrum,
whereas they should be simultaneously considered in a proper
description of their influence on the shape of the photon spectrum.
Such a discussion can be found in Ref.~\cite{Galitsky:1964}.  Finally,
it should also be clear that the above considerations are only meant
to supply a general guide to the shape of the photon spectrum to be
expected in the bremsstrahlung process for high-energy electrons
crossing an amorphous medium, while detailed theories are needed for a
quantitative comparison with experimental data. The description of
some of them is summarized in the following sections.

\subsection{\label{sec:Migdal}The Migdal approach}

In the papers by Landau and
Pomeranchuk~\cite{Landau:1953a,Landau:1953b}, starting from the
classical expression for the energy radiated (per unit solid angle and
unit frequency interval) by a charged particle moving along a path
$\textbf{r}(t),-\infty<t<\infty$ (see e.g.~page~676 in
Ref.~\cite{Jackson:1998}) approximate expressions were given for the
modification to the BH photon energy spectrum induced by the multiple
scattering of the particle in crossing the formation length
$l_{\mathrm{f}0}$ in a medium. Shortly after,
Migdal~\cite{Migdal:1954}, still in a completely classical framework,
derived a kinetic equation of the Fokker-Planck type to be obeyed
(when the momentum of the radiated photon is neglected) by the
distribution function of electrons suffering multiple scattering in
atomic collisions determined by a screened Coulomb potential. The
solution of this equation (see page~164 in
Ref.~\cite{TerMikaelian:1972}) leads to an expression describing the
reduced intensity of the low-energy section of the photon spectrum
fully consistent with the reduction factor given in Eq.~(\ref{eq:ms})
above.

\subsubsection{Effect of multiple scattering}

The decisive step in obtaining a fully quantum-mechanical description
of the bremsstrahlung process (taking into account the momentum of the
radiated photon and the influence of the medium) and one valid over
the whole radiated spectrum was taken by Migdal in
Ref.~\cite{Migdal:1955}. Indeed, he realized that the state of the
radiating particle in presence of multiple scattering is of mixed type
and therefore properly described by a density matrix in the momentum
representation of the positive energy spinor eigenfunctions of the
free-particle Hamiltonian. A quantum kinetic equation obeyed by the
density matrix (including the spin degree of freedom and averaged over
the spatial coordinates of the scattering centers) was set up in
Ref.~\cite{Migdal:1955} using the Born approximation. This was then
utilized in Ref.~\cite{Migdal:1956} where the transition probability
from the initial to the final state was connected to the density
matrix. The integral equation satisfied by the average density matrix
is then simplified by using a series expansion up to second order in
the small parameter given by the change in the momentum component of
the electron, transverse to the photon momentum, in a single
scattering process obtaining again a differential equation of the
Fokker-Planck type (see Eq.~(24) in Ref.~\cite{Migdal:1956}).

After a long series of transformations, this equation is solved in
terms of the functions $\Phi(s)$ and $G(s)$ (to be discussed below)
where the basic quantity $s$, defined by (see Eq.~(47) in
Ref.~\cite{Migdal:1956})
\begin{equation}
  s=\sqrt{\frac{m^{2}c^{4}\,X_{0}\,\alpha\,k}
  {4\pi\,\hbar c\,8\,E\,(E-k)}}
  \;\;,
  \label{eq:M_s1}
\end{equation}
essentially expresses (when polarization is neglected) the square root
of the ratio of the minimum longitudinal momentum given in
Eq.~(\ref{eq:qpar}) to the additional longitudinal momentum
transferred to the nucleus due to multiple scattering (see also
pages~157-158 in Ref.~\cite{TerMikaelian:1972}). It follows that
$s\gg1$ corresponds to the high-energy section of the photon spectrum,
where the effect of the multiple scattering is negligible, while
$s\ll1$ characterizes the suppressed low-energy section of the
spectrum.

A further refinement, aimed at taking into account the weak energy
dependence of the ratio of the maximum to minimum scattering angle in
the single bremsstrahlung process, led Migdal to define implicitly a
parameter $s_{\mathrm{M}}$ slightly different from that given in
Eq.~(\ref{eq:M_s1})
\begin{equation}
  s_{\mathrm{M}}\equiv\frac{s}{\sqrt{\xi(s_{\mathrm{M}})}}=
  \sqrt{\frac{m^{2}c^{4}\,X_{0}\,\alpha\,k}
  {8\,E\,(E-k)\,4\pi\,\hbar c\,\xi(s_{\mathrm{M}})}}
  \;\;.
  \label{eq:M_s}
\end{equation}
The function $\xi(s_{\mathrm{M}})$ is a slowly varying monotonic
function of $s_{\mathrm{M}}$ defined by
\begin{equation}
  \xi(s_{\mathrm{M}})=
  \begin{cases}
    2                            & \text{if $s\leq s_{1}$} \\
    1+\ln(s_{\mathrm{M}})/\ln(s_1) & \text{if $s_{1}<s_{\mathrm{M}}<1$} \\
    1                            & \text{if $s_{\mathrm{M}}\geq 1$}
  \end{cases}
  \;\;.
  \label{eq:M_xi}
\end{equation}

In Migdal's paper the quantity $s_{1}$ was defined as
$s_{1}=(Z^{\frac{1}{3}}/190)^{2}$, while in order to comply as far as
possible with our constraint to consistently refer to a common value
for the radiation length, we must adopt as explained in detail in
Ref.~\cite{Mangiarotti:2011} the value defined by
$\ln(s_{1})=-\left(2\,\alpha\,r_\mathrm{e}^{2}\,Z^{2}\,X_{0}\,n\right)^{-1}$.
Notice that the first derivative of $\xi(s_{\mathrm{M}})$ is
discontinuous for $s_{\mathrm{M}}=1$ and $s_{\mathrm{M}}=s_{1}$.
Equation~(\ref{eq:M_s}) should in principle be solved recursively for
$s_{\mathrm{M}}$.  As a final expression for the probability per unit
target length of emitting a photon differential in the fractional
photon energy $x=k/E$, the following relation is adopted:
\begin{equation}
  \frac{dp^{\mathrm{M}}_{\gamma}}{dx} =
  \frac{\xi (s_{\mathrm{M}})}{3\,X_{0}\,k}\,
  \left\{x^{2}\,G(s_{\mathrm{M}})
    +2\,\left[1+(1-x)^{2}\right]\,
    \Phi(s_{\mathrm{M}})\right\}
  \;\;.
  \label{eq:M_dsdk_x0}
\end{equation}
The functions $\Phi(s)$ and $G(s)$, introduced by
Migdal~\cite{Migdal:1956}, describe the influence of the LPM effect:
\begin{subequations}
  \begin{eqnarray}
    \Phi(s)=&&6\,s^{2}\left(-\frac{\pi}{4}+
      \int_{0}^{\infty}dz\,
      e^{-s\,z}\,\frac{\sin(s\,z)}
      {\tanh(z)}\right)
    \;\;,\\
    G(s)=&&12\,s^{2}\left(\frac{\pi}{4}-
      \int_{0}^{\infty}dz\,
      e^{-s\,z}\,\frac{\sin(s\,z)}
      {\sinh(z)}\right)
    \;\;.
  \end{eqnarray}
  \label{eq:M_PhiG}
\end{subequations}
It can be seen that $G(s)$ and $\Phi(s)$ approach unity for high
values of the variable $s$ and in this limit Eq.~(\ref{eq:M_dsdk_x0})
goes over to Eq.~(\ref{eq:BH}).

As a matter of fact, the original treatment by Migdal, whose solution
of the quantum kinetic equation for the density matrix is based on the
Born approximation, did not include the Coulomb correction, which is,
however, automatically brought in when a modern value of $X_{0}$ is
adopted in Eq.~(\ref{eq:M_dsdk_x0}). Such a procedure has been applied
since the first detailed comparison with experimental data performed
by Anthony at al.~\cite{Anthony:1995,Anthony:1997}, but received a
full theoretical support only recently in the work by Voskresenskaya
et al.~\cite{Voskresenskaya:2014}. As a first step, the Moli\'{e}re
multiple scattering theory was improved to take into account the
Coulomb correction in the elastic cross section~\cite{Kuraev:2014}.
This was then used for the calculation of the LPM suppression. From
Fig.~2 of Ref.~\cite{Voskresenskaya:2014}, it is possible to see that
the approximation of a Coulomb correction independent from $x$ (or
$s$), as implied by Eq.~(\ref{eq:M_dsdk_x0}), is only approximately
true. However, for gold, the largest variation shown is few percent of
the asymptotic value of the Coulomb correction, which is only several
percent of the total cross section, well below present experimental
uncertainties.

\subsubsection{\label{Migdal_pol}Effect of the polarization of the medium}

In his fundamental paper~\cite{Migdal:1956}, Migdal tackled also the
problem of including the polarization effect at low photon
energies. To this end, he introduces the quantity
$\Gamma=1+(k_{\mathrm{p}}/k)^{2}$ where
$k_{\mathrm{p}}=\gamma\,\hbar\,\omega_{\mathrm{p}}$.  It was then
shown that to account for the suppression due to the polarization
effect, it suffices to replace $\Phi(s_{\mathrm{M}})$ in
Eq.~(\ref{eq:M_dsdk_x0}) with $\Phi(s_{\mathrm{M}}\Gamma)/\Gamma$
neglecting the term containing the function $G$ which is of minor
importance at low photon energies.  When the substitution
$s_{\mathrm{M}}\rightarrow s_{\mathrm{M}}\,\Gamma$ is applied
consistently to all the $s_{\mathrm{M}}$-dependent quantities
appearing in Eq.~(\ref{eq:M_dsdk_x0}), we obtain
\begin{equation}
  \begin{split}
    \frac{dp^{\mathrm{M}}_{\gamma}}{dx} =&
    \frac{\xi(s_{\mathrm{M}}\,\Gamma)}{3\,X_{0}\,x}\,
    \bigg\{x^{2}\,\frac{G(s_{\mathrm{M}}\,\Gamma)}{\Gamma^{2}}\\
    &+2\,\left[1+\left(1-x\right)^{2}\right]\,
    \frac{\Phi(s_{\mathrm{M}}\,\Gamma)}{\Gamma} \bigg\} \;\;.
  \end{split}
  \label{eq:M_dsdk_diel}
\end{equation}

\subsection{\label{sec:BK}The Baier--Katkov approach}

Baier and Katkov (BK in the following) also studied the LPM effect in
a series of
papers~\cite{Baier:1998,Baier:1999a,Baier:1999b,Baier:2000a}, trying
to improve on the original treatment by Migdal~\cite{Migdal:1956} and
including several additional features like boundary radiation and
multiphoton effects. In order to simplify the comparison with other
theoretical approaches, we restrict the description of their work to
the case of a semi-infinite target. In addition, the polarization
effect will be neglected for the moment and introduced at a later
stage.

\subsubsection{\label{sec:BK_nodiel}Effect of multiple scattering}

Starting from the general expression (based on the quasiclassical
operator method~\cite{BKS:1998}) for radiation emission by high-energy
electrons along a given classical trajectory, Baier and Katkov solve
the problem of averaging over all possible trajectories in an
amorphous medium by using the classical kinetic equation method
arriving at the following expression for the mean transition
probability differential in the fractional photon energy $x=k/E$:
\begin{equation}
  \begin{split}
    \frac{dp^{\mathrm{BK}}_{\gamma}}{dx}=&\frac{2\alpha}{\gamma^{2}}
    \,\Re\bigg[\int_{0}^{\infty}dt\,\exp(-i\,t)\\
    &\Big(R_{1}\,\varphi_{0}(0,t)
    +R_{2}\,\vect{p}\cdot\vect{\varphi}(0,t)\Big)\bigg] \;\;,
  \end{split}
  \label{initBK}
\end{equation}
where
\begin{subequations}
  \begin{eqnarray}
    R_{1}=&&\frac{x^{2}}{1-x}
    \;\;,\\
    R_{2}=&&\frac{1+(1-x)^{2}}{1-x}
    \;\;,
  \end{eqnarray}
  \label{eq:BK_R1R2}
\end{subequations}
and $\vect{p}$ is the momentum operator in the two-dimensional
transverse (to the original electron momentum) space and the functions
$\varphi_{\mu}\equiv(\varphi_{0},\vect{\varphi})$ satisfy a
Schr{\"o}dinger-type equation in the two-dimensional space of the
impact parameter $\vect{\rho}=\vect{b}/\gamma$ where $\vect{b}$ is
measured in units of the reduced Compton wavelength
$\lambdabar_{\mathrm{c}}=\hbar/(mc)$.

The Schr{\"o}dinger-type equation is arrived at by introducing the
Fourier transform of the scattering cross section on a screened
Coulomb potential evaluated in the Born approximation and reads (using
nondimensional variables as detailed in Ref.~\cite{Baier:2005})
\begin{equation}
  \begin{split}
    i\frac{\partial\varphi_{\mu}}{\partial t}=
    &\cal{H}\,\varphi_{\mu}
    \;\;,\\
    \mathcal{H}=&\vect{p}^{2}-i\,V(\rho)
    \;\;,\\
    \vect{p}=&-i\,\nabla_{\rho}
    \;\;,
  \end{split}
  \label{eq:BK_sch}
\end{equation}
to be solved using the initial conditions
$\varphi_{0}(\vect{\rho},0)=\delta(\vect{\rho})$ and
$\vect{\varphi}(\vect{\rho},0)=\vect{p}\,\delta(\vect{\rho})$.

In Eq.~(\ref{eq:BK_sch}), the ``potential'' $V$ (which, due to the
azimuthal symmetry around the initial electron momentum, depends only
on the magnitude of $\vect{\rho}$) reads
\begin{equation}
  V(\rho)=-Q\rho^{2}
  \left[\ln\left(\gamma^{2}\,\theta_{1}^{2}\right)+
  \ln\left(\frac{\rho^{2}}{4}\right)+2\,C-1\right]
  \;\;,
  \label{eq:BK_v}
\end{equation}
where $Q=(2\pi\,(\hbar
c)^{3}\,Z^{2}\,\alpha^{2}\,n\,E(1-x))/(m^{4}c^{8}\,x)$, $C$ is the
Euler-Mascheroni constant ($C=0.577..$), and $\theta_{1}=(\hbar
c)/(E\,a_{\mathrm{s}})$ is the angular parameter appearing in the
scattering cross section ($a_{\mathrm{s}}=0.81\,Z^{-1/3 }$\AA\ is the
screening radius). The potential $V$ is then refined to take into
account the Coulomb correction by defining a parameter
$\theta_{2}=\theta_{1}\,e^{f(Z\,\alpha)-1/2}$ where $f$ is the
Bethe-Maximon Coulomb correction rederived, by a different method, in
Appendix A of Ref.~\cite{Baier:1998}.  The authors then determine the
effective impact parameter $\rho_{\mathrm{c}}$ giving the main
contribution to the cross section. As it turns out, when the
root-mean-square scattering angle over a distance of a formation
length is much less than $1/\gamma$, one can assume
$\rho_{\mathrm{c}}=1$, whereas, in the opposite case,
$\rho_{\mathrm{c}}$ can be found by solving the transcendental
equation
\begin{equation}
  4\,Q\,\rho_{\mathrm{c}}^{4}\,\ln
  \left(\frac{1}{\gamma^{2}\,\theta_{2}^{2}\,
  \rho_{\mathrm{c}}^{2}}\right)=1
  \;\;,
  \label{rhoc}
\end{equation}
which always results in a value $\rho_{\mathrm{c}}\leq1$. The authors
then define the all-important function $L_{\mathrm{c}}\equiv
L(\rho_{\mathrm{c}})=-\ln(\gamma^{2}\,\theta_{2}^{2}\,\rho_{\mathrm{c}}^{2})$.
This function has an obvious minimum value $L_{1}\equiv
L(1)=-\ln(\gamma^{2}\,\theta_{2}^{2})$ which is estimated as
\begin{equation}
  L_{1}=2\ln(B Z^{-\frac{1}{3}})-f(Z\alpha)
  \;\;.
  \label{eq:BK_L1}
\end{equation}
The function $L_{\mathrm{c}}$ has also a maximum value since, if
$\rho_{\mathrm{c}}$ becomes comparable to the nuclear radius
$R_{\mathrm{n}}$, the form of the potential $V(\rho) $ changes
completely (it acquires a harmonic oscillator form~\cite{Baier:2005})
and the expressions derived below would no longer be valid. The
maximum value of $L_{\mathrm{c}}$ is currently estimated as $2L_{1}$.
The next step is to decompose the potential $V(\rho)$ in
Eq.~(\ref{eq:BK_v}) (with the substitution $\theta_{1}\rightarrow
\theta_{2}$) as a sum of two terms, a main one given by
$V_{\mathrm{c}}=Q\,L(\rho_{\mathrm{c}})\rho^{2}$ and a minor one (that
will be treated as a perturbation) given by
$v(\rho)=-Q\,\rho^{2}(2\,C+\ln(\rho^{2}/(4\,\rho_{\mathrm{c}}^{2})))$
(notice that a term $\ln(1/\rho_{\mathrm{c}}^{2})$ has been added and
subtracted in Eq.~(\ref{eq:BK_v}) so that the final result for the
total radiation probability will be independent of $\rho_{\mathrm{c}}$
up to the first order in $v_{\mathrm{c}}$).  Through a lengthy and
complex calculation, Baier and Katkov were able to solve
Eq.~(\ref{eq:BK_sch}), where only the main term $V_{\mathrm{c}}$ is
considered, and after inserting this solution into Eq.~(\ref{initBK})
arrive at the following expression for the probability of radiation
per unit length, differential in the fractional photon energy
$x=k/E$~\footnote{A factor $\pi$ is missing in the denominator of
  Eq.~(2.12) in Ref.~\cite{Baier:2005} as well as a factor
  $\phi_{\mu}(x,\tau)$ on the right hand side of Eq.~(A22) in the same
  paper.},
\begin{equation}
  \frac{dp^{\mathrm{BKM}}_{\gamma}}{dx}=\frac{\alpha\,\nu_{0}^{2} \,mc^{2}}
  {12\pi\,\gamma\,\hbar c}\,
  \left[R_{1}\,G\left(\frac{s_{\mathrm{BK}}}{2}\right)+
  2\,R_{2}\Phi\left(\frac{s_{\mathrm{BK}}}{2}\right)\right]
 \;\;,
 \label{eq:BK_main}
\end{equation}
where $\nu_{0}^{2}=4\,Q\,L_{\mathrm{c}}$,
$s_{\mathrm{BK}}=1/(\sqrt{2}\,\nu_{0})=1/(2\,\sqrt{2\,Q\,L_{\mathrm{c}}})$,
$R_{1}$ and $R_{2}$ have been defined in Eqs.~(\ref{eq:BK_R1R2}), and
the functions $G$ and $\Phi$ are those defined in
Eqs.~(\ref{eq:M_PhiG}).  A discontinuity in the slope of the photon
energy spectrum is expected for the minimum value of $k$ resulting in
$\rho_{\mathrm{c}}=1$.

\begin{figure}[t!!!!]
  \centering
  \begin{tabular}{c}
    \includegraphics[width=.425\textwidth]{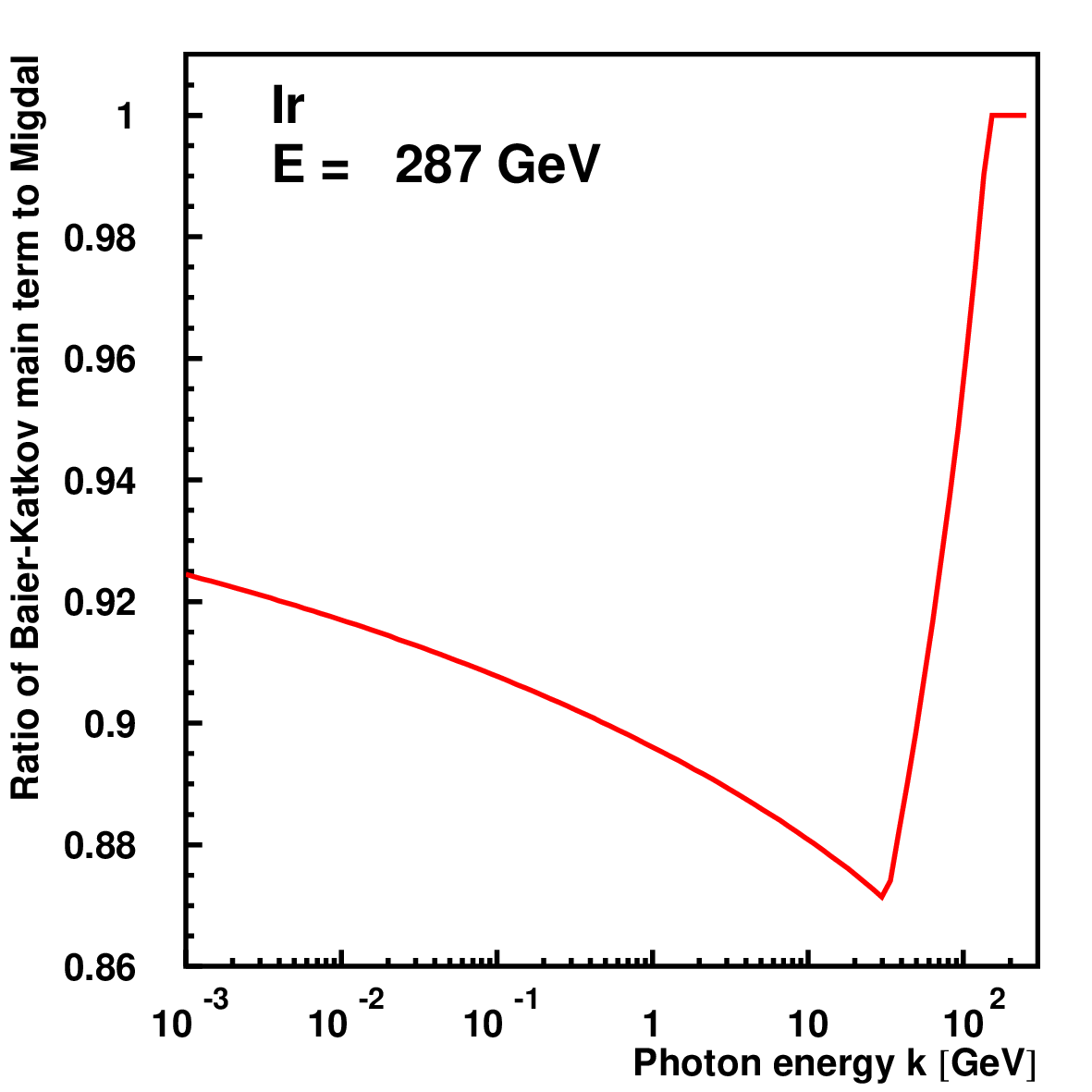}
  \end{tabular}
  \caption{\label{fig:BK_ratio_M}Ratio of the main term of the BK
    radiation intensity to the Migdal one, see Eq.~(\ref{eq:erre}). To
    compare both approaches on the same footing, the Coulomb
    correction is omitted from the BK expression and
    $X^{\mathrm{c}}_{0}$ is used for $X_{0}$ in both
    Eq.~(\ref{eq:M_s}) and Eq.~(\ref{eq:M_dsdk_x0}).}
\end{figure}

It is interesting to compare the radiation intensity given in
Eq.~(\ref{eq:M_dsdk_x0}) by Migdal with that given in
Eq.~(\ref{eq:BK_main}) by Baier and Katkov. However, since the Coulomb
correction and inelastic scattering were not included in the Migdal
theory, a proper comparison is only obtained by setting $f(Z\alpha)=0$
in Eq.~(\ref{eq:BK_L1}) and substituting the previously defined (in
Sec.~\ref{sec:theo}) $X^{\mathrm{c}}_{0}$ for $X_{0}$ in both
Eq.~(\ref{eq:M_s}) [which influences the definition of
$\xi(s_{\mathrm{M}})$] and Eq.~(\ref{eq:M_dsdk_x0}).  The ratio $R$ of
the main term in the BK approach to the probability of the Migdal
theory is given by
\begin{equation}
  R=\frac{L_{1}-\ln(\rho_{\mathrm{c}}^{2})}
  {\xi(s_{\mathrm{M}})\,\ln(B/Z^{\frac{1}{3}})}
  \;\;.
  \label{eq:erre}
\end{equation}
As an example, the value of $R$ as a function of the photon energy $k$
for a $287$-GeV electron beam crossing an iridium medium is shown in
Fig.~\ref{fig:BK_ratio_M}.  It is seen that differences up to $13$~\%
occur, the Migdal approach always lying above the main term of the BK
theory. The discontinuities of the first derivative of $R$ around
$k=30$~GeV and $k=153$~GeV are due to the discontinuities of the
derivative of $\rho_{\mathrm{c}}(k)$ (at $\rho_{\mathrm{c}}=1$) and of
$\xi(s_{\mathrm{M}})$ (at $s_{\mathrm{M}}=1$), respectively.

As anticipated above, the correction $v(\rho)$ to the potential is
treated perturbatively by expanding the electron propagator and
keeping terms up to first order in $v(\rho)$.

The final expression for the correction term to the radiation
probability requires the definition of several functions such as
\begin{equation}
  \begin{split}
    G_{\mathrm{BK}}(z)=&\int_{0}^{z}(1-y\coth(y))dy\\
    =&z-\frac{z^{2}}{2}-\frac{\pi^{2}\,z}{12}
    \ln\left(1-e^{-2z}\right)+
    \frac{\mathrm{Li}_{2}(e^{-2z})}{2}
    \;\;,\\
    \\
    g_{\mathrm{BK}}(z)=&z\cosh(z)-\sinh(z)
    \;\;,\\
    \\
    d(z)=&\mathrm{H}(1-\nu_{0})\ln(\nu_{0})- \ln\big(\sinh(z)-C\big)\,
    g_{\mathrm{BK}}(z)\\
    &-2\,G_{\mathrm{BK}}(z)\cosh(z) \;\;,
  \end{split}
\end{equation}
where $\mathrm{H}$ and $\mathrm{Li}_{2}$ represent the Heaviside unit
step function and the dilogarithmic function, respectively.

In addition, it is necessary to introduce the functions
$D_{1}(\nu_{0})$ and $D_{2}(\nu_{0})$ via the following integral
representations
\begin{subequations}
  \begin{eqnarray}
    D_{1}(\nu_{0})=&&\int_{0}^{\infty}dz\,
    \frac{\exp(-p\,z)}{\sinh^{2}(z)}\nonumber\\
    &&\left[d(z)\sin(p\,z)+\frac{\pi}{4}g(z)\cos(p\,z)\right]
    \label{eq:BK_D1}
    \;\;,\\
    D_{2}(\nu_{0})=&&\int_{0}^{\infty}dz\,
    \frac{\exp(-p\,z)}{\sinh^{3}(z)}\nonumber\\
    &&\bigg\{\left[d(z)-\frac{1}{2}\right]
    \big[\sin(p\,z)+\cos(p\,z)\big]+\nonumber\\
    &&\frac{\pi}{4}g(z)
    \big[\cos(p\,z)-\sin(p\,z)\big]\bigg\}
    \;\;,
    \label{eq:BK_D2}
  \end{eqnarray}
  \label{eq:BK_D1D2}
\end{subequations}
where $p=1/(\sqrt{2}\,\nu_{0})$.

The final expression for the correction term of the radiation
probability per unit length then reads
\begin{equation}
  \frac{dp^{\mathrm{BKC}}_{\gamma}}{dx}=
  \frac{\alpha\,mc^{2}}
  {4\pi\,\gamma\,\hbar c\, L_{\mathrm{c}}}
  \big(D_{1}(\nu_{0})\,R_{1}+D_{2}(\nu_{0})\,
  R_{2}\,\sqrt{2}\,\nu_{0}\big)
  \;\;.
 \label{eq:BK_corr}
\end{equation}

\subsubsection{\label{sec:BK_diel}Effect of the polarization of the medium}

Baier and Katkov also estimated the influence of the medium
polarization on the radiation probability. They first introduce the
basic quantity $\kappa$ (similar to the quantity $\Gamma$ of the
Migdal approach introduced in Sec.~\ref{Migdal_pol}),
\begin{equation}
  \kappa=1+\frac{E}{E-k}\left(\frac{k_{\mathrm{p}}}{k}\right)^{2}
  \;\;.
\end{equation}
As it turns out, to include the polarization effect, it suffices to
implement the following substitutions (listed on page~298 of
Ref.~\cite{Baier:2005}) in Eq.~(\ref{eq:BK_main}) and
Eq.~(\ref{eq:BK_corr}) of the previous subsection:
\begin{subequations}
  \begin{eqnarray}
    \rho &\rightarrow& \tilde{\rho}=\rho\,\sqrt{\kappa}
    \;\;,\\
    Q&\rightarrow&\tilde{Q}=\frac{Q}{\kappa^{2}}
    \;\;,\\
    R_{1}&\rightarrow&R_{1}
    \;\;,\\
    R_{2}&\rightarrow&\tilde{R}_{2}=\kappa\,R_{2}
    \;\;,\\
    \tilde{L}_{\mathrm{c}}&\equiv&
    \tilde{L}(\tilde{\rho}_{\mathrm{c}})=
    \ln\left(\frac{\kappa}{\gamma^{2}\,\theta_{2}^{2}\,
        \tilde{\rho}_{\mathrm{c}}^{2}}\right)
    \;\;,\\
    \nu_{0}&\rightarrow&
    \tilde{\nu}_{0}=\sqrt{4\,\tilde{Q}\,\tilde{L}\,
    \tilde{\rho}_{\mathrm{c}}}
    \;\;.
  \end{eqnarray}
\end{subequations}
Notice in particular that $L_{1}\equiv
L(\rho_{\mathrm{c}}=1)=\tilde{L}(\tilde{\rho}_{\mathrm{c}}=1)$.  The
main consequence of these substitutions is that
$\tilde{\rho}_{\mathrm{c}}$ can assume the value $1$ not only for high
enough values of $x=k/E$ but also for small enough values of $x$.  In
addition, the quantity $\tilde{\nu_{0}}$ is no longer a monotonic
decreasing function as $k$ increases, but starting from zero (due to
the fact that $\tilde{Q}\rightarrow 0$ as $k \rightarrow 0$) reaches a
maximum and then decreases monotonically. This in turn will imply that
the slope of the photon energy spectrum could possibly exhibit,
depending on the energy $E$, two discontinuities. As stated in
Sec.~\ref{sec:fl}, the polarization effect results in a strong
suppression of the radiation only for photon energies below $\simeq
10^{-4}E$.

\subsubsection{\label{sec:BK_norm}Normalization to an arbitrary radiation length}

In presenting the BK approach for the spectral distribution of the
radiation in an infinite amorphous medium, no reference has yet been
made to the concept of radiation length, since the really basic
quantity of the theory is $L_{1}$.  However, as remarked at the
beginning of Sec.~\ref{sec:theo}, in order to compare different
theories on the same basis, we need to refer them, for any given
material, to the same radiation length which we have chosen to be the
one reported in Ref.~\cite{Tsai:1974}. Fortunately, the authors of
Ref.~\cite{Baier:2005}, when discussing the integral characteristic of
bremsstrahlung, provide a very useful relation between the radiation
length and $L_{1}$ (see Eq.~(2.36) in Ref.~\cite{Baier:2005}) which is
here reported using our notation
\begin{equation}
  X_{0}=\frac{9\,L_{1}}{1+9\,L_{1}}\,L_{\mathrm{rad}}^{0}
  \;\;,
  \label{BK236}
\end{equation}
where
$L_{\mathrm{rad}}^{0}=(m^{2}c^{4})/(2\,Z^{2}\,\alpha^{3}\,n\,(\hbar c)^{2}\,L_{1})$.

Indeed one can invert this relation and derive $L_{1}$ from the value
of the radiation length given in Ref.~\cite{Tsai:1974}. This value of
$L_{1}$ is then substituted for the one given by Eq.~(\ref{eq:BK_L1}),
which defines the minimum value assumed by the function
$L(\rho_{\mathrm{c}})$, whenever appropriate to obtain the final
expressions actually used in our calculations (see
Eqs.~(\ref{eq:BK_main}) and~(\ref{eq:BK_corr})).

\begin{figure}[t!!!!]
  \centering
  \begin{tabular}{c}
    \includegraphics[width=.425\textwidth]{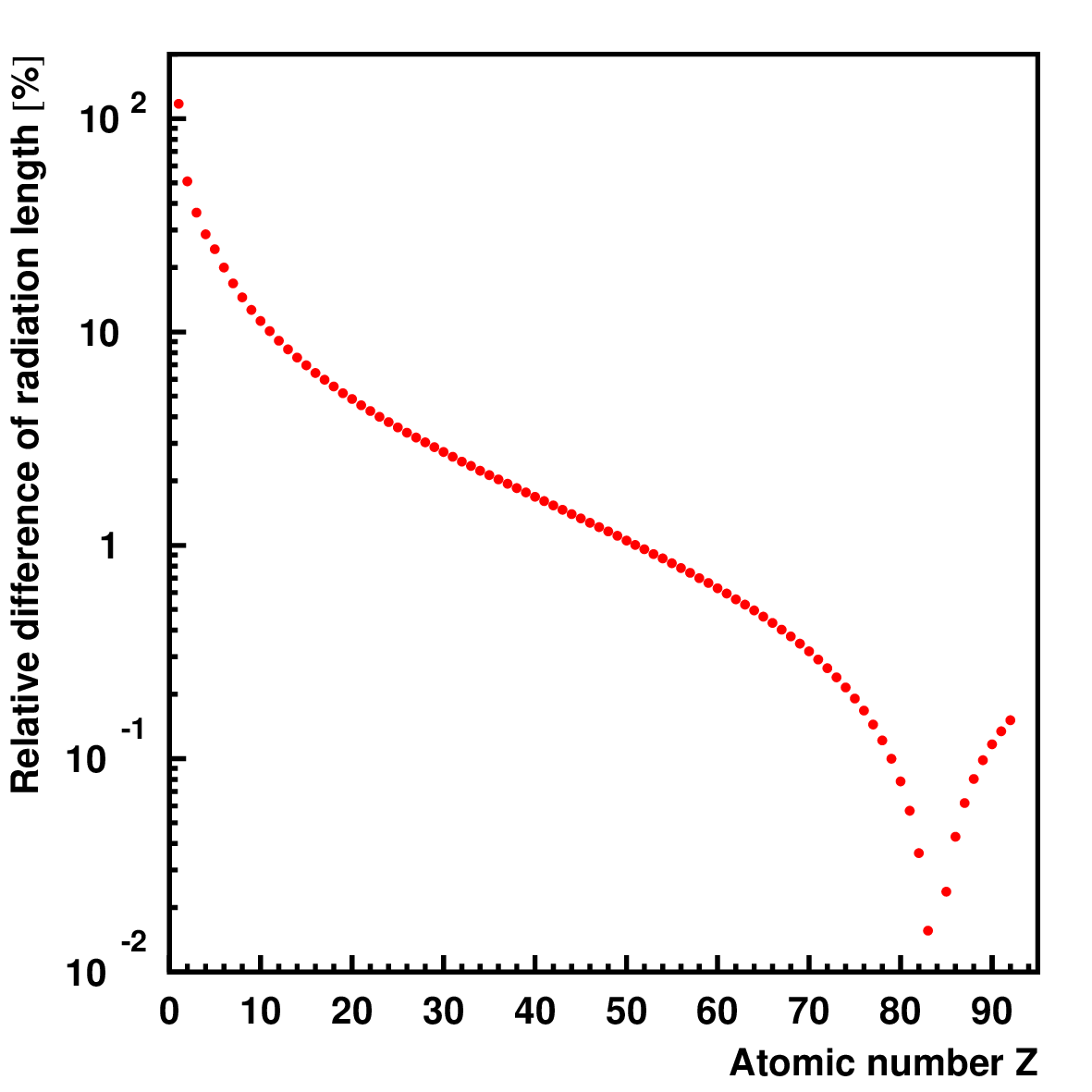}
  \end{tabular}
  \caption{\label{fig:x0_ratio}Relative difference of the radiation
    length, $X_{0}$, obtained from the BK expression,
    Eq.~(\ref{BK236}), using for $L_{1}$ the value of
    Eq.~(\ref{eq:BK_L1}) from the value tabulated in
    Ref.~\cite{Tsai:1974} and used in the present work by a
    redefinition of $L_{1}$, as explained in the text.}
\end{figure}

It is anyway interesting to compare (for each atomic number $Z$) the
radiation length obtained from Eq.~(\ref{BK236}), using for $L_{1}$
the value of Eq.~(\ref{eq:BK_L1}), and the one given in
Ref.~\cite{Tsai:1974}.  The percentage deviation between the two
values is reported in Fig.~\ref{fig:x0_ratio} as a function of $Z$. It
is seen that important differences arise for low-$Z$ materials so that
a direct comparison of the BK approach with different theories would
have been misleading had we avoided to ``normalize'' to a common
radiation length. The discrepancy is most probably due to the
different atomic form factors adopted. In Ref.~\cite{Baier:2005}, a
simple parametrization of the atomic screening function with a single
exponential was used resulting in the Schiff approximation for the
form factor (see Eq.~(2.14) in Ref.~\cite{Baier:2005}). The parameter
of the exponential is adjusted to reproduce on average all atoms with
a $Z$ dependence taken from the Thomas-Fermi model. In
Ref.~\cite{Tsai:1974}, a sum of three exponentials is used to
approximate the screening function leading to the Moli\`{e}re
approximation for the form factor (see Eq.~(3.66) in
Ref.~\cite{Tsai:1974}). Again, the parameters of three exponentials
are adjusted to reproduce on average all atoms with a $Z$ dependence
taken from the Thomas-Fermi model. It is in general expected that the
Moli\`{e}re approximation is superior to the Schiff
one~\footnote{There is another difference: the contribution of
  electron-electron bremsstrahlung is included in the radiation length
  given in Ref.~\cite{Tsai:1974}, but it is only considered by Baier
  and Katkov in the definition of an effective value
  $Q_{\mathrm{eff}}$ and of an effective angular parameter
  $\theta_{\mathrm{e}}$. The influence of $Q_{\mathrm{eff}}$ and
  $\theta_{\mathrm{e}}$ on the previously defined $L_{1}$ remains
  obscure and has not been taken into account in the present
  work. This difference is, however, unimportant except for very
  low-$Z$ materials.}.

\subsection{The Blankenbecler-Drell approach}

A completely different approach to the problem of the LPM suppression
was taken by Blankenbecler and Drell~\cite{Blankenbecler:1996} (BD in
the following). They started by solving a Klein--Gordon equation in an
eikonal approximation including terms up to first order (in the
inverse electron initial and final momentum) in the total phase of the
wave function while neglecting these terms in the expression for the
amplitude.  This approximation enables also to include electron spin
effects in a simple way at a later stage of the calculation (see
Ref.~[15] in Ref.~\cite{Blankenbecler:1996}).  Such an approach
basically takes into account, in the stated approximation, the total
phase difference and momentum variation accumulated by the wave
function as the electron propagates between all possible pair of
points having coordinates $(z_{1},z_{2})$ (with $z_{1}\leq z_{2}$)
along the axis of cylindrical symmetry provided by the initial
direction of the impinging electron; these points can even lie outside
the physical dimensions of the target. The approach does not include
the concept of cross section per atom as it considers the whole target
as a single unit.  The theory is valid for any target thickness $t$
and even for inhomogeneous amorphous targets including the case of
several slabs of any composition separated by
gaps~\cite{Blankenbecler:1997a}.  For the simplest case of a
homogeneous target of thickness $t$ and radiation length $X_{0}$ (for
which no ``adjustment'' is necessary so that the above-mentioned
tabulated values of Tsai~\cite{Tsai:1974} can safely be used) the
final result for the probability of emission per unit length,
differential in the fractional photon energy $x=k/E$ can be concisely
written as
\begin{equation}
  \frac{dp^{\mathrm{BD}}_{\gamma}}{dx} =
  \frac{4\,(1-x)\,J(t,x)}{3\,X_{0}\,x}
  \;\;.
  \label{BD}
\end{equation}
Here the calculation of the function $J(t,x)$ requires, for each value
of $x$, the careful evaluation of a double integral of oscillating
functions of two variables
$(b_{1}=z_{1}/l_{\mathrm{f}},b_{2}=z_{2}/l_{\mathrm{f}})$ (with
$b_{1}\leq b_{2}$) each ranging from $-\infty$ to $+\infty$. All the
details can be found in Ref.~\cite{Blankenbecler:1997a}. The proper
evaluation of the integral [which is actually conveniently split over
four integration regions in the $(b_{1},b_{2})$ plane] requires the
introduction of a convergence factor and has been implemented in a
dedicated code.  It must also be mentioned that in a subsequent
paper~\cite{Blankenbecler:1997b}, Blankenbecler added a correction
term (usually referred to as $\delta$-term) to the previous formula in
an attempt to take into account correlations between amplitude and
phase change in the wave function during the electron
propagation. This generally leads to a significant predicted reduction
(up to about $20$~\%; see Ref.~\cite{Blankenbecler:1997b}) of the
radiation intensity. The question whether the inclusion of the
$\delta$-term could result in an improved matching of the theory to
the experimental data remained somewhat controversial for some
time~\cite{Klein:1999,Andersen:2012,Andersen:2013}.  However, the
authors of Ref.~\cite{Andersen:2014} eventually not only presented
convincing experimental data which clearly support the superiority of
the approach developed in Ref.~\cite{Blankenbecler:1996} without
corrections, but also raised some doubts about a possible
inconsistency in the procedure of adding the $\delta$-term, given that
terms of the first order in the inverse electron momentum in the
amplitude of the wave function had been neglected in the original
eikonal approximation.  This is why the results of our calculations
including the $\delta$-term (which also point to an excessive
suppression) are not reported in this paper.  Unfortunately, the BD
approach does not include the dielectric suppression at low photon
energies and, while taking into account in a natural way the surface
effects, cannot easily be corrected for multiphoton emission, which
plays an important role in almost all experimental conditions as
remarked by Klein~\cite{Klein:1999}. Finally, no discontinuity in the
slope of calculated radiation spectra appears anywhere in this
approach.

\subsection{Other theoretical approaches}

The accurate measurement of the LPM effect at SLAC~\cite{Anthony:1997}
encouraged theoreticians to find alternative and possibly more
accurate methods to describe the suppression effects. We very briefly
describe the ones not mentioned above and give some explanations why
they are not considered any further in the present work.

Baier et al.~\cite{Yu:1996} based their theory on the multiple
scattering of high-energy electrons on a large number of scattering
centers fixed in the medium. The LPM suppression was ascribed to the
destructive interference between radiation amplitudes from a number of
scattering centers. However, only the soft photon limit was considered
and no general formula was given for an accurate description of the
radiation process valid over the whole photon energy range. Moreover,
different mechanisms for suppression, like that due to the
polarization of the medium, were not considered.

Zakharov~\cite{Zakharov:1996a} introduced since 1996 the so-called
``light-cone path integral approach'' to describe the LPM suppression
effect. The intensity of the photon radiation is expressed in terms of
the Green function of a two-dimensional Schr{\"o}dinger
equation. Simple expressions for the radiation probability are only
obtained for the soft photon limit and an infinite medium (see
Eq.~(45) in Ref.~\cite{Zakharov:1999}) and, in such a case, the
results are almost coincident with those by Migdal. In the general
case, very complicated expressions for the radiation probability,
involving multiple integrals, are given (see e.g.~Eq.~(52) in
Ref.~\cite{Zakharov:1999}). These were successfully compared both to
the SLAC E-146 data in Ref.~\cite{Zakharov:1999} and to the
high-energy CERN LPM data in Ref.~\cite{Zakharov:2003}. However, the
lack of adequate information in the description of the relevant
formulae, as well as the absence of any consideration of dielectric
suppression, led us to renounce the implementation of a Monte Carlo
code for this approach.

As a last issue, it should be mentioned that Baier and
Katkov~\cite{Baier:2005}, as well as Zakharov~\cite{Zakharov:1999},
developed versions of their approaches capable of dealing with the
presence of boundary surfaces for finite targets like the BD
approach. These effects are very interesting and lead to a direct
measurement of the formation length with structured
targets~\cite{Andersen:2013}. However, these formulations are
incompatible with the basic Monte Carlo philosophy, since they do not
localize the emission of a photon in a particular region of the
target~\cite{Klein:1999}.

\section{\label{sec:th_comp}General  comparison of the  theories }

\begin{figure}[t!!!!]
  \centering
  \begin{tabular}{c}
    \includegraphics[width=.425\textwidth]{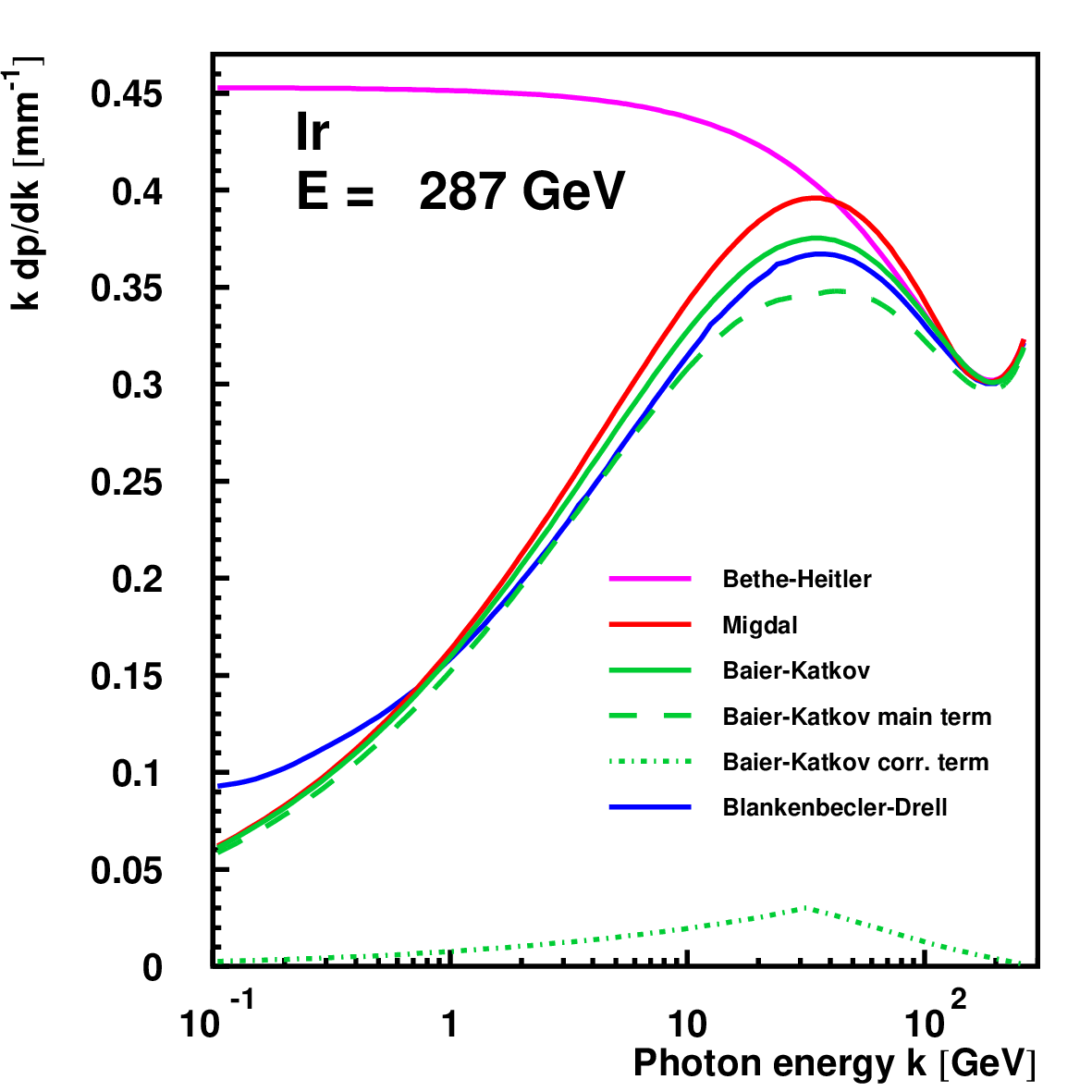}
  \end{tabular}
  \caption{\label{fig:comp_th_Ir} Comparison of the predicted power
    spectra for a $287$-GeV electron beam impinging on a
    $128$-$\mu$m-thick iridium target. Different curves are calculated
    according to the BH, Migdal, BD, and BK approaches as specified in
    the legend. For the latter, the main term as well as the
    correction term are reported separately. Polarization effects are
    completely negligible for the photon energy range considered.}
\end{figure}

We are now in a position to compare the results of the three
approaches including quantum coherence effects for a few specific
cases. A general preliminary remark is in order: in the BD theory, one
has to introduce a specific value for the target thickness and no
restriction on the validity of the results is introduced by the
consideration of the ($k$-dependent) coherence length. On the
contrary, an infinite medium is assumed both in the Migdal approach
and in the version of the BK theory which is considered in the present
work, so that the comparison with the BD approach is only meaningful
for the section of the spectral probability distribution for which the
coherence length is smaller than the assumed target thickness in the
BD theory. It must also be noted that a comparison of this type was
previously reported in Fig.~9 in Ref.~\cite{Andersen:2013} for a
$178$-GeV electron beam impinging on a thin ($1.97$~\%~$X_{0}$) carbon
target. However, in that study no attempt was made to compare the
three approaches for the same value of the radiation length, as is
done in the present work, so that the drawn conclusions might be
different in some details from those presented hereafter.
Figure~\ref{fig:comp_th_Ir} shows the spectral distribution of the
quantity $x\,dp_{\gamma}/dx$ in units of mm$^{-1}$ for a $287$~GeV
electron beam impinging on a $128$-$\mu$m-thick iridium foil (one of
the cases covered by the CERN LPM experiment). The BH result is
compared to the three theories considered here for the LPM suppression
(for the BK approach both the contribution of the main and correction
terms are shown separately together with their sum). In this case, the
coherence length, as given in Eq.~(\ref{eq:lf}), is larger than the
target thickness for photon energies lower than $1$~GeV. The
polarization effect is quite negligible for the photon energy range
displayed in the figure so that a valid comparison with the results of
the BD approach, which does not include this effect, is possible. Here
is a list of comments to this figure.

\begin{enumerate}
\item All approaches agree with the BH limit at the SWL, where the
  formation length is smallest.
\item The Migdal theory overshoots the BH approach in the
  $45$-$125$-GeV range with a maximum deviation of $2.6$~\% at
  $72$~GeV. This was already pointed out at lower energies in
  Ref.~\cite{Anthony:1997}. The most probable origin are the numerous
  approximations in the theory even though it is almost impossible to
  pinpoint the reason for the discrepancy in that particular energy
  range.
\item The Migdal approach exceeds the BK theory (main+correction term)
  by $5$~\% in the photon energy range $15$--$45$~GeV, suggesting the
  possibility of an experimental check to discriminate between the two
  theories.
\item The correction term in the BK approach reaches a maximum of
  $8$~\% of the total probability at $31$~GeV where a discontinuity in
  the derivative of the spectrum is apparent. A less apparent
  discontinuity (of the opposite sign) is also present at the same
  energy in the main term, so that the total spectrum appears smooth
  at the $1$~\% level.
\item Remarkably, for photon energies larger than $1$~GeV the BK and
  BD approaches, based on quite different approaches, agree to within
  $4$~\%.
\end{enumerate}

\begin{figure}[t!!!!]
  \centering
  \begin{tabular}{c}
    \includegraphics[width=.425\textwidth]{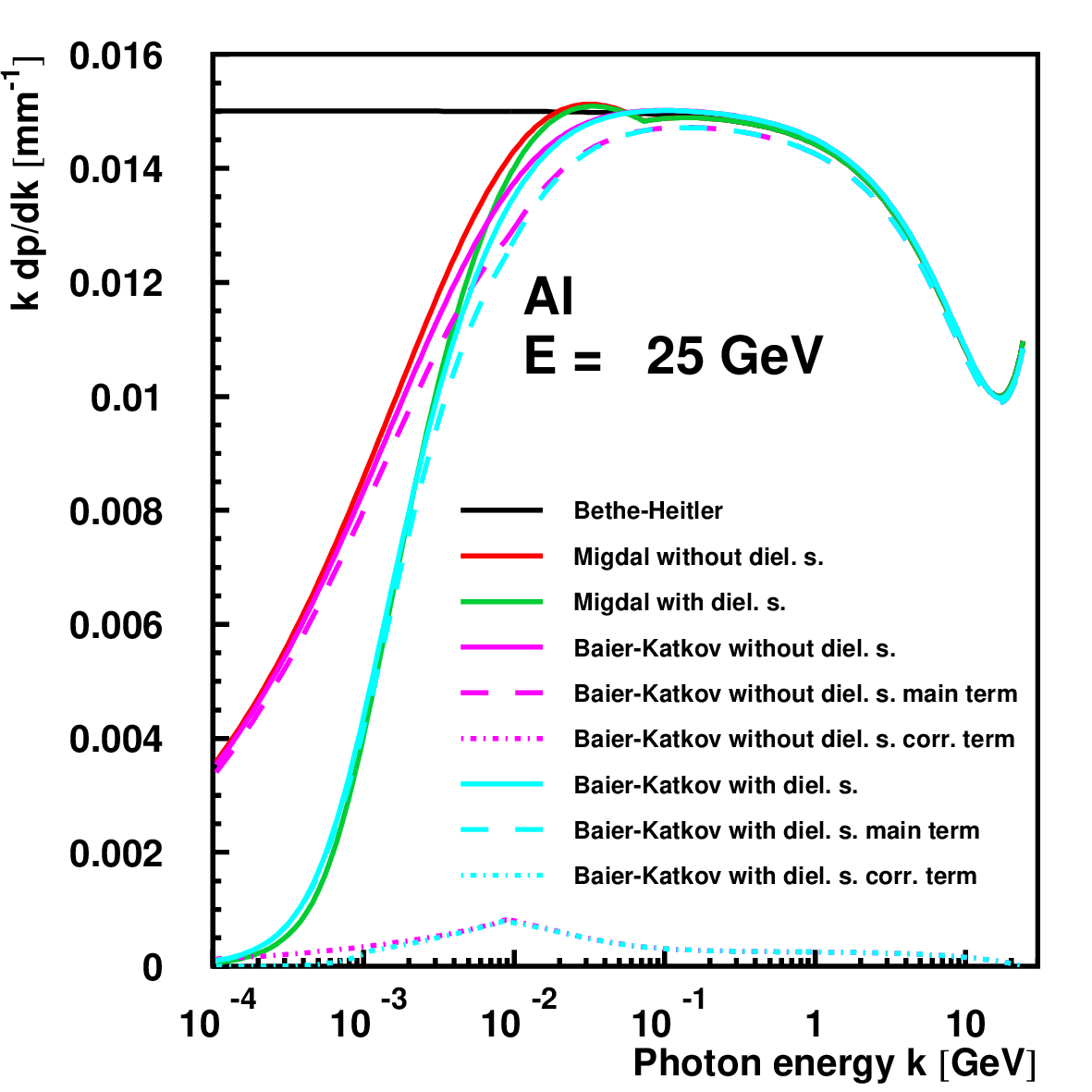}
  \end{tabular}
  \caption{\label{fig:comp_th_Au}Comparison of the predicted power
    spectra for a $287$-GeV electron beam impinging on a $3$~\%
    $X_{0}$ aluminum target. Different curves are calculated according
    to the BH, Migdal, and BK approaches both including and excluding
    dielectric suppression as specified in the legend. For the latter,
    the main term as well as the correction term are reported
    separately.}
\end{figure}

As an example which illustrates the influence of the polarization
effect on the predicted theoretical photon spectrum, we report in
Fig.~\ref{fig:comp_th_Au} the calculated spectra for a $25$-GeV
electron beam impinging on a $3$~\%~$X_{0}$ aluminum target (one of
the cases covered by the SLAC E-146 data) showing separately the
Migdal and BH spectra and the main term, the correction term, and
their sum for the BK approach, including and excluding the
polarization effect. The above-mentioned condition for a formation
length (reduced by the LPM and polarization effect) smaller than the
target thickness is satisfied for photon energies larger than $\approx
0.3$~MeV. One notes that the polarization effects are indeed most
important in the $k\lesssim 5$-MeV region.

\section{\label{sec:multiph}Considerations on multiphoton effects}

The theoretical calculations of the bremsstrahlung cross section
described in Sec.~\ref{sec:theo} cannot be directly compared to
experimental data because of the thickness of the targets considered
here. In particular, a lower limit on the thickness is imposed by the
neglect of the presence of boundary surfaces, as mentioned in
Sec.~\ref{sec:th_comp}. Three main sources of distortions have then to
be taken into account. They are discussed in turn in the following
subsections. The most straightforward way to handle all of them, and
the one followed in the present work, is to implement the theoretical
cross sections in a Monte Carlo program, as described in
Sec.~\ref{sec:imp}. To justify the need for such an effort, the
magnitude of the three distortions is illustrated under typical
conditions by using such a code.

\subsection{\label{sec:dmultiph}Direct multiphoton effects}

\begin{figure}[t!!!!]
  \centering
  \begin{tabular}{c}
    \includegraphics[width=.425\textwidth]{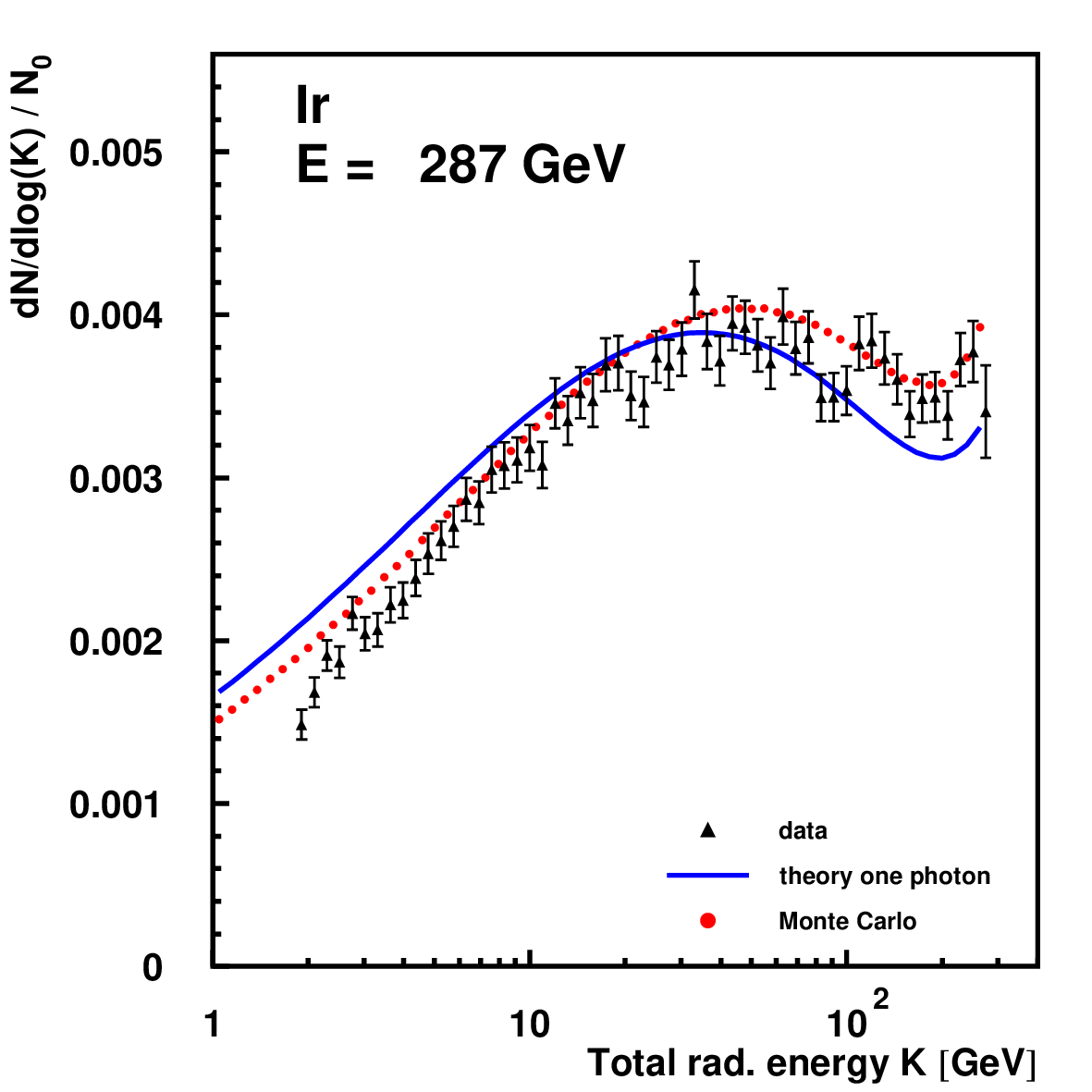}
  \end{tabular}
  \caption{\label{fig:multiph_th_MC}Comparison of the BK approach
    (continuous line) [see Eqs.~(\ref{eq:BK_main})
    and~(\ref{eq:BK_corr})] with the CERN LPM data for a $287$-GeV
    electron beam impinging on a $128$-$\mu$m-thick iridium target
    (solid triangles). The result of a Monte Carlo simulation, based
    on the same theory, and with all other interaction processes
    disabled, is also shown (solid circles).}
\end{figure}

A single crossing of the target by one electron can result in the
emission of multiple photons, which, reaching the calorimeter, give
rise to a signal proportional to the sum of their energies. For such a
reason, it is necessary to clearly distinguish between the energy of
the photon radiated in a single bremsstrahlung act, indicated by $k$,
and the total energy deposited in the calorimeter as the sum of all
the emitted photons by a single electron, indicated by $K$ hereafter.
The area of the spectrum is not altered by multiphoton emission (if
not at least one photon is emitted, no energy is deposited in the
calorimeter), but its shape is. In particular, photon pile-up tends to
deplete the low-energy part of the spectrum and enrich the high-energy
one~\cite{Anthony:1997}. A typical example is shown in
Fig.~\ref{fig:multiph_th_MC} for the BK approach and the CERN LPM
data. It is apparent that the bremsstrahlung cross section is far from
the data, while the Monte Carlo, accounting for the emission of
multiple-photons, is much closer.

The need to include multiphoton emission was recognized both in the
experimental~\cite{Anthony:1997,Hansen:2004} and theoretical
works~\cite{Baier:1999a,Baier:2005}. However, while the former
employed Monte Carlo simulations, which are essentially exact, the
latter developed analytic approximations giving explicit correction
formulae. The accuracy of such expressions was investigated in detail
in our previous publication~\cite{Mangiarotti:2012} for the BH and
Migdal cross sections. Here we extend that study by considering the BK
approach.

A multiplicative multiphoton correction factor is defined in analytic
calculations~\cite{Baier:1999a,Baier:2005,Zakharov:2003} as
\begin{equation}
  f(K)=\frac{K\,\frac{dN}{dk}}
  {N\,t\,n\,\left(k\,\frac{d\sigma}{dk}\right)_{k=K}}
  \;\;,
  \label{eq:fmph}
\end{equation}
where $N$ is the total number of events in the $dN/dK$ spectrum (not
to be confused with the total number of impinging electrons $N_{0}$
used in Sec.~\ref{sec:exp}) and $t$ is the target thickness.  The same
quantity can be obtained from the Monte Carlo simulations
as~\cite{Mangiarotti:2012}
\begin{equation}
  f_{\mathrm{MC}}(K)=\frac
  {\left(\frac{dN}{dK}\right)_{\mathrm{mph}}}
  {\left(\frac{dN}{dk}\right)_{\mathrm{fph}\;k=K}}\,
  \frac{1-\exp(-t/\lambda)}{t/\lambda}
  \;\;,
  \label{eq:fmc}
\end{equation}
where multiphoton ``mph'' and first photon only ``fph'' spectra refer
to the same set of events and $\lambda$ is the mean free-path for the
bremsstrahlung interaction. All other processes are disabled for the
simulations considered in the present subsection.

Baier and Katkov~\cite{Baier:1999a,Baier:2005} started from the Landau
solution of the kinetic equation valid under the assumption that the
particle energy loss is much smaller than the particle energy. Then
they were able to find an expression for $f$ valid in the case of the
BH cross section even when an arbitrary number of hard photon is
emitted, namely
\begin{equation}
  f_{\mathrm{BK}}^{\mathrm{BH}}(K)=
  (1+\beta)^{1/4}\,
  \left(1+\frac{\beta}{4}\right)^{3/4}\,
  \left(\frac{K}{E}\right)^{\beta}
  \;\;,
  \label{eq:f_BK_BH}
\end{equation}
where $\beta=4\,t/(3\,X_{0})$. Baier and Katkov claim that this
expression is valid for all photon energies and all target
thicknesses, but this could not be confirmed by Monte Carlo
simulations~\cite{Mangiarotti:2012}. However, it is a reasonable
approximation for the thicknesses considered in the present work.

\begin{figure}[t!!!!]
  \centering
  \begin{tabular}{c}
    \includegraphics[width=.425\textwidth]{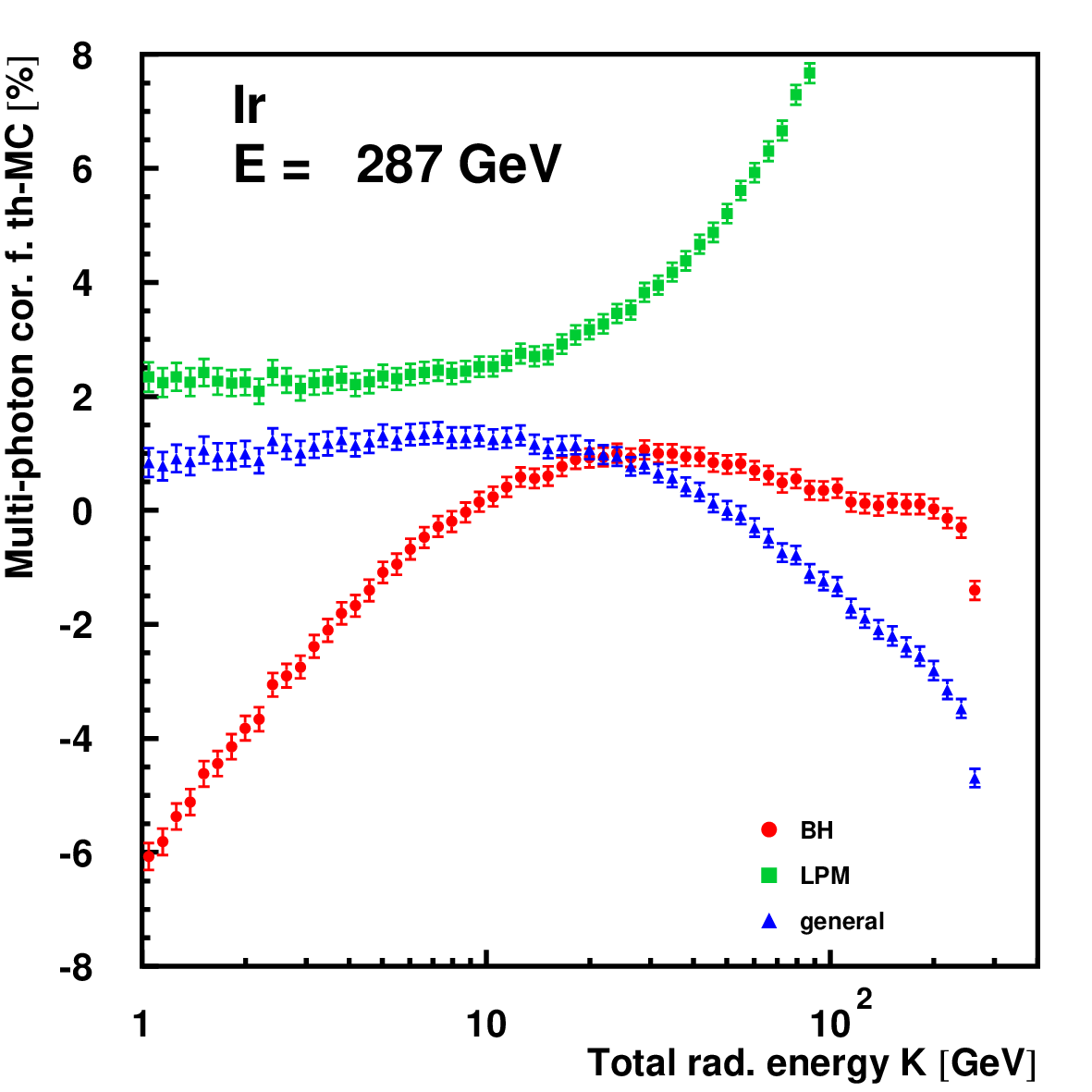}
  \end{tabular}
  \caption{\label{fig:multiph_anal_MC}Difference of the approximated
    analytic multiphoton correction factor $f$ [see
    Eq.~(\ref{eq:fmph})] with respect to the exact Monte Carlo
    calculations [see Eq.~(\ref{eq:fmc})] for a $287$-GeV electron
    beam impinging on a $128$-$\mu$m-thick iridium target. The three
    approximate expressions developed by Baier and Katkov are
    considered: Eq.~(\ref{eq:f_BK_BH}), (\ref{eq:f_BK_LPM}),
    and~(\ref{eq:f_BK_1}) (solid circles, squares, and triangles,
    respectively).}
\end{figure}

Baier and Katkov~\cite{Baier:1999a,Baier:2005}, starting again from
the Landau solution, found also the expression for $f$ in the case of
strong LPM suppression (described by their approach) valid for
$\beta\ll 1$:
\begin{equation}
  \begin{split}
    f_{\mathrm{BK}}^{\mathrm{LPM}}(K)=&
    \left(\frac{k_{\mathrm{c}}}{E}\right)^{\beta}\,
    \Gamma(1+\beta)\,(1+\beta)^{1/4}\\
    &\left(1+\frac{\beta}{2}\right)^{3/4}\,
    e^{-\beta(C+c_{1})}\\
    &\left(1+\frac{3\pi}{2\sqrt{2}}\,\beta
    \sqrt{\frac{K}{k_{\mathrm{c}}}}\right)
    \;\;,
    \label{eq:f_BK_LPM}
  \end{split}
\end{equation}
where $\Gamma$ is the Euler Gamma function, the constant $c_{1}$ can
be calculated, once and for all, from the integral
\begin{equation}
  c_{1}=12\,\int_{0}^{\infty}\,\ln\,z\,
  \left(\frac{1}{z^{3}}-\frac{\cosh\,z}
  {\sinh^{3}\,z}\right)\,dz
\end{equation}
and
\begin{equation}
  k_{\mathrm{c}}=\frac{4\pi}{\alpha}\,\hbar c\,
  \frac{\gamma^{2}}{L_{\mathrm{rad}}^{0}}
\end{equation}
is the photon energy below which the LPM suppression shows up. By
assumption, Eq.~(\ref{eq:f_BK_LPM}) is valid only for $k\ll
k_{\mathrm{c}}$, where the LPM suppression is strong.

Finally, Baier and Katkov~\cite{Baier:1999a,Baier:2005}, still
employing the Landau solution, give an expression for a thin target
$\beta\ll 1$ and an arbitrary cross section,
\begin{equation}
  f_{\mathrm{BK}}^{1}(K)=\exp\left(
  -t\,n\,\int_{K}^{E}
  \frac{d\sigma}{dk}(E,k)\,
  dk\right)
  \;\;.
  \label{eq:f_BK_1}
\end{equation}
Here, this integral has been evaluated numerically employing the cross
section corresponding to the sum of Eqs.~(\ref{eq:BK_main})
and~(\ref{eq:BK_corr}).

All the three previous approximations, Eqs.~(\ref{eq:f_BK_BH}),
(\ref{eq:f_BK_LPM}), and~(\ref{eq:f_BK_1}), are compared to Monte
Carlo simulations in Fig.~\ref{fig:multiph_anal_MC} for the case of a
$287$-GeV electron beam impinging on a $128$-$\mu$m-thick iridium
target. The expression for the LPM case [Eq.~(\ref{eq:f_BK_LPM})]
works with an error of $\approx 2$~\% for $K\lesssim 10$~GeV, but then
fails more and more deeply close to the SWL. The expression for the BH
case [Eq.~(\ref{eq:f_BK_BH})] works essentially in the complementary
region $K\gtrsim 10$~GeV. Finally, Eq.~(\ref{eq:f_BK_1}) performs
better than Eq.~(\ref{eq:f_BK_LPM}) for $K\lesssim 30$~GeV, but still
fails more than Eq.~(\ref{eq:f_BK_BH}) close to the SWL. Baier and
Katkov~\cite{Baier:2005} state that to compare their approach with the
CERN LPM data they employed Eq.~(\ref{eq:f_BK_1}). For the SLAC E-146
measurements, they mention~\cite{Baier:1999a,Baier:2005} an
interpolation of Eqs.~(\ref{eq:f_BK_BH}) and~(\ref{eq:f_BK_LPM}) with
no further details. Given the present comparison with exact Monte
Carlo simulations, it is difficult to imagine that their accuracy is
better than $\approx 2$~\%, getting worse close to the SWL.

\subsection{\label{sec:multiph_att}Multiphoton effects coupled to self-absorption}

\begin{figure}[t!!!!]
  \centering
  \begin{tabular}{c}
    \includegraphics[width=.425\textwidth]{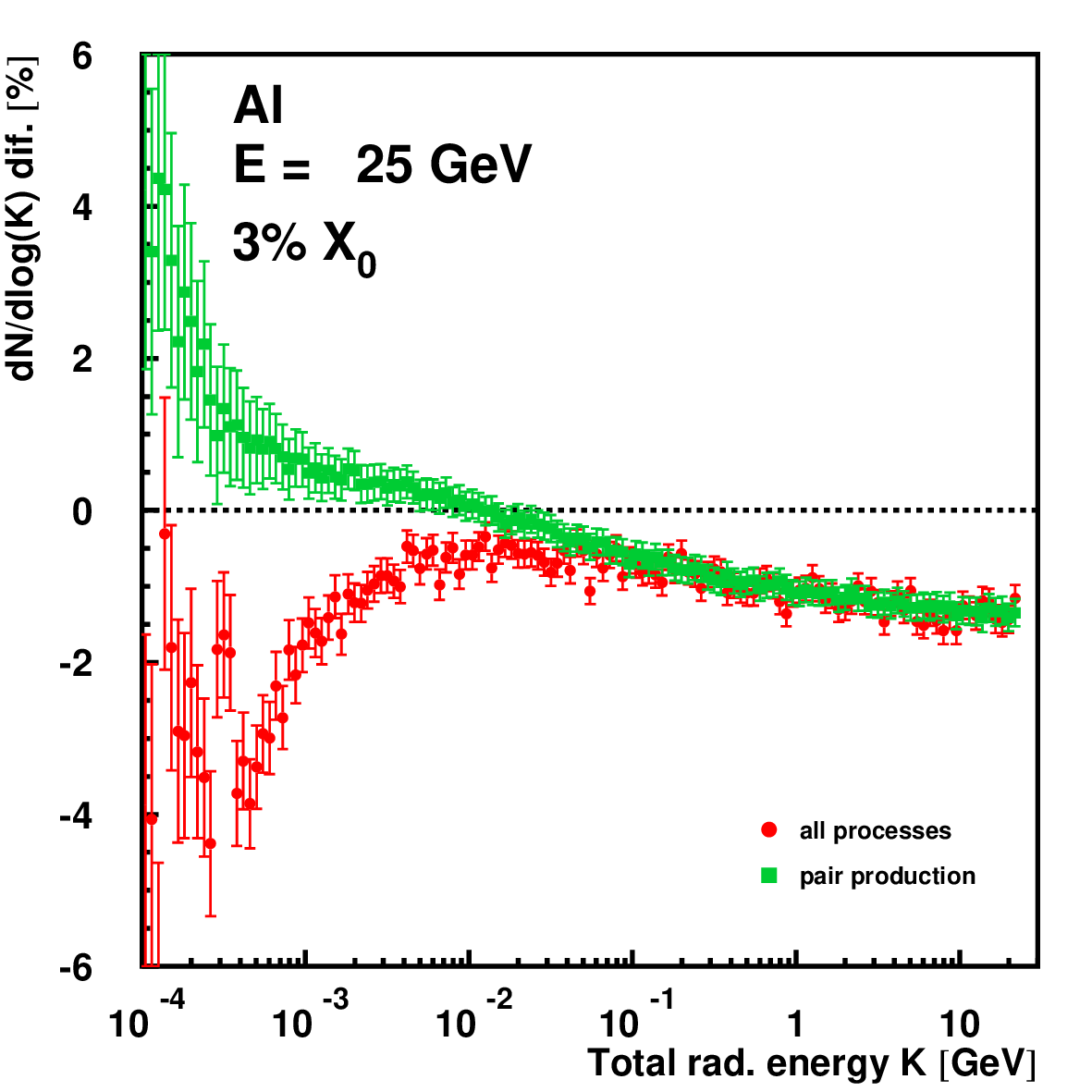}
  \end{tabular}
  \caption{\label{fig:multiph_pair}Difference $(s-r)/s$ of a
    simulation $s$ with respect to a reference one $r$ for two
    cases. In the first (solid circles), $s$ is the complete
    simulation with all processes and $r$ the simulation with only
    bremsstrahlung always considering $25$-GeV electrons impinging on
    a $3$~\%~$X_{0}$ aluminum target. In the second (solid squares),
    $s$ is the simulation with bremsstrahlung and pair production
    enabled and $r$ the simulation with only bremsstrahlung.}
\end{figure}

Beyond multiphoton emission, a Monte Carlo code allows to take into
account other sources of distortion in the shape of the spectrum that
are not included (and hardly could be) in the approximate analytic
approaches~\cite{Baier:1999a,Baier:2005}: namely pair production,
Compton scattering, and photoelectric absorption in the target. To
illustrate their importance in one typical case, a $25$-GeV electron
beam impinging on a $3.12$-mm-thick aluminum target is considered in
Fig.~\ref{fig:multiph_pair}. The difference of a simulation with
bremsstrahlung (BK approach) and pair production enabled with respect
to one where only the former is taken into account (solid squares),
clearly indicates that for $K\gtrsim 20$~MeV an attenuation is
present, which increases up to to $\approx 2$~\% at the SWL. The
enhancement for $K\lesssim 20$~MeV is less expected and it is an
example of a hard to predict consequence of multiphoton
emission~\cite{Mangiarotti:2012}. Consider an event where one
low-energy and one high-energy photon have been emitted by the same
electron crossing the target. The calorimeter only registers the sum
of the two energies, resulting in the depletion of the low-energy part
of the spectrum. However, if the high-energy photon is absorbed in the
target by pair production, the surviving low-energy one can reach the
calorimeter producing a signal (remember that electrons and positrons
are swept away by the magnets present in the setup and do not reach
the calorimeter).  If beyond pair production, also Compton scattering
and photoelectric absorption are enabled, the difference with respect
to a simulation with only bremsstrahlung (solid circles) is now a
reduction of the spectrum for all values of $K$ with a steep increase
in attenuation at the lower end. It can be mentioned that few
simulations have also been performed taking into account the influence
of the Fermi motion of bound electrons inside atoms on Compton
scattering with the GLECS extension by Kippen~\cite{Kippen:2004}. No
differences were found for the photon energy range of interest in the
present work.

We note that the photon path was kept in vacuum in the SLAC E-146
setup to allow to reach down to photon energies of $200$~keV. The only
material present between the target and the calorimeter was a thin
aluminum window with a thickness of
$0.7$~\%~$X_{0}$~\cite{Anthony:1997}. However, since pair production
in this material hardly results in a loss of signal (the window acting
like a preshower) no account in the simulation is necessary. The
photon path was not kept in vacuum in the CERN LPM experiment, but the
lowest photon energy considered was $2$~GeV~\cite{Hansen:2004}. Thus,
the only relevant attenuation process is pair production, which leads
to a loss of collected energy in the calorimeter only if it happens in
the small distance between the target and the deflecting
magnet. Background estimates~\cite{Hansen:2004} suggest that all the
crossed materials are equivalent to
$0.7$~\%~$X_{0}$~\cite{Hansen:2004}. The setup of the CERN LOW-$Z$
experiment was also not kept in vacuum and the minimum photon energy
considered was about $0.05$~GeV. The background intensity was somewhat
worse, corresponding to $3$~\%~$X_{0}$~\cite{Andersen:2013}. The
attenuation in such media is approximately taken into account when the
secondary target, representing the background, is placed after the
primary one in the Monte Carlo simulations (a detailed account would
require inclusion of the full setup).

To summarize, the neglect of secondary processes can produce an error
in the theoretical predictions at least around $1$ to $2$~\%,
depending on the value of $K$ and the amount of background. Such a
correction was taken into account in previous comparison of the
experiments with the Migdal
approach~\cite{Hansen:2004,Mangiarotti:2008,Mangiarotti:2011,Mangiarotti:2012},
but not in the only available confrontation with the BK
one~\cite{Baier:1999a,Baier:2005}, based on the analytic
approximations. Thus, the uncertainty in the inclusion of the
multiphoton and secondary processes in the results presented by Baier
and Katkov are possibly rather close to the discrepancy between theirs
and the Migdal approach. The first comparison of the Migdal and BK
approaches on the same footing is the one presented in the present
work.

\subsection{\label{sec:multiph_bksub}Multiphoton effects coupled to background subtraction}

\begin{figure}[t!!!!]
  \centering
  \begin{tabular}{c}
    \includegraphics[width=.425\textwidth]{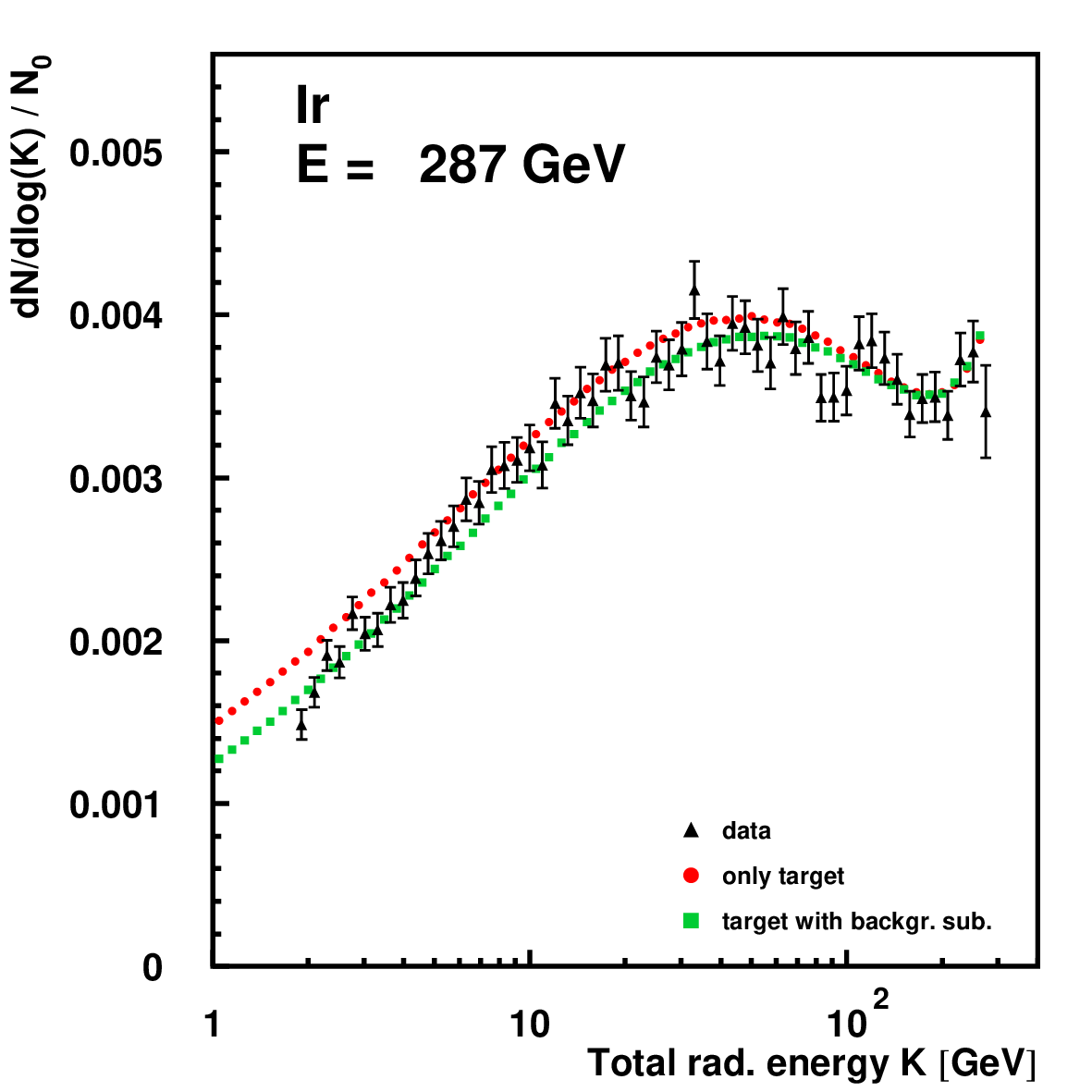}
  \end{tabular}
  \caption{\label{fig:multiph_bckg}Comparison of a full Monte Carlo
    simulation, employing the BK approach and taking into account all
    secondary processes, with the CERN LPM data for a $287$-GeV
    electron beam impinging on a $128$-$\mu$m-thick iridium
    target. The target alone (solid circles) and the target and
    background with background subtracted (solid squares) are shown.}
\end{figure}

The CERN LPM experiment covered the highest energies reached so far in
measurements of the LPM suppression. As a result of the more difficult
experimental conditions, despite all
efforts~\cite{Hansen:2003,Hansen:2004}, the photons from the
background amounted to $0.7$~\%~$X_{0}$, as mentioned. The employed
thicknesses of the targets were $\approx 4$~\%~$X_{0}$, leading to a
non-negligible probability of photons from bremsstrahlung interactions
in the target and the background media simultaneously reaching the
calorimeter. As a consequence, their energy is summed causing a
distortion of the measured spectrum, closely resembling that
originated by multiphoton emission in a single event. As first shown
in Ref.~\cite{Andersen:2012}, the simple subtraction of a
``no-target'' run is not a full solution to this problem and does not
allow to completely remove all distortions resulting from pile-up of
photons emitted from the target with the background. Thus, it is
necessary to apply the same background subtraction in the simulation:
one run is performed with the nominal target (e.g.~iridium with a
thickness of $128$~$\mu$m) and a secondary target (e.g.~carbon because
most of the elements producing the background have low $Z$) and
another run with the secondary target alone. Then the two results are
subtracted, to reproduce the procedure employed in the experiment. The
comparison with measurements of the target-only Monte Carlo and the
more correct procedure just described is shown in
Fig.~\ref{fig:multiph_bckg}. The distortion produced by the background
subtraction is not negligible and well comparable to the difference
between the Migdal and BK approaches. It was not taken into account in
all previous
publications~\cite{Hansen:2003,Hansen:2004,Baier:2005,Mangiarotti:2008,Mangiarotti:2011,Mangiarotti:2012}
and is included for the first time in the present work.

On account of the distortions produced by background subtraction, as
realized in Ref.~\cite{Andersen:2012}, it was chosen not to apply this
procedure in the analysis of the CERN LOW-$Z$
data~\cite{Andersen:2013}. Consequently, the data can only be compared
with Monte Carlo simulations including the pile-up with the background
and cannot be handled by the analytic approximations discussed in
Sec.~\ref{sec:dmultiph}.

\section{\label{sec:imp}Dedicated Monte Carlo}

First numerical implementations of approaches describing the LPM
effect were developed for Monte Carlo codes in the seventies by three
groups: one in Japan, one in the U.S., and one in Bulgaria. The main
emphasis was the study of electromagnetic showers initiated by
ultra-high-energy cosmic rays. The first group worked alone and
validated the results against analytic
approximations~\cite{Konishi:1978}, while the other two
joined~\cite{Stanev:1982} arriving at a common consensus that has been
the basis of all subsequent efforts. Nowadays, the Migdal approach is
available in Monte Carlo codes for the simulation of extensive air
showers, like AIRES~\cite{Cillis:1999}, CORSIKA~\cite{Heck:1998}, or
COSMOS~\cite{Kasahara:Cosmos} and even in general-purpose packages,
like GENAT4~\cite{Schaelicke:2008}, EGS5~\cite{Kirihara:2010}, or
EPICS~\cite{Kasahara:Epics}. We also developed our version of the
Migdal approach~\cite{Mangiarotti:2008,Mangiarotti:2011}, which
improves on all the others for the careful handling of the
discontinuity present in the first derivative of the cross section. It
is based on the GEANT3 code~\cite{Geant3:1994}, since, when the work
begun, GEANT4 had not yet gained widespread acceptance. It was used to
compare with the CERN LPM data~\cite{Mangiarotti:2011}. To the best of
our knowledge, no implementation of the BK approach in a Monte Carlo
code has yet been reported. Thus we extend our previous work on the
Migdal approach~\cite{Mangiarotti:2011} to the BK one by reusing, in
particular, the careful handling of the discontinuity of the first
derivative of the cross section.

As discussed in Sec.~\ref{sec:multiph_att}, the emitted photon can be
absorbed inside the target due to the pair production process. This
has the consequence that the program needs to handle positrons as
well. In the present implementations, the same cross sections for
electrons are applied to positrons. Due to the different
kinematics~\cite{Klein:1999}, the LPM suppression in pair production
is small for the energies of interest here (e.g.~for $1$-TeV gamma
rays in iridium, the maximum reduction of the cross section over the
full electron-positron energy range is $\approx
1$~\%~\cite{Mangiarotti:2011}) and is therefore not taken into
account.

In general, dielectric suppression is only important for low photon
energies (see Sec.~\ref{sec:fl}) and its inclusion is not necessary
under some conditions, like those of the comparison with the CERN LPM
data. Therefore, the program has the option to disable dielectric
suppression to simplify the logic and speed up the computation.

\subsection{\label{sec:cimp}Implementation}

Two main difficulties have to be overcome for an efficient
implementation of the BK approach in a Monte Carlo code.

First, the functions $G$ and $\Phi$ [see Eqs.~(\ref{eq:BK_R1R2})] and
the functions $D_{1}$ and $D_{2}$ [see Eqs.~(\ref{eq:BK_D1D2})] need
to be evaluated each time the photon energy $k$ is sampled, so that
direct use of their definition is very inefficient. Approximate
methods for the evaluation of $G$ and $\Phi$ were given by Stanev et
al.~\cite{Stanev:1982} employing rational functions, and were widely
utilized in essentially all programs written
afterwards~\cite{Schaelicke:2008,Mangiarotti:2008,Kirihara:2010,Mangiarotti:2011},
one notable exception being Ref.~\cite{Cillis:1999}, where other
approximations, still based on rational functions, were
proposed. Here, the original one by Stanev et al.~\cite{Stanev:1982}
is maintained. A fit to the functions $D_{1}$ and $D_{2}$ in terms of
rational functions of $\nu_{0}$ has been developed for different
ranges of the variable to ease the repetitive evaluation of $D_{1}$
and $D_{2}$ in the calculations presented here; an analysis of its
accuracy together with the values of the coefficients employed can be
found in Appendix~\ref{sec:imp_D1_D2}.

Second, for a given material, electron energy $E$, and photon energy
$k$, the corresponding value of $\tilde{\rho}_{\mathrm{c}}$ has to be
determined by solving Eq.~(\ref{rhoc}). The simple bisection method
has been employed because, once proper bracketing points are selected,
its convergence is always guaranteed. Typically $\approx 20$
iterations are needed to achieve a relative precision better than
$10^{-8}$, which can be very fast given the simplicity of the function
that needs to be evaluated at each step. As remarked,
$\tilde{\rho}_{c}$ cannot exceed $1$. Note that the efficiency can be
improved by checking the value of the variable
$\tilde{\nu}_{1}^{2}=4\,\tilde{Q}\,\Tilde{L}_{1}$: if it is greater
than $1$, the variable $\tilde{\rho}_{c}$ can be set to $1$ and the
bisection procedure skipped.

As discussed in Sec.~\ref{sec:BK_diel}, a discontinuity in the first
derivative of the cross section is present both in the main term and
in the correction term for the photon energy $k=k_{\mathrm{d}}$
corresponding to $\tilde{\rho}_{\mathrm{c}}=1$. Although the effect on
the total cross section is much less severe than in the case of Migdal
approach (see Sec.~\ref{sec:th_comp}) an explicit handling has still
been implemented following closely Ref.~\cite{Mangiarotti:2011}. More
details on the determination of the location of the discontinuity are
give in Appendix~\ref{sec:BK_ldisc}, on evaluation of the total cross
section in Appendix~\ref{sec:BK_sigmatot}, and on the efficient
sampling of the photon energy in Appendix~\ref{sec:BK_samp}.

\subsection{\label{sec:sim_par}Simulation parameters}

The developed program has been run with $2\cdot10^{8}$ events with the
GEANT3 \cal{ABAN} flag set to 0 (the default being 1), as recommended
in Ref.~\cite{Razzaque:2004}, to enforce effective tracking of all
electrons and positrons down to the chosen low-energy cut
$T_{\mathrm{min}}=50$~MeV. The low-energy cut for photons,
$\mathrm{TCUT}$, has been set to $10$~keV, $50$~MeV, and $10$~keV in
the simulations for the SLAC E-146, CERN LPM, and CERN LOW-$Z$ data,
respectively. Dielectric suppression is enabled, disabled, and enabled
for the previous sets, respectively. The angular cuts have been set to
$1.4$, $1$, and $0.6$~mrad for the SLAC E-146~\cite{Anthony:1997},
CERN LPM~\cite{Hansen:2004}, and CERN LOW-$Z$~\cite{Andersen:2013}
data, respectively, according to the geometry of the setup described
in the corresponding publications. Although the program can include
the effect of the resolution of the calorimeter, this has been found
to be negligible for all
datasets~\cite{Mangiarotti:2008,Mangiarotti:2011} and has thus been
switched off. The complete GEANT3 configuration can be found in
Ref.~\cite{Mangiarotti:2012} (see, in particular, the rightmost column
of Table~III). Typical execution times on a modern multicore 64-bit
unit running at $2.0$ GHz (opteron 6128 HE manufactured by
AMD\textsuperscript{\textregistered} processing one simulation per
core) are of the order of $21.0$ and $21.2$~$\mu$s per event for the
implementation of the Migdal and BK approaches, respectively, in the
case of the simulations for the SLAC E-146 data (these timings include
tracking with one cylindrical radiator volume and histogram filling).

\subsection{\label{sec:tests}Tests of the implementation}

\begin{figure}[t!!!!]
  \centering
  \begin{tabular}{c}
    \includegraphics[width=.425\textwidth]{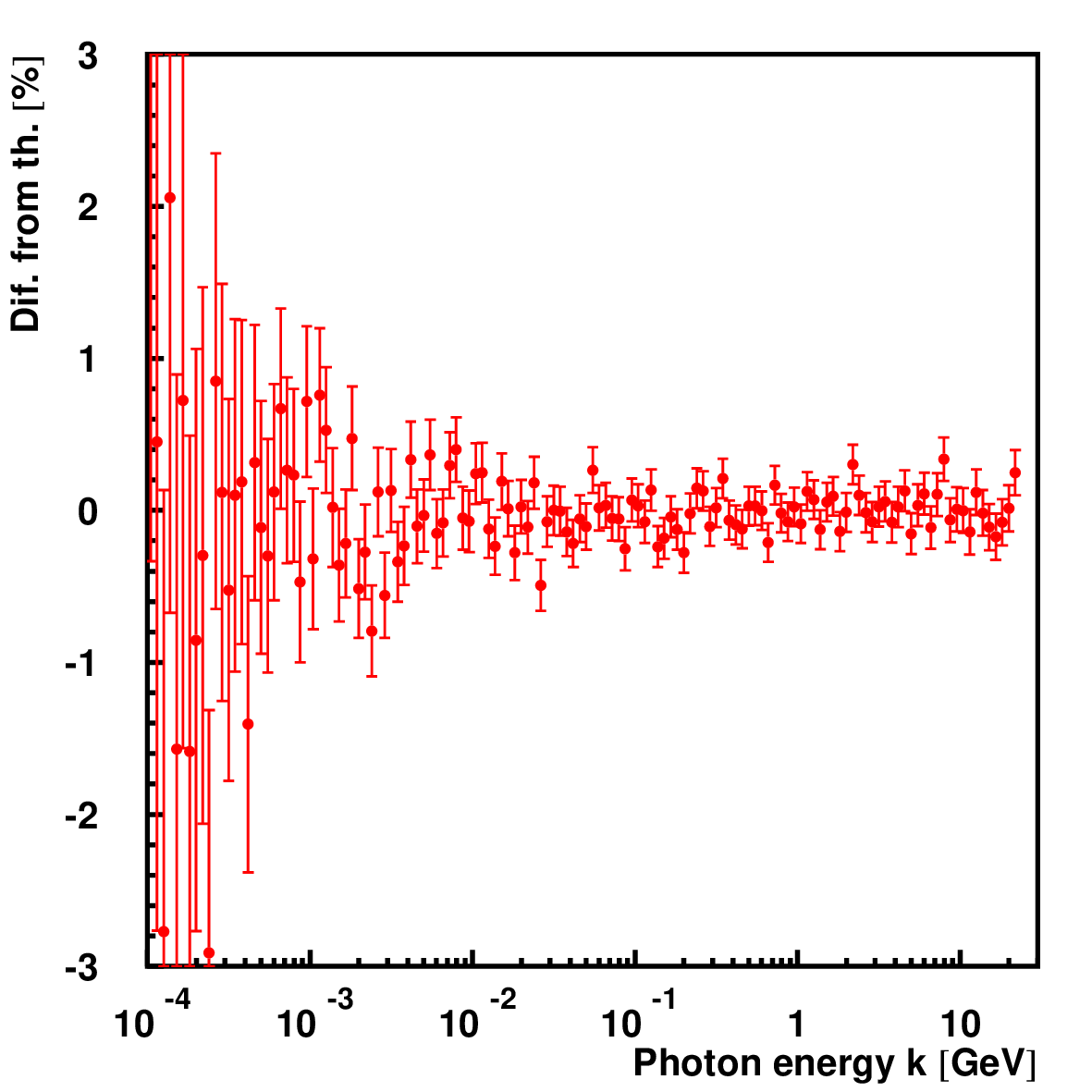}
  \end{tabular}
  \caption{\label{fig:sim_dev}Difference between the simulated first
    photon spectrum and the BK cross section for an electron impinging
    on iridium with an energy $E=25$~GeV. The dielectric suppression
    is taken into account.}
\end{figure}

The program contains a sophisticated logic to monitor the tracking
status, including inconsistencies due to multiple entering of an
electron into a volume without an intermediate exiting. This can
happen due to the finite precision of the tracking steps; it is
however limited, with the selected tracking parameters, to a few cases
out of $2\cdot 10^{8}$ events. Multiphoton emission can be
investigated constructing a series of histograms like those in the
detailed studies presented in Ref.~\cite{Mangiarotti:2012}. The
first-photon spectrum can be employed to validate the present
implementation: in fact, a direct match with the BK approach is
expected. The case of a $25$-GeV electron impinging on iridium is
shown in Fig.~\ref{fig:sim_dev}. The differences of the simulation
from theory are well within the estimated statistical uncertainties
(the bars indicate one standard deviation) in almost all cases. More
than five orders of magnitude in photon energy are covered in
Fig.~\ref{fig:sim_dev}. A problem in the evaluation of the total cross
section would result in a systematic shift of all points, while a
mistake in the sampling would produce local deviations from zero
incompatible with the statistical fluctuations. Due to simultaneous
LPM and dielectric suppressions, the number of events at low energies
decreases dramatically, resulting in larger error bars. To
definitively confirm that the total cross section is properly
calculated, it is interesting to compare the average deviation for all
bins with the corresponding statistical uncertainty. For iridium,
copper, and carbon, under the same conditions of
Fig.~\ref{fig:sim_dev}, these values are $(-0.034\pm 0.067)$~\%,
$(0.020\pm 0.050)$~\%, and $(0.027\pm 0.027)$~\%, respectively. The
same test has been repeated for all the simulations employed for the
comparison with data in Sec.~\ref{sec:exp}, finding always very
similar results.

Finally, before accepting the results of the simulations, it is
fundamental to verify their stability. The most important test
concerns the effect of the electron or photon energy cuts
$E_{\mathrm{min}}$ and $\mathrm{TCUT}$. For the SLAC E-146 case,
$\mathrm{TCUT}$ and $E_{\mathrm{min}}$ have been increased from $10$
to $50$~keV and from $50$ to $250$~MeV, finding no change within the
statistical uncertainty for all the cases compared to the data in
Sec.~\ref{sec:exp}. We do not show the corresponding figures here, but
they are very similar to those displayed in
Ref.~\cite{Mangiarotti:2012} for the Migdal approach (see in
particular Figs.~13 and~14). For the CERN LPM case, $\mathrm{TCUT}$
and $E_{\mathrm{min}}$ have been decreased from $50$ to $5$~MeV
finding again no difference, within the statistical uncertainty, for
all the cases considered in Sec.~\ref{sec:exp}. The cuts for CERN
LOW-$Z$ simulations are identical to those of the SLAC E-146 ones, but
the beam energy is $178$ versus $25$~GeV and the lowest photon energy
covered is $50$~MeV against $100$~keV; thus any further testing has
been deemed unnecessary. The targets used in the SLAC E-146, CERN LPM,
and CERN LOW-$Z$ experiments are rather thin, so that a check of the
impact of the boundary crossing precision $\mathrm{EPSIL}$ on the
final result is in order. Full stability has been found when
$\mathrm{EPSIL}$ is below $0.1$ $\mu$m. The adopted value for all the
simulations compared with data in Sec.~\ref{sec:exp} is
$\mathrm{EPSIL}=0.1$~$\mu$m.

\section{\label{sec:exp}Comparison with all experimental data}

The direct measurements of the LPM suppression are essentially limited
to the SLAC E-146~\cite{Anthony:1995,Anthony:1996,Anthony:1997}, CERN
LPM~\cite{Hansen:2003,Hansen:2004}, and CERN
LOW-$Z$~\cite{Andersen:2013} experiments. The main differences are the
following.

\begin{figure*}[hp!!!!]
  \centering
  \begin{tabular}{cc}
    \includegraphics[width=.425\textwidth]{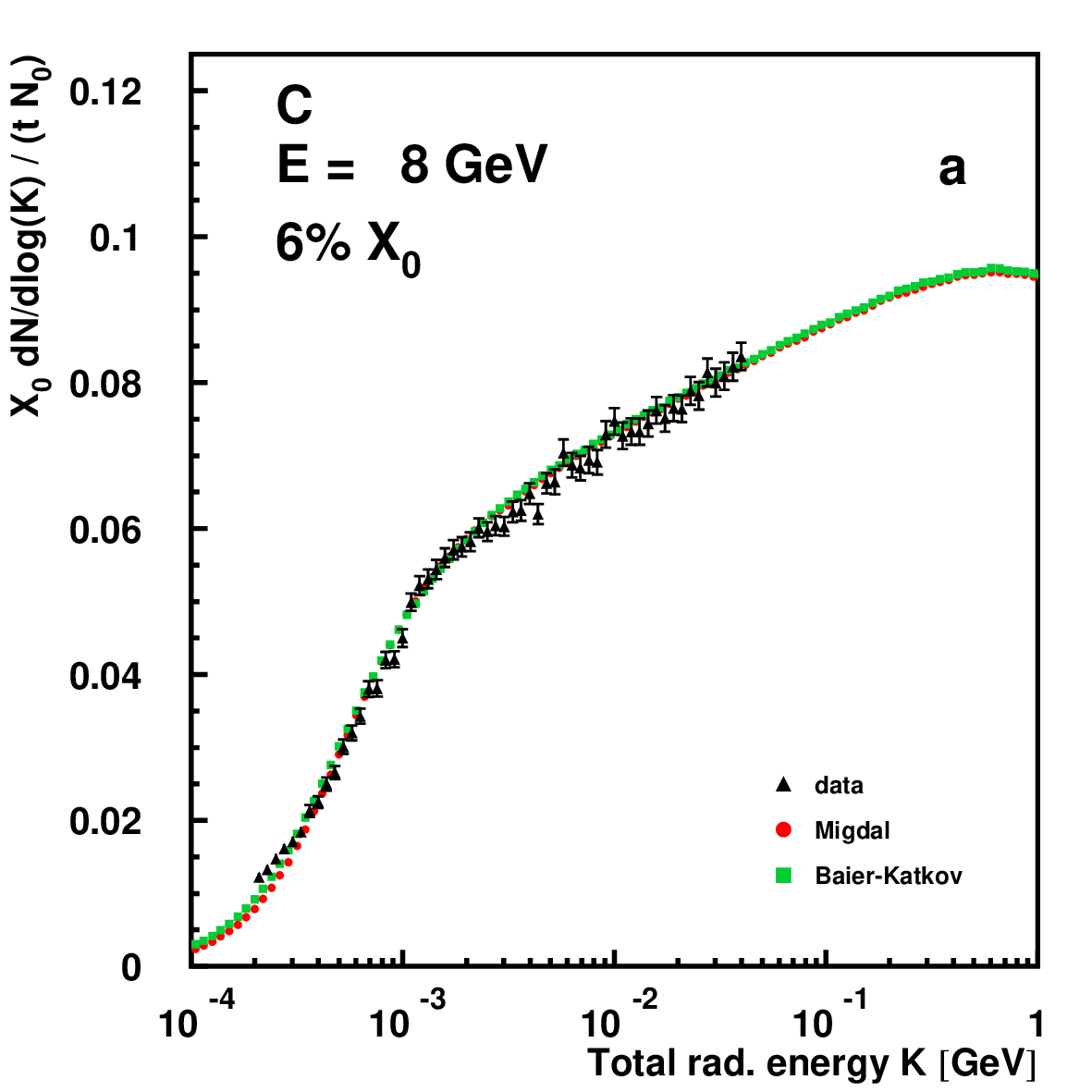} &
    \includegraphics[width=.425\textwidth]{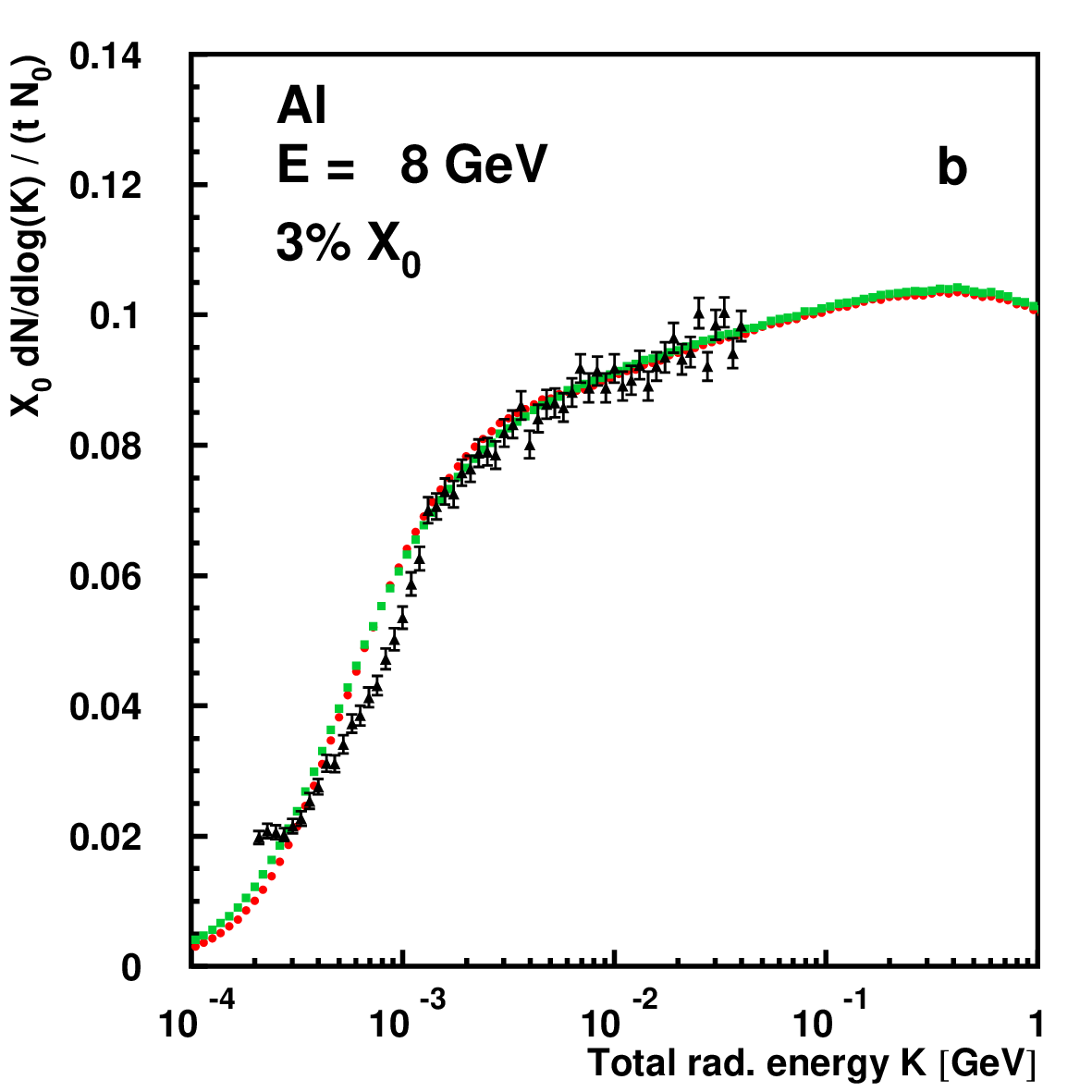} \\
    \includegraphics[width=.425\textwidth]{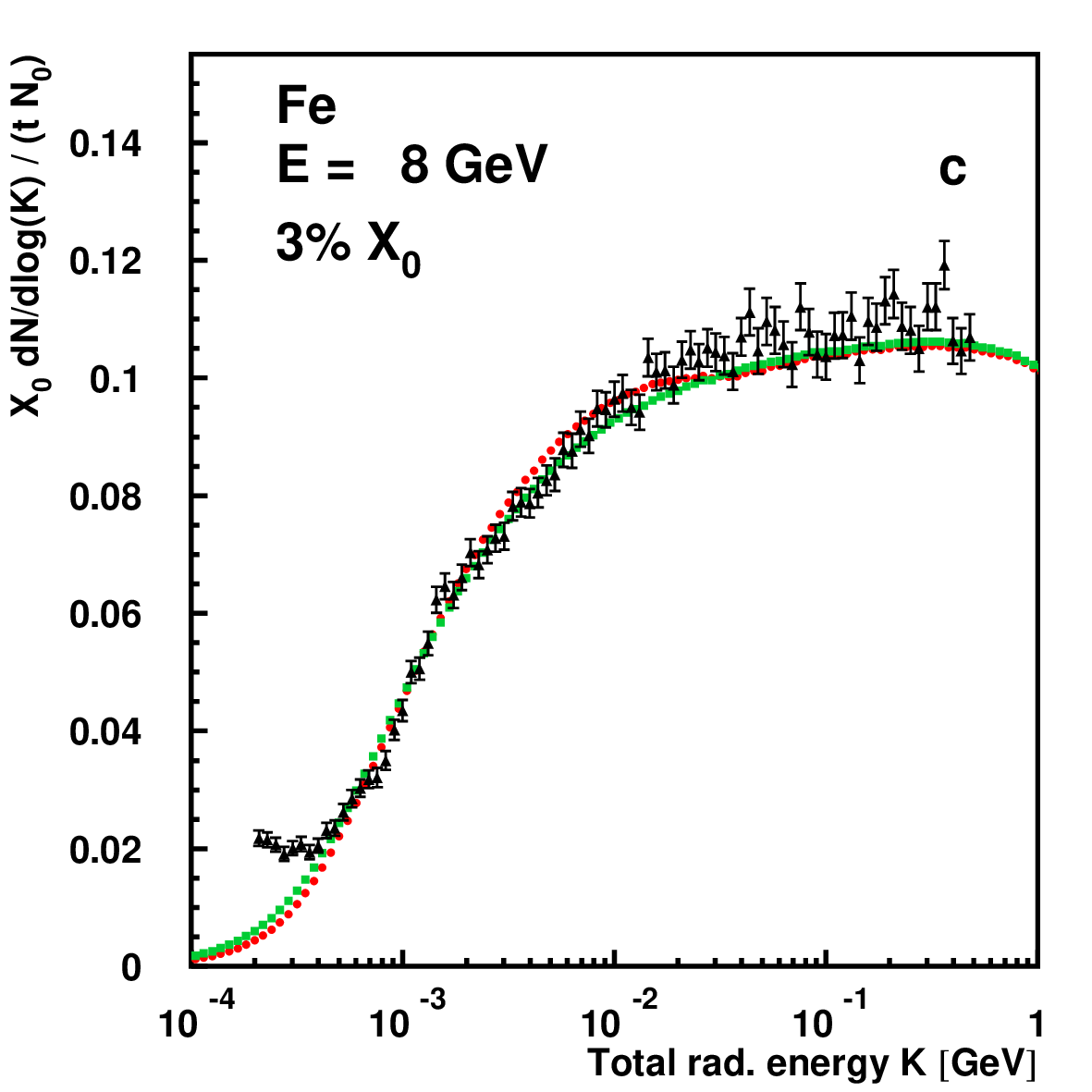} &
    \includegraphics[width=.425\textwidth]{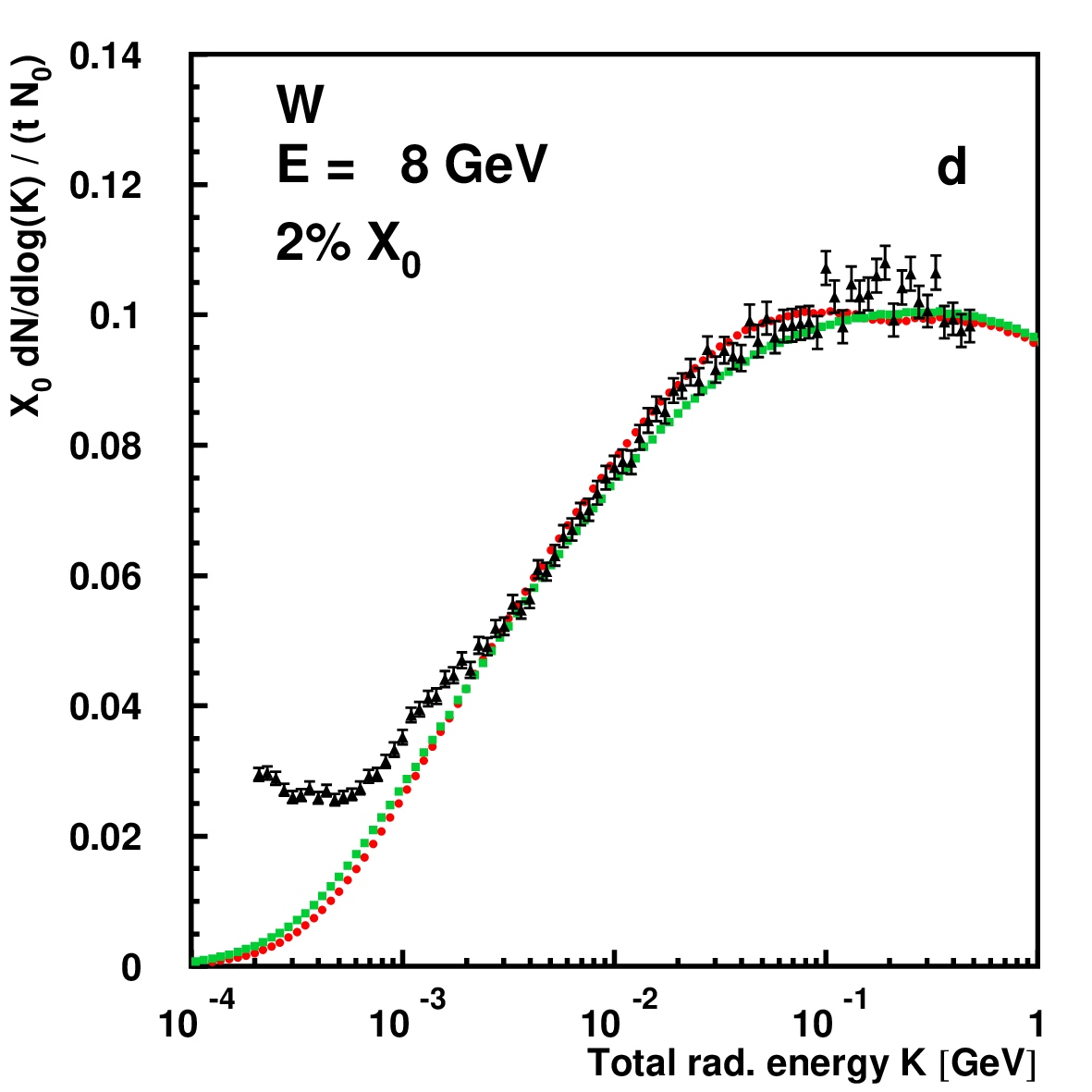} \\
    \includegraphics[width=.425\textwidth]{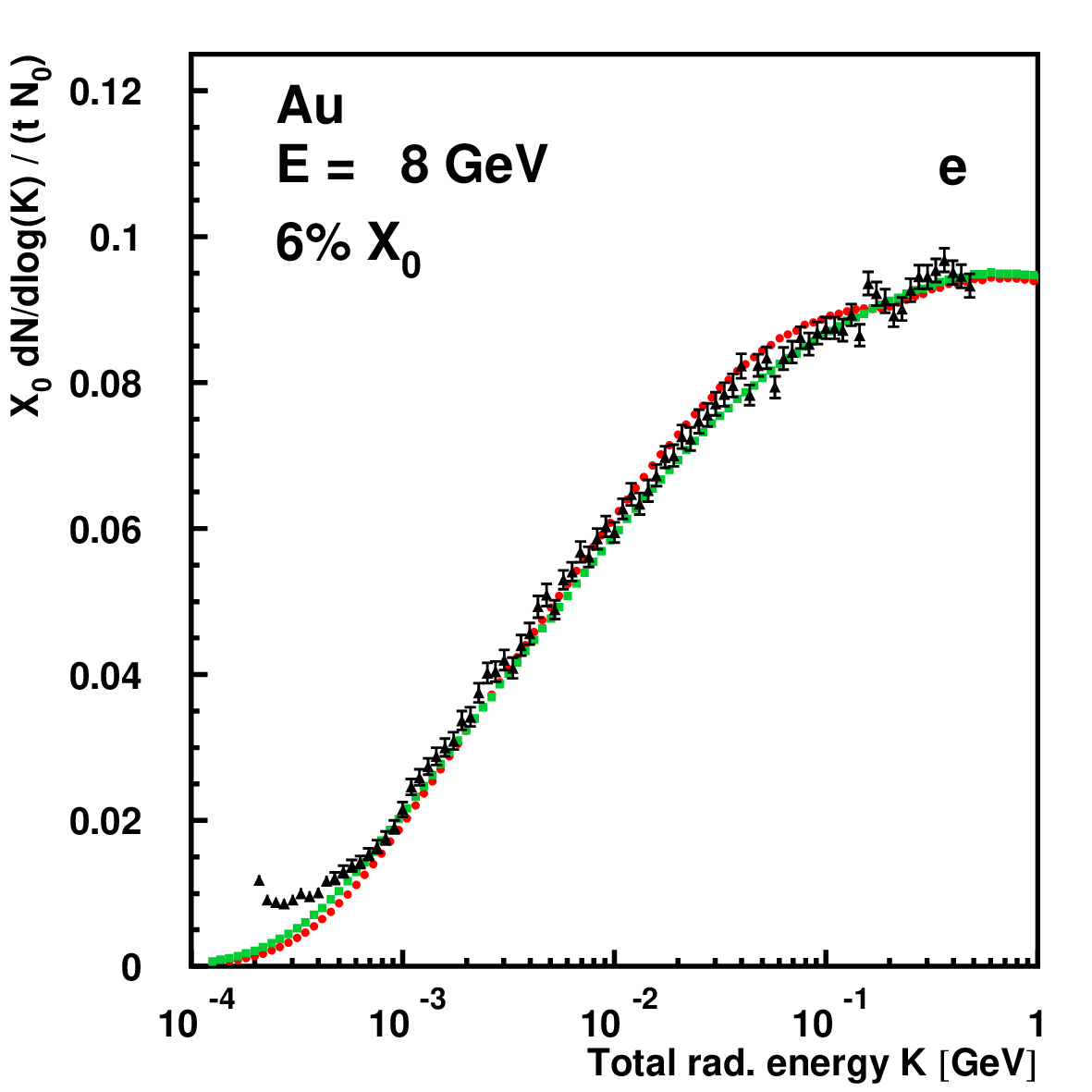} &
    \includegraphics[width=.425\textwidth]{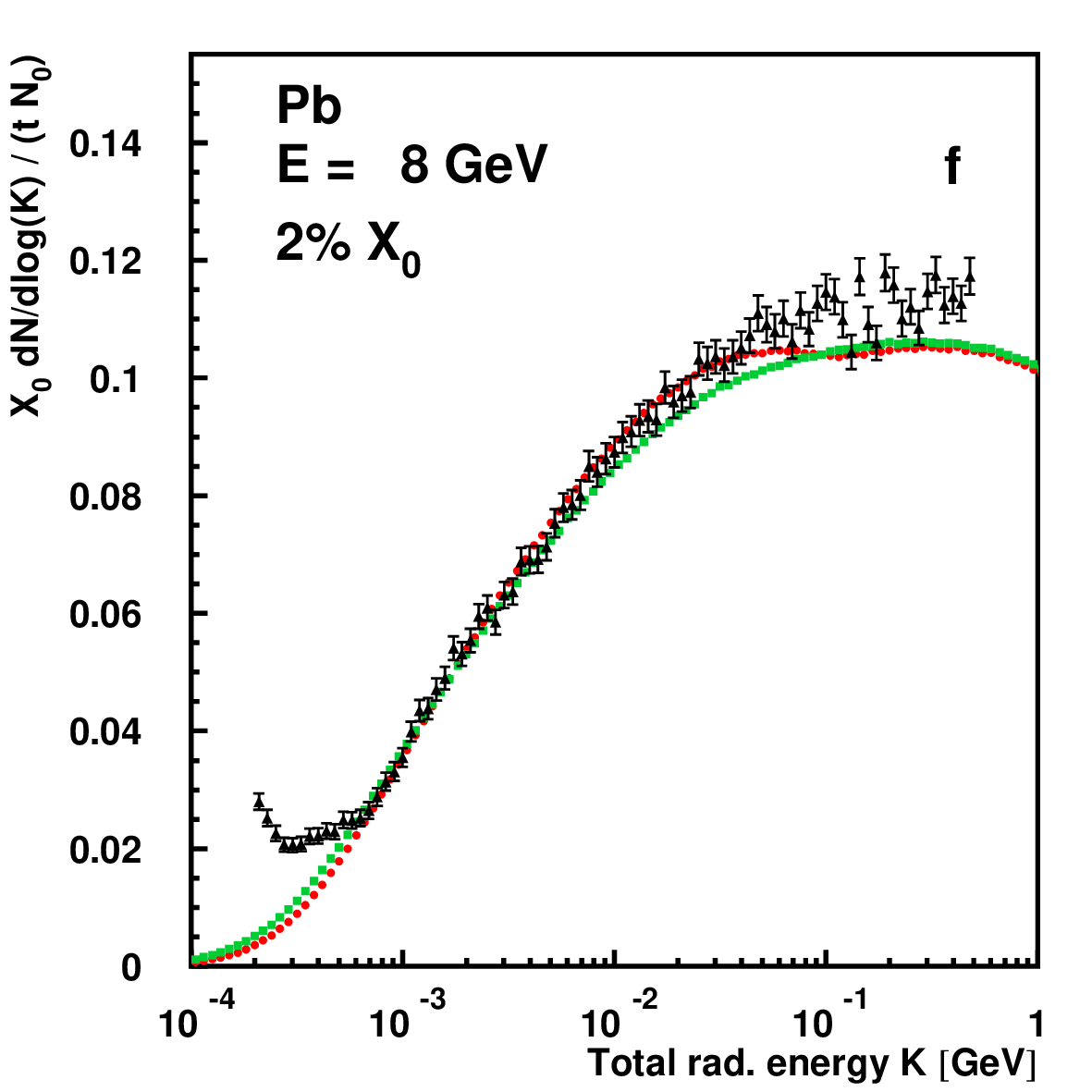}
  \end{tabular}
  \caption{\label{fig:slac_8GeV}Comparison of the simulations based on
    the Migdal (solid circles) and BK approaches (solid squares) with
    the SLAC E-146 data at $8$~GeV (solid triangles).}
\end{figure*}

\begin{figure*}[hp!!!!]
  \centering
  \begin{tabular}{cc}
    \includegraphics[width=.425\textwidth]{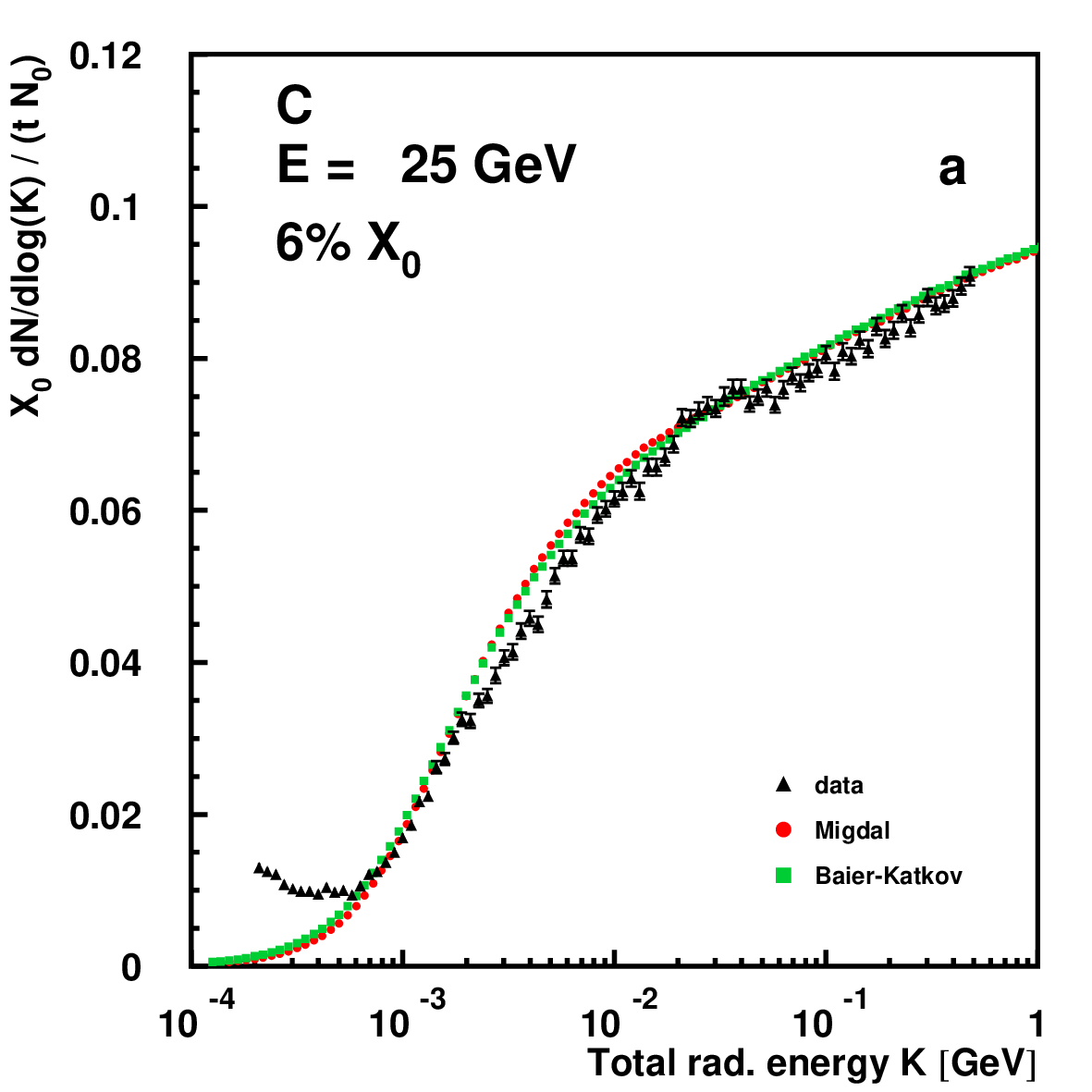} &
    \includegraphics[width=.425\textwidth]{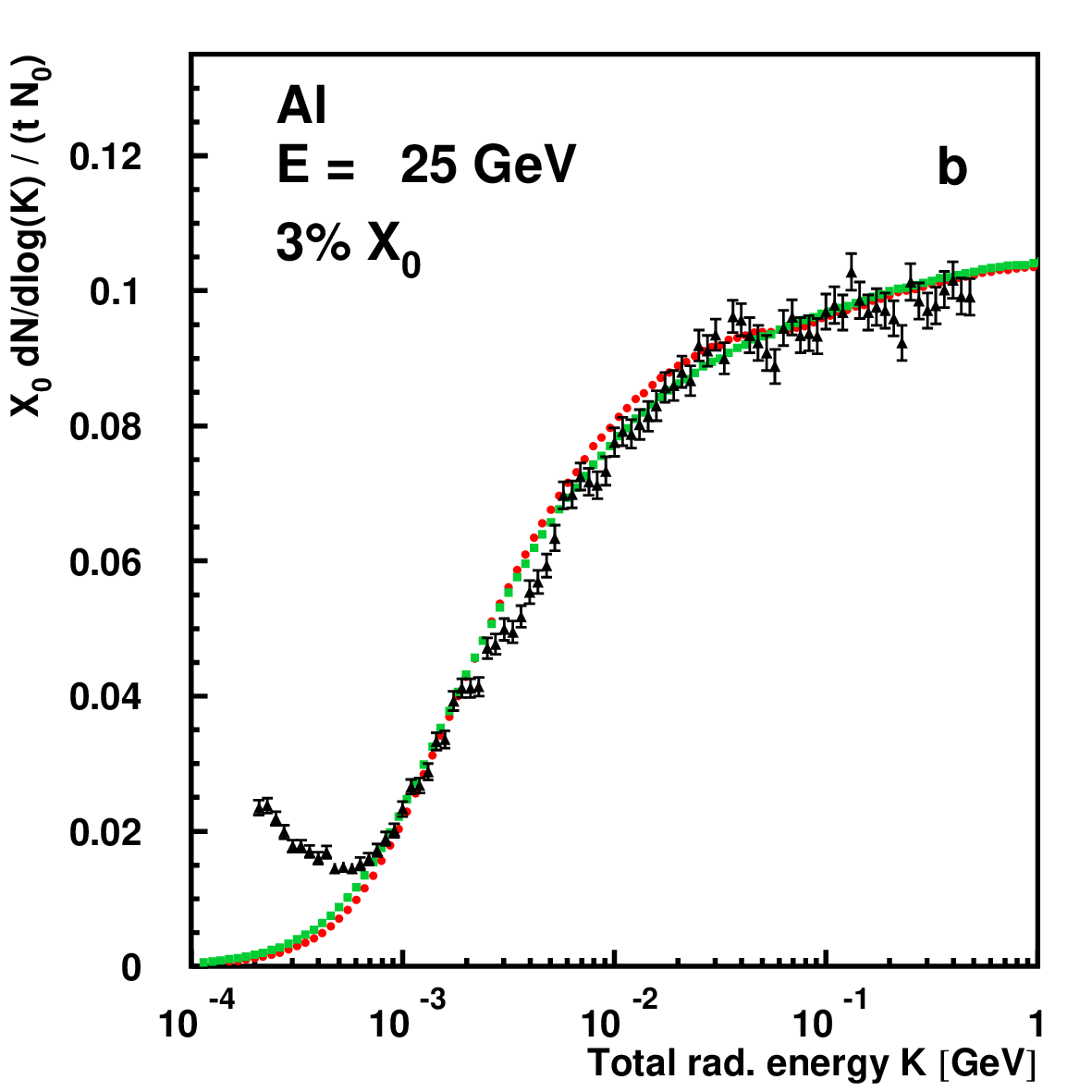} \\
    \includegraphics[width=.425\textwidth]{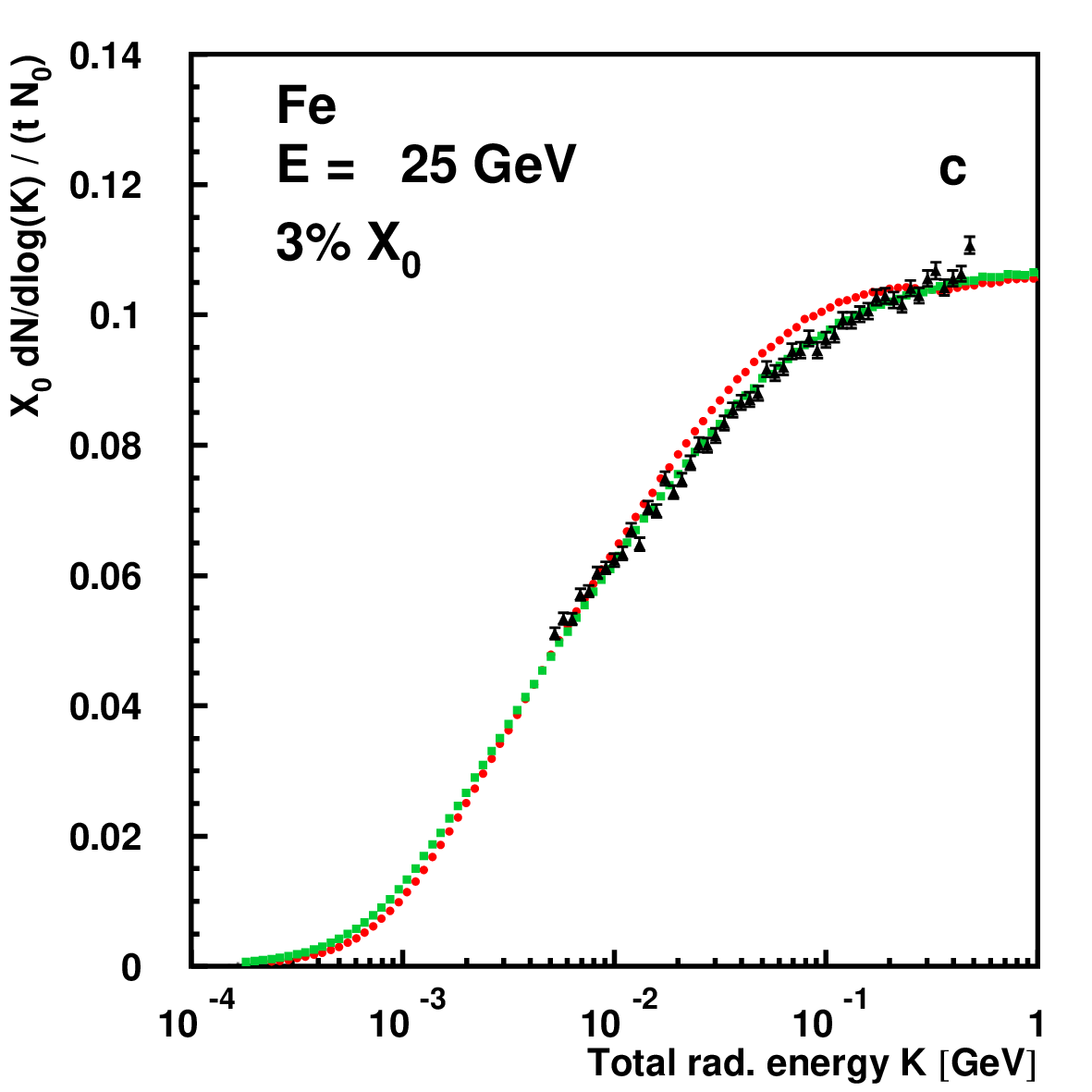} &
    \includegraphics[width=.425\textwidth]{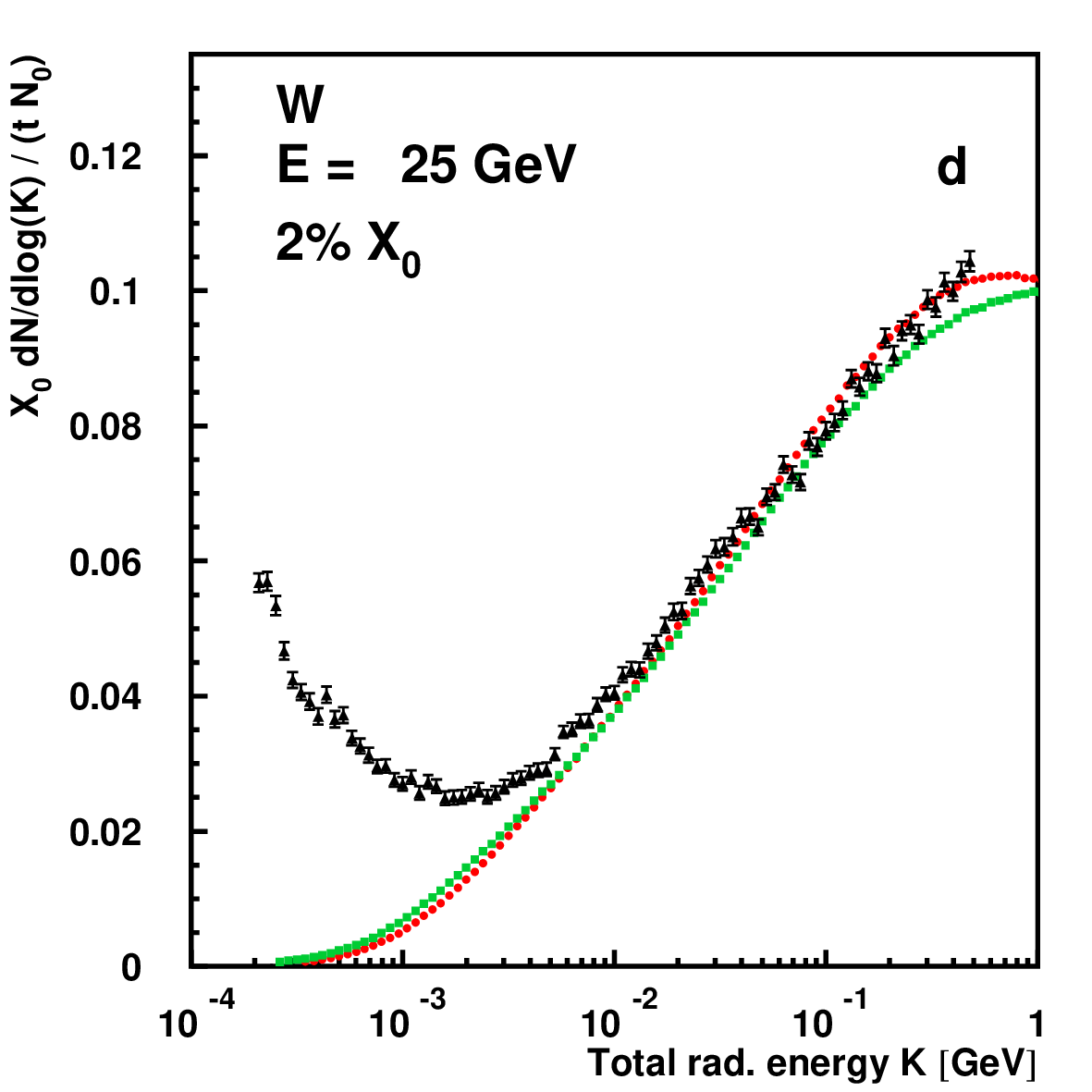} \\
    \includegraphics[width=.425\textwidth]{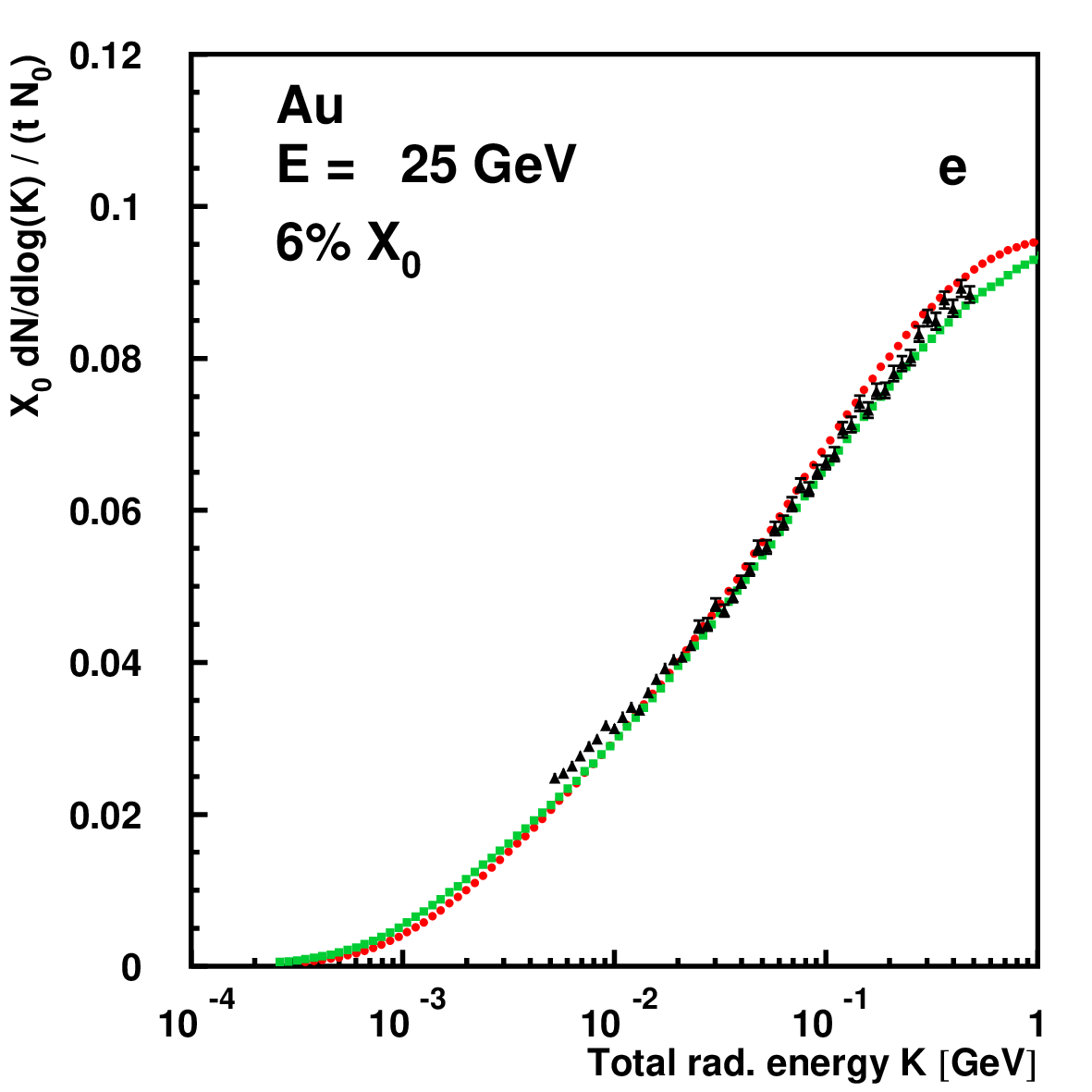} &
    \includegraphics[width=.425\textwidth]{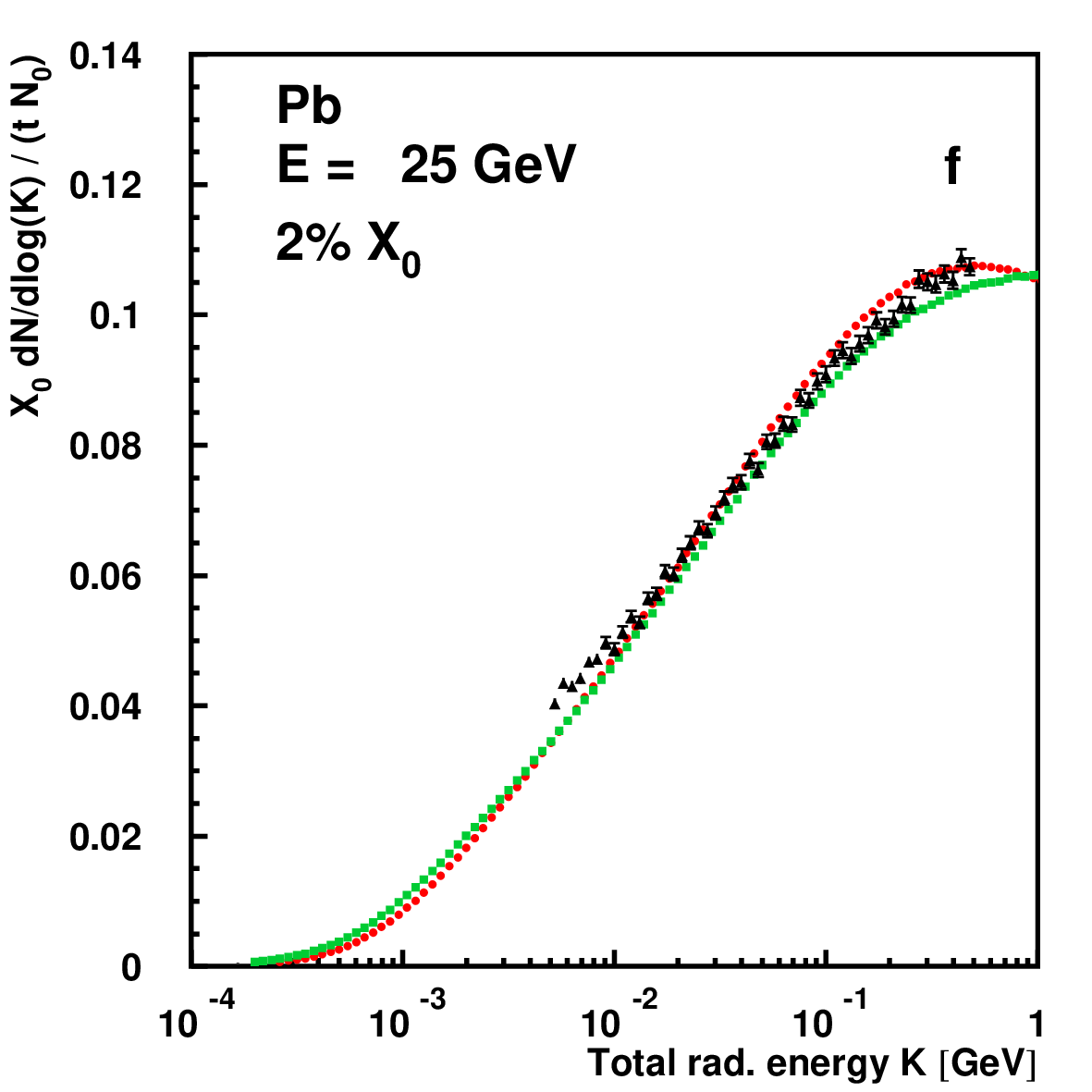}
  \end{tabular}
  \caption{\label{fig:slac_25GeV}Comparison of the simulations based
    on the Migdal (solid circles) and BK approaches (solid squares)
    with the SLAC E-146 data at $25$~GeV (solid triangles).}
\end{figure*}

\begin{enumerate}
\item The SLAC E-146 experiment used $8$- and $25$-GeV electron beams
  covering more than $3$ orders of magnitude in photon energy but
  limited below $500$~MeV, thus leaving the region close to the SWL
  unmeasured. The CERN LPM measurements reached the highest beam
  energies so far, up to $287$~GeV, and covered the spectral region of
  the SWL, but was downward limited to a photon energy above
  $2$~GeV. The CERN LOW-$Z$ data employed $178$-GeV electrons but
  again covered only a part of the spectrum from $50$~MeV to
  $3.8$~GeV. From the considerations exposed in
  Sec.~\ref{sec:th_comp}, the highest discriminating power between the
  Migdal and BK approaches is expected for the CERN LPM results.
\item No inclusion of dielectric suppression is necessary for the CERN
  LPM data so that it has been disabled in the simulations. On the
  contrary, it is always taken into account for comparing with the
  SLAC E-146 and CERN LOW-$Z$ results.
\item The SLAC E-146 experiment reached photon energies down to
  $200$~keV, where transition radiation is present and should be
  incorporated into the theoretical description. The dominance of such
  a contribution can easily be spotted by the upturn of the
  spectrum. The Monte Carlo employed in the present work does not
  include transition radiation, so that the part of the spectrum at
  lower photon energies must be excluded from the comparison to the
  data.
\item The SLAC E-146 study included rather thin gold targets
  corresponding to $0.1$~\% and $0.7$~\%~$X_{0}$ to investigate
  surface effects. They cannot be compared with the present Monte
  Carlo simulations based on approaches valid for infinite targets
  (see Sec.~\ref{sec:th_comp}).
\item The photon flight-path, from the target to the calorimeter, was
  $\approx 50$~m in the SLAC E-146 experiment and was kept in vacuum,
  except for a very thin exit window in front of the
  calorimeter~\cite{Anthony:1997}, as mentioned. Note that even if
  this length can be decreased using a larger field magnet to deflect
  the electrons that crossed the target, this has the price of a
  stronger and higher-energy synchrotron radiation. As a matter of
  fact, synchrotron radiation was present in the SLAC E-146 setup and
  had to be removed by a special cut on the reconstructed position of
  photon impact on the calorimeter (which was
  segmented)~\cite{Anthony:1997}. The final amount of background is
  quoted in Ref.~\cite{Anthony:1997} as, on average, $1$ photon in the
  $200$-keV to $500$-MeV energy range per $1000$ electrons. The
  spectra shown in Fig.~4 of Ref.~\cite{Anthony:1997} do not appear to
  have a definite BH profile, but by assuming such a shape as a
  reasonable approximation, an equivalent background of
  $0.01$~\%~$X_{0}$ can be estimated. Thus, it is not necessary to
  handle background subtraction in the Monte Carlo code for the SLAC
  E-146 setup. As noted in Sec.~\ref{sec:multiph_bksub}, the same
  solution could not easily be applied to the CERN LPM or CERN LOW-$Z$
  setups employed at energies about an order of magnitude
  larger. Indeed, the length of the photon path in the CERN LPM case
  was $\approx 80$~m, part in helium and part in
  air~\cite{Hansen:2003,Hansen:2004}. The corresponding background,
  mostly due to the aluminized mylar windows of the first drift
  chamber located between the target and the deflecting magnet, was
  found in Refs.~\cite{Hansen:2003,Hansen:2004} to be equivalent to
  $0.7$~\%~$X_{0}$. As demonstrated in Sec.~\ref{sec:multiph_bksub},
  such a background level must be explicitly handled in the processing
  of the Monte Carlo results by following the same subtraction
  procedure adopted to derive the data. As shown in Fig.~5 of
  Ref.~\cite{Hansen:2004}, the background spectrum has, in quite good
  an approximation, a BH shape.  For all the results shown here, we
  use a secondary carbon target (placed downstream from the primary
  one) with a thickness of $0.7$~\%~$X_{0}$. To speed up the
  simulation, a BH spectrum without LPM suppression is assumed for the
  secondary target. The CERN LOW-$Z$ case was similar, except that
  background subtraction was not performed, to avoid any distortion,
  and thus the background has to be explicitly added in the
  simulations to compare with data. The background was equivalent to
  $3$~\%~$X_{0}$~\cite{Andersen:2013} and again had a BH shape to a
  quite good approximation for the energy range covered by the final
  data (see Fig.~11 of Ref.~\cite{Andersen:2013}). Again, a secondary
  target made of carbon with a thickness of $3$~\%~$X_{0}$ and a BH
  spectrum has been placed downstream the primary one in all
  simulations.
\end{enumerate}

It should be emphasized that all comparisons to be presented here are
absolute, i.e. no free overall normalization factor is
present. Moreover, both the Migdal and BK approaches have been
reformulated, as described in Secs.~\ref{sec:Migdal} and~\ref{sec:BK},
respectively, to adopt the same radiation lengths tabulated by
Tsai~\cite{Tsai:1974,Tsai:1977}.

The vertical scales of the data agree with the original publications,
to ease the comparison. Unfortunately, different choices were made by
the original authors. In general, the number of events in each channel
$dN$ has always been normalized to the total number of impinging
electrons $N_{0}$ (thus including those that do not radiate). In the
case of the SLAC E-146 values, a normalization to the target
thicknesses $t$ in units of $X_{0}$, i.e.~$t/X_{0}$, has been included
and, in that of the CERN LOW-$Z$ ones, the bin width $w_{\mathrm{b}}$
has been taken into account.

\begin{figure}[t!!!!]
  \centering
  \begin{tabular}{c}
    \includegraphics[width=.425\textwidth]{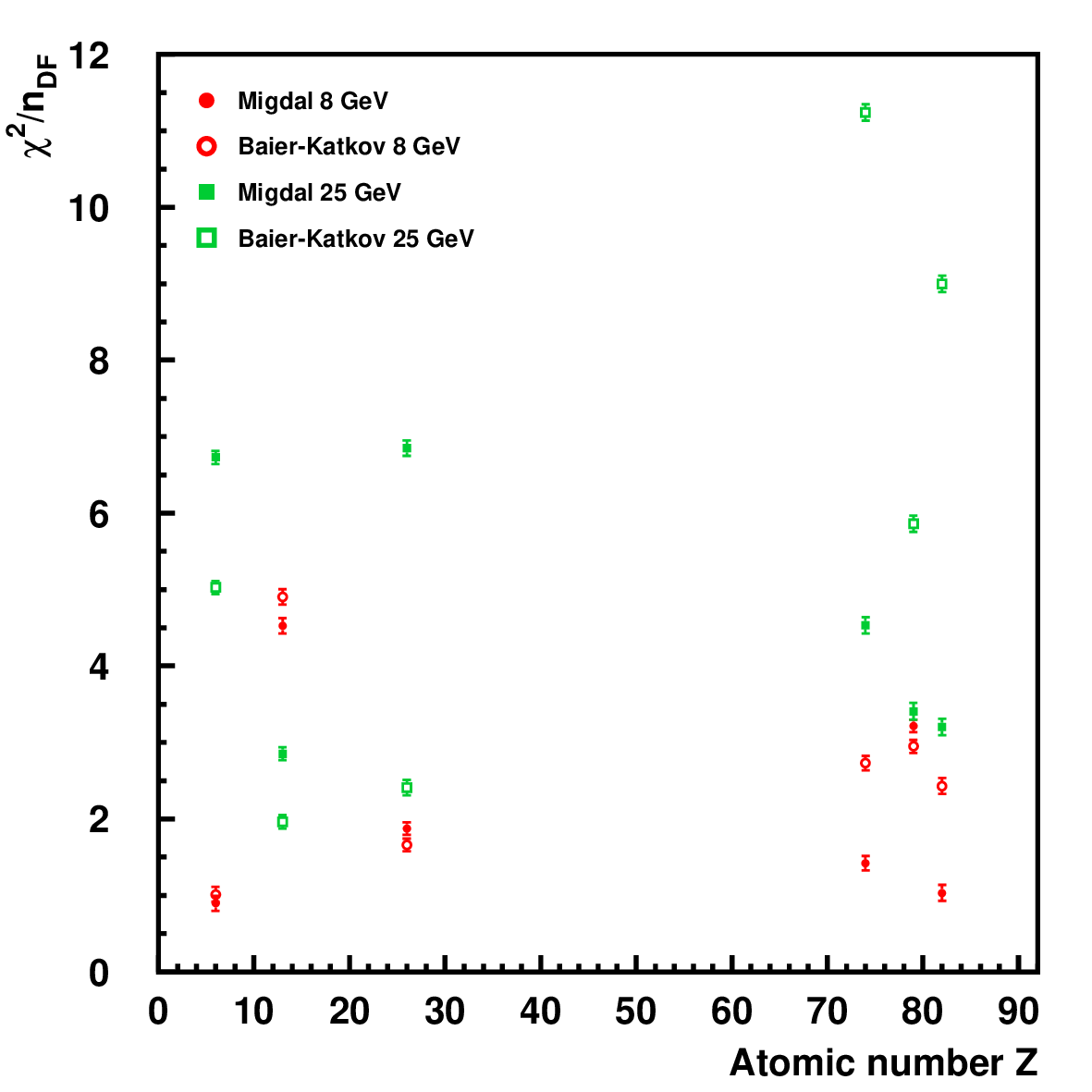}
  \end{tabular}
  \caption{\label{fig:slac_chi2}Values of $\chi^{2}/n_{\mathrm{DF}}$
    for the comparison of the simulations based on the Migdal (solid
    symbols) and BK approaches (open symbols) with the SLAC E-146 data
    at $8$ (circles) and $25$~GeV (squares) as a function of the
    atomic number of the target $Z$. The bars represent the standard
    deviation of the reduced $\chi^{2}$.}
\end{figure}

The comparison between the final full Monte Carlo simulations (see
Sec.~\ref{sec:sim_par} for the parameters) and the SLAC E-146 data,
collected for $E=8$- and $25$-GeV electron beams, are shown in
Figs.~\ref{fig:slac_8GeV} and~\ref{fig:slac_25GeV}, respectively. The
main apparent feature is that the Migdal and BK spectra are rather
close, especially at low $Z$ and for the lowest beam energy $E=8$~GeV.
As mentioned, the discriminating power of the experiment is small,
given also the statistical and systematic
uncertainties~\cite{Anthony:1997}, so that we deemed unnecessary to
compare with all the measured targets; we have excluded uranium and
kept only one thickness for each element. In agreement with the
discussion of the basic cross sections (see Sec.~\ref{sec:th_comp}),
it remains true, even after the full Monte Carlo, that the Migdal
predictions are above the BK one in the photon energy region of the
transition between the BH limit at the SWL and the strong LPM
suppression on the opposite side of the spectrum. On the other hand,
the discontinuity in the first derivative of the basic cross section
does not appear in the simulations due to the smearing brought about
by the multiphoton emission. The effect of transition radiation is
also very clear, note in particular the tungsten target for
$E=25$~GeV. Since it is not included in the Monte Carlo, the simulated
spectra steady drop towards low photon energies and do not show the
characteristic upturn of the experiment.

\begin{table}[t!!!!]
  \centering
  \begin{tabular}{||r|r|c|c|c||}\hline
&
\multicolumn{1}{||c|}{$K_{\mathrm{min}}$} &
$\chi^{2}/n_{\mathrm{DF}}$ &
$\chi^{2}/n_{\mathrm{DF}}$ &
$n_{\mathrm{DF}}$ \\
&
\multicolumn{1}{||c|}{cut} &
Migdal &
BK &
\\
&
\multicolumn{1}{||c|}{[MeV]} &
&
&
\\\hline
\multicolumn{5}{||c||}{8~GeV}\\\hline
 C 6 \% & $0.4$ &
  $  0.9$ &
  $  1.0$ &
  $ 51$ \\
Al 3 \% & $0.4$ &
  $  4.5$ &
  $  4.9$ &
  $ 51$ \\
Fe 3 \% & $0.5$ &
  $  1.9$ &
  $  1.7$ &
  $ 76$ \\
 W 2 \% & $3.0$ &
  $  1.4$ &
  $  2.7$ &
  $ 57$ \\
Au 6 \% & $0.7$ &
  $  3.2$ &
  $  2.9$ &
  $ 72$ \\
Pb 2 \% & $0.7$ &
  $  1.0$ &
  $  2.4$ &
  $ 47$ \\
\hline
\multicolumn{5}{||c||}{25~GeV}\\\hline
 C 6 \% &  $0.8$ &
  $  6.7$ &
  $  5.0$ &
  $ 71$ \\
Al 3 \% &  $1.0$ &
  $  2.9$ &
  $  2.0$ &
  $ 68$ \\
Fe 3 \% &  $6.0$ &
  $  6.8$ &
  $  2.4$ &
  $ 49$ \\
 W 2 \% & $10.0$ &
  $  4.5$ &
  $ 11.2$ &
  $ 43$ \\
Au 6 \% & $10.0$ &
  $  3.4$ &
  $  5.9$ &
  $ 43$ \\
Pb 2 \% & $10.0$ &
  $  3.2$ &
  $  9.0$ &
  $ 43$ \\
\hline
\end{tabular}

  \vspace{0.5cm}
  \caption{Values of $\chi^{2}/n_{\mathrm{DF}}$ for the comparison of
    the simulations with the SLAC E-146 data. The uppermost energy
    is in all cases the maximum reported in Ref.~\cite{Anthony:1997},
    i.e.~$500$~MeV. The indicated low cut, $K_{\mathrm{min}}$, on the
    radiated energy has been applied to avoid the contribution from
    transition radiation, which is not taken into account in the
    Monte Carlo.}
  \label{tab:slac_chi2}
\end{table}

A more objective comparison of the simulation with the SLAC E-146 data
can be made on the basis of the $\chi^{2}$ normalized to the number of
degrees of freedom $n_{\mathrm{DF}}$. The simulation has the same
logarithmic binning of the data ($25$ per decade) and a direct
calculation of $\chi^{2}$ is possible without interpolation. The
contribution of the statistical error of the simulations is negligible
when compared to the experimental one. The values of
$\chi^{2}/n_\mathrm{DF}$ are reported in Fig.~\ref{fig:slac_chi2} and
Table~\ref{tab:slac_chi2}. Unfortunately, no clear-cut general
preference for either approach is apparent.

Considering first the data for $E=8$~GeV, most of the targets show
rather similar values of $\chi^{2}/n_{\mathrm{DF}}$, except for
tungsten and lead, which favor the Migdal and BK theory,
respectively. Direct examination of Fig.~\ref{fig:slac_8GeV} confirms
these cases: in particular for lead, there is no doubt that the Migdal
approach is closer to the data. However, it is also necessary to note
that for iron, tungsten, and most notably lead, the experiment has a
tendency to overshoot all simulations at the high-energy end of the
spectrum. Even considering all targets at $E=8$~GeV, half has a lower
$\chi^{2}/n_{\mathrm{DF}}$ for the Migdal (carbon, aluminum, and
tungsten) and half for the BK (iron, gold, and lead) approach,
respectively.

\begin{figure}[ph!!!!]
  \centering
  \begin{tabular}{c}
    \includegraphics[width=.425\textwidth]{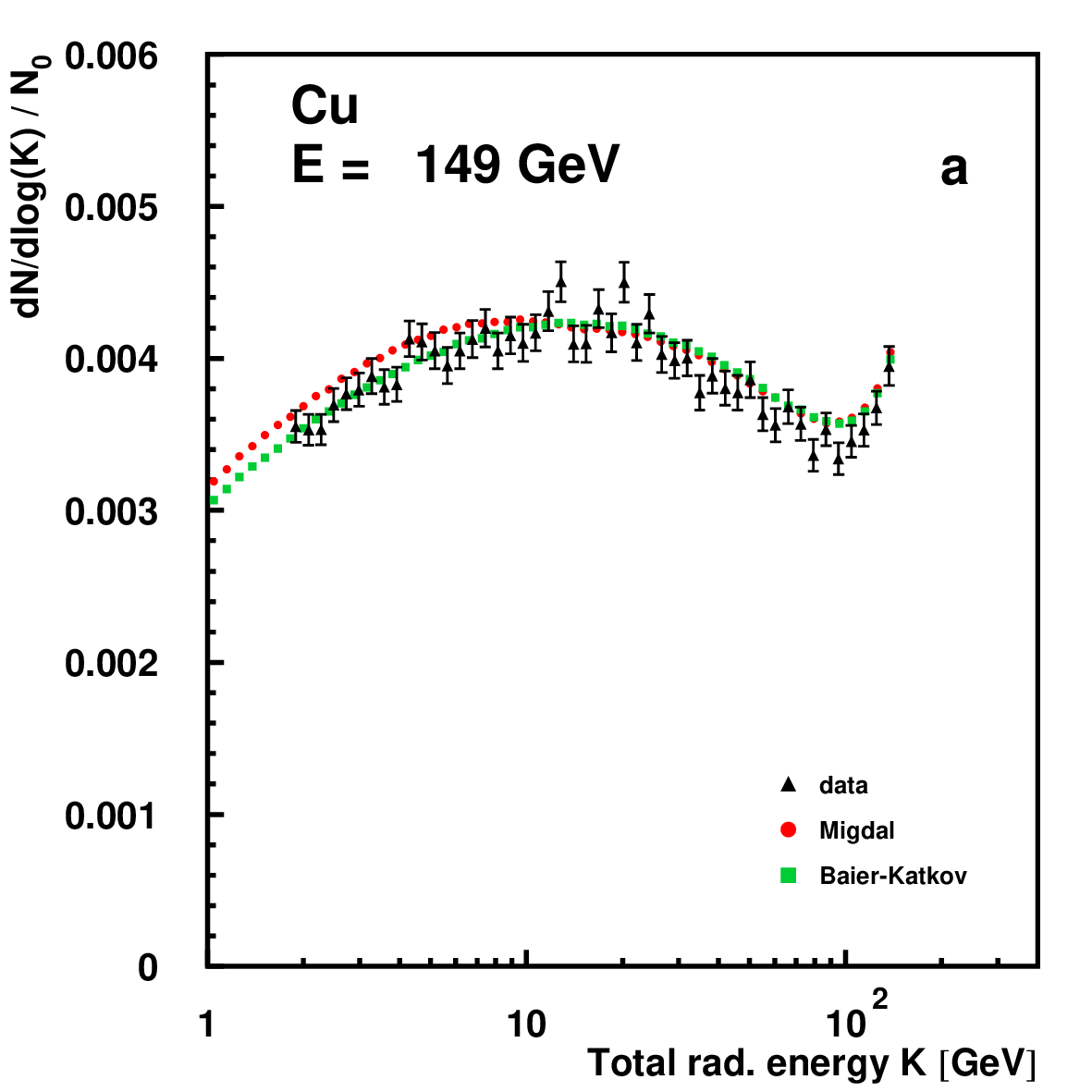} \\
    \includegraphics[width=.425\textwidth]{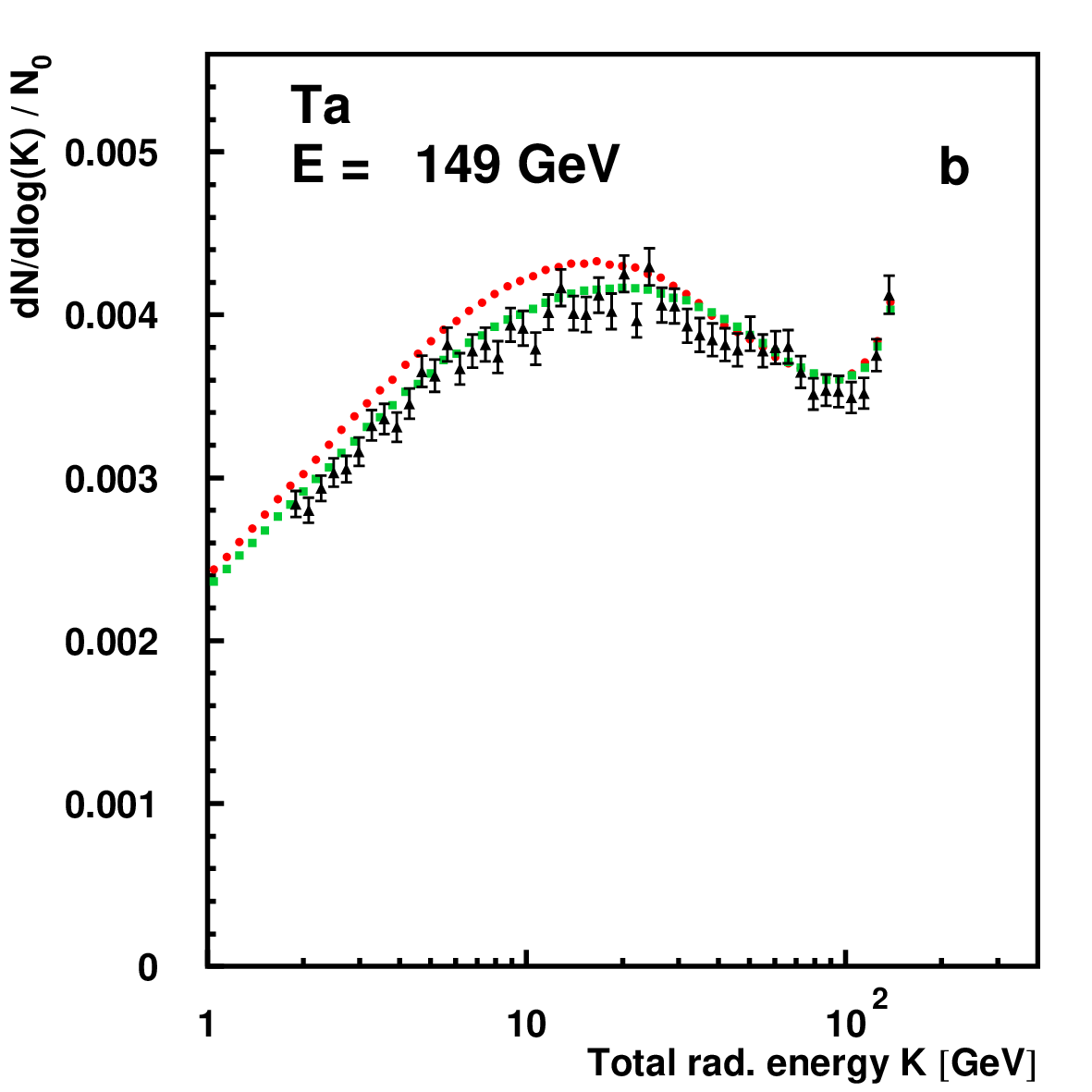} \\
    \includegraphics[width=.425\textwidth]{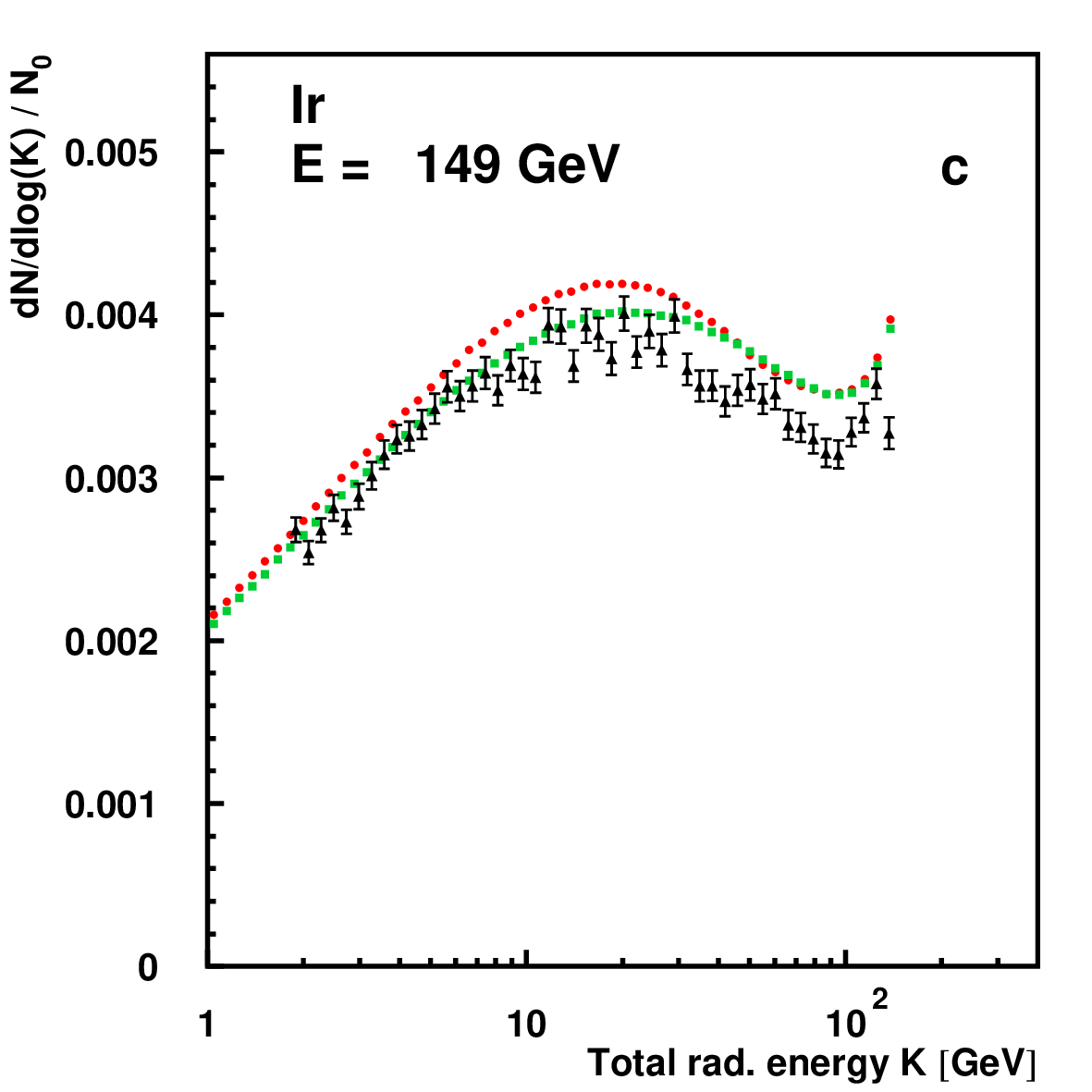}
  \end{tabular}
  \caption{\label{fig:cern_149GeV}Comparison of the simulations based
    on the Migdal (solid circles) and BK approaches (solid squares)
    with the CERN LPM data at $149$~GeV (solid triangles).}
\end{figure}

\begin{figure}[ph!!!!]
  \centering
  \begin{tabular}{c}
    \includegraphics[width=.425\textwidth]{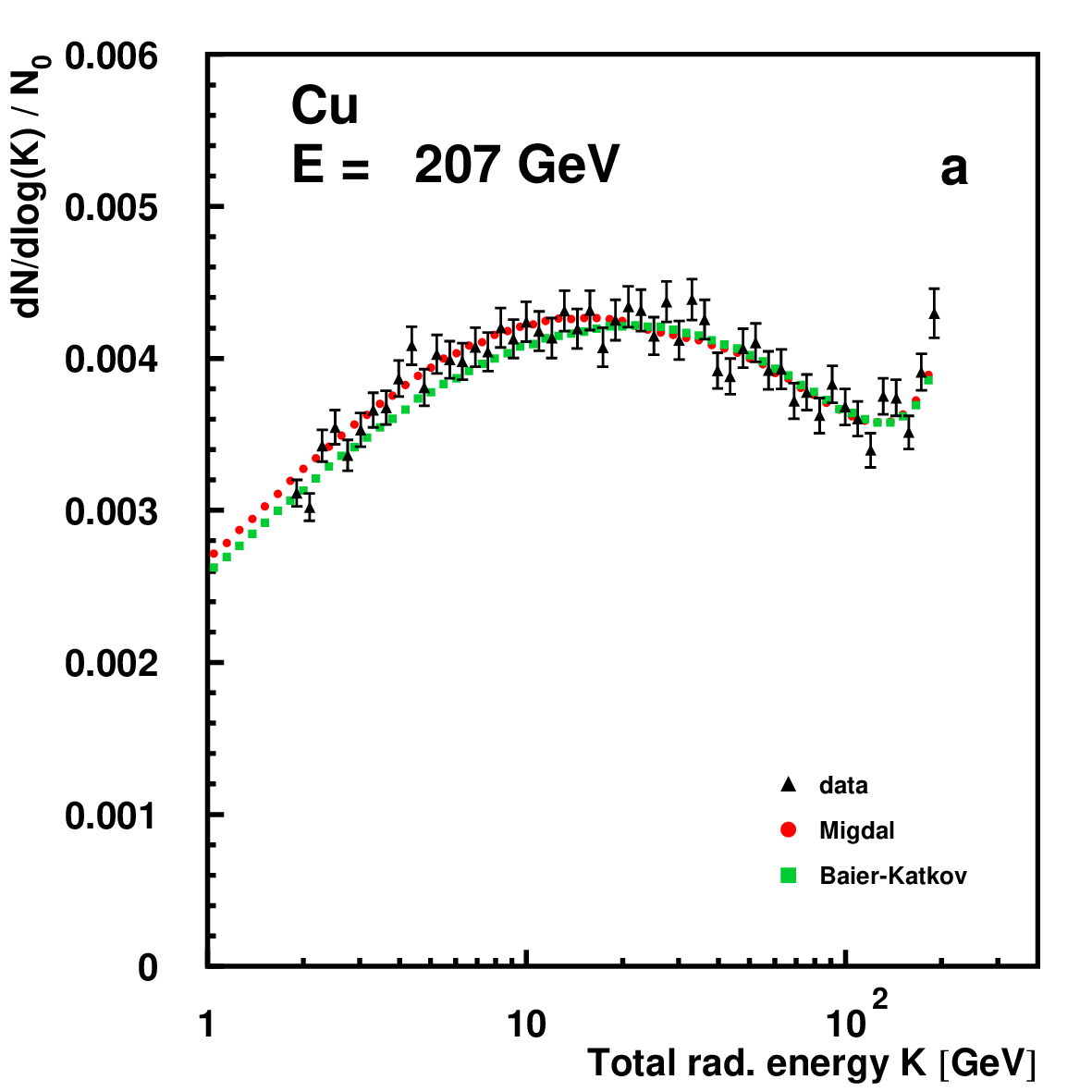} \\
    \includegraphics[width=.425\textwidth]{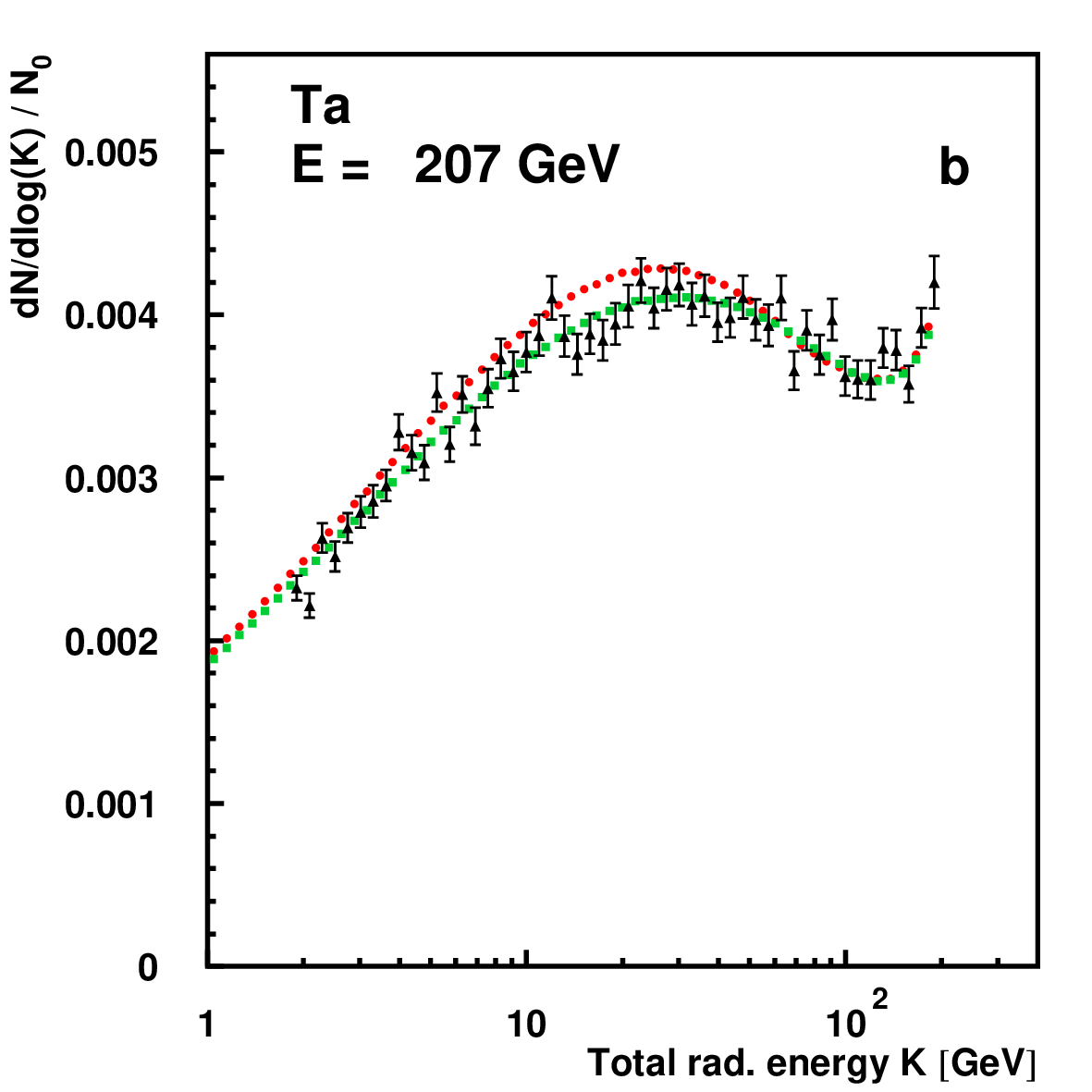} \\
    \includegraphics[width=.425\textwidth]{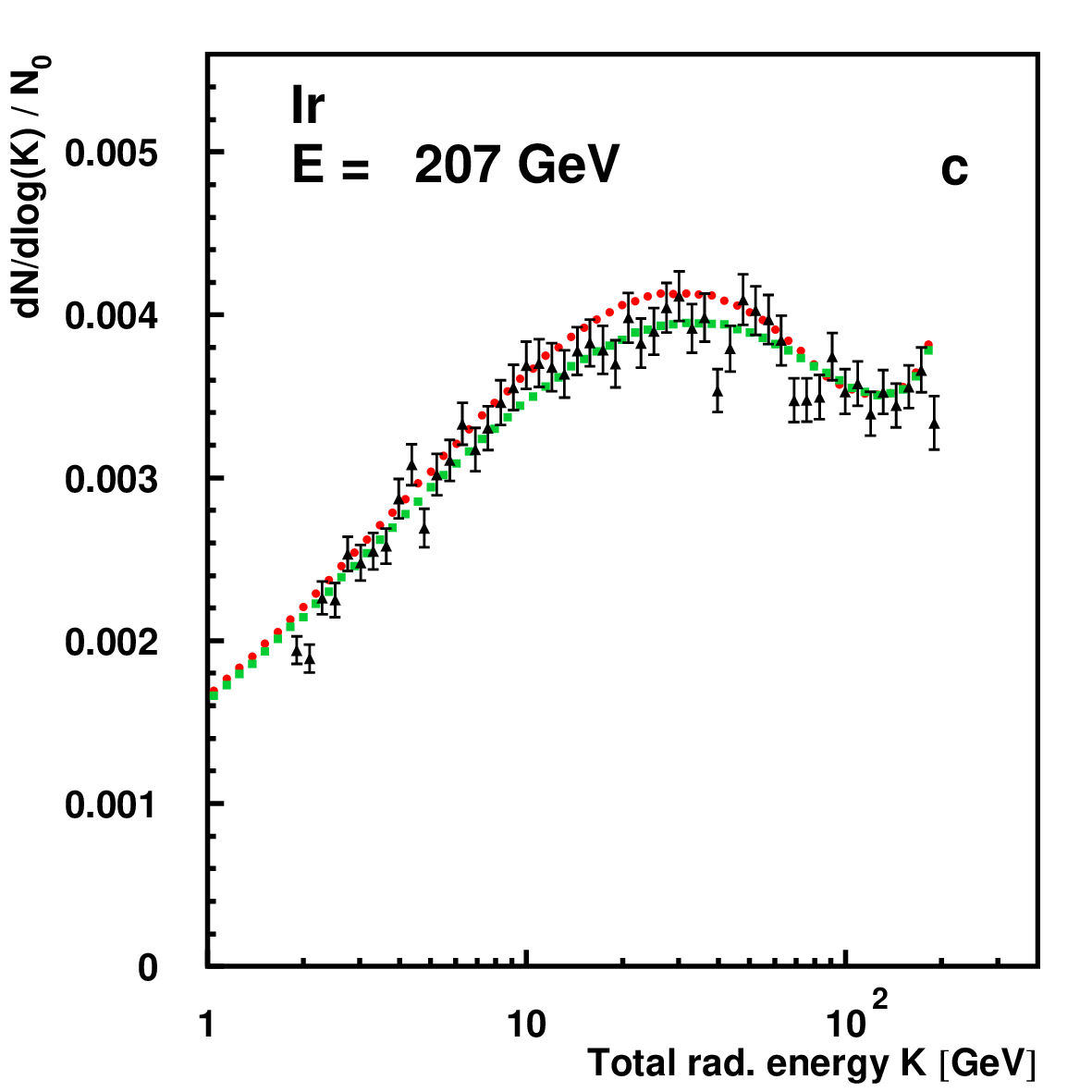}
  \end{tabular}
  \caption{\label{fig:cern_207GeV}Comparison of the simulations based
    on the Migdal (solid circles) and BK approaches (solid squares)
    with the CERN LPM data at $207$~GeV (solid triangles).}
\end{figure}

\begin{figure}[ph!!!!]
  \centering
  \begin{tabular}{c}
    \includegraphics[width=.425\textwidth]{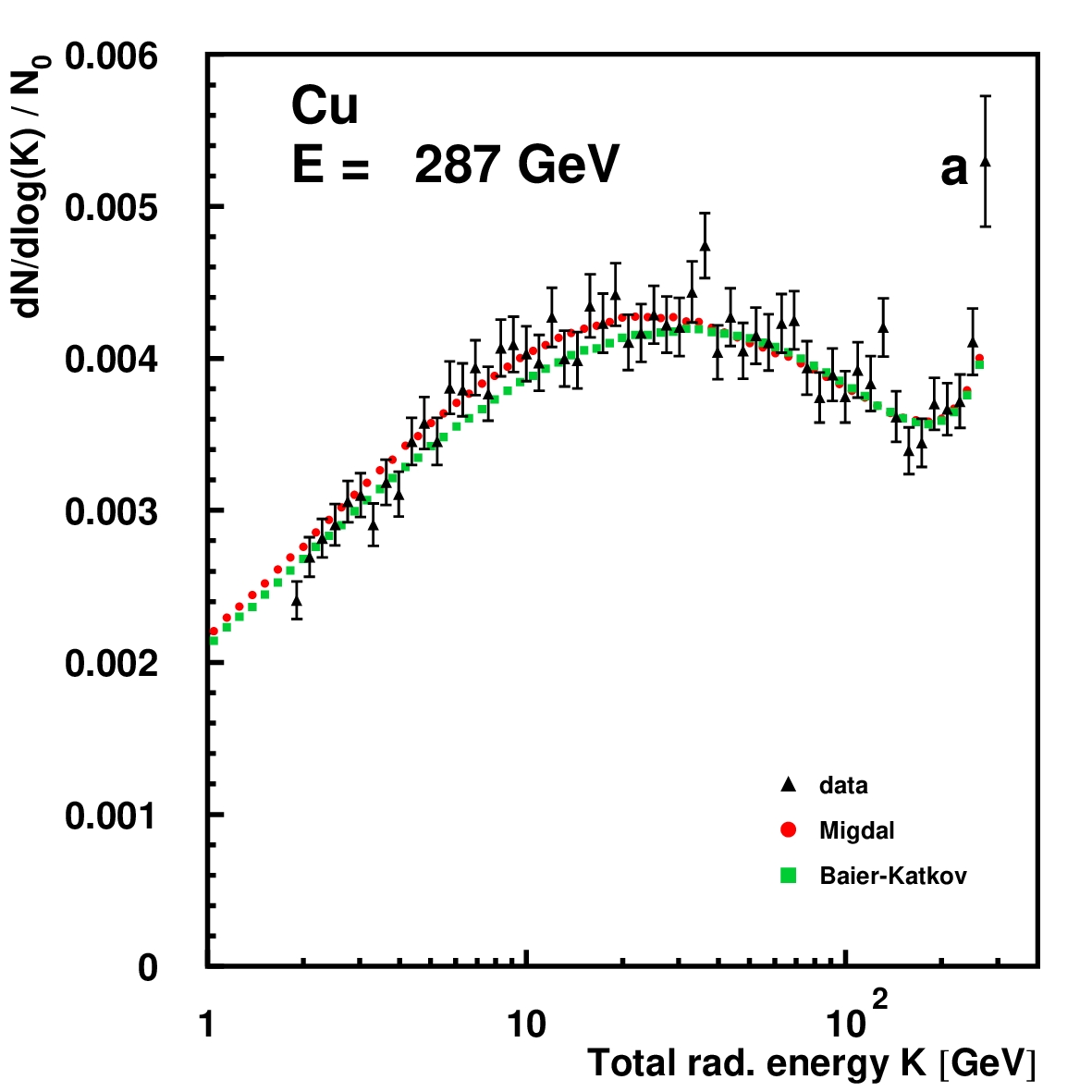} \\
    \includegraphics[width=.425\textwidth]{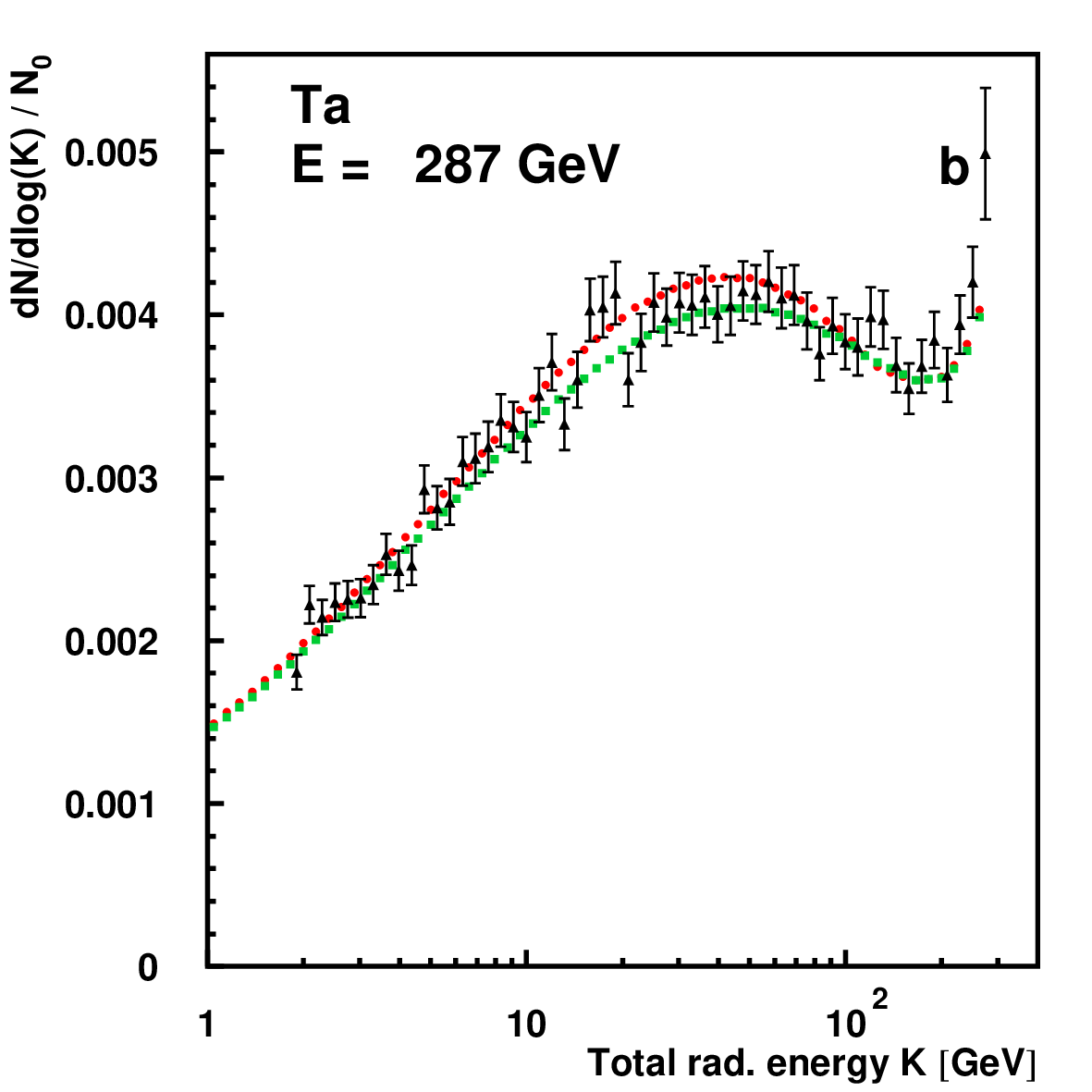} \\
    \includegraphics[width=.425\textwidth]{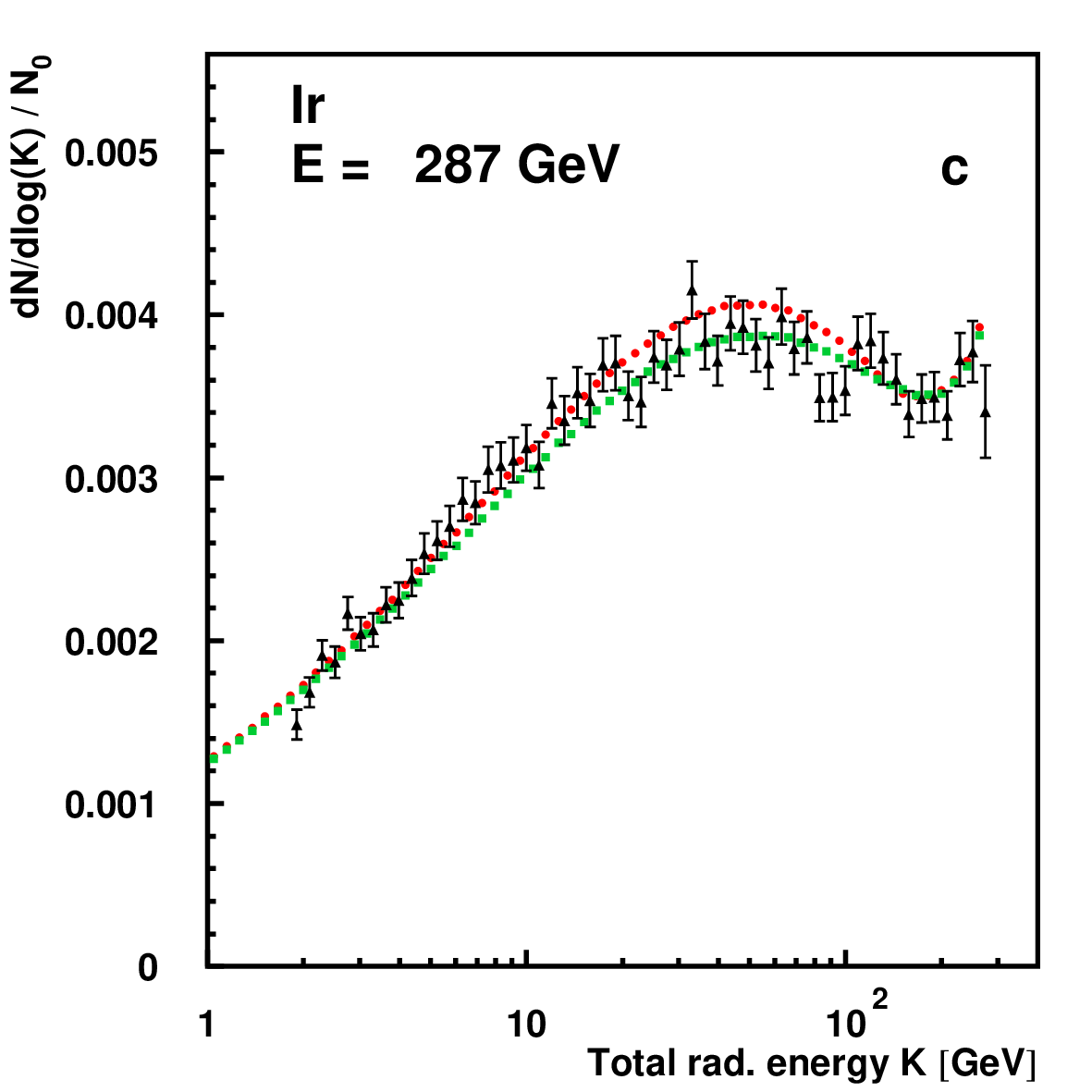}
  \end{tabular}
  \caption{\label{fig:cern_287GeV}Comparison of the simulations based
    on the Migdal (solid circles) and BK approaches (solid squares)
    with the CERN LPM data at $287$~GeV (solid triangles).}
\end{figure}

For $E=25$~GeV, the clearest preferences appear for iron, tungsten,
gold, and lead, of which only the first is in favor of the BK
approach. Inspection of Fig.~\ref{fig:slac_25GeV}, clearly shows that,
in the case of iron, the BK approach is particularly
successful. Again, the tendency of the measurements to stay higher in
the uppermost energy section of the spectrum results in the other
three cases being better reproduced by the Migdal
formulae. Considering all targets, still half has a lower
$\chi^{2}/n_{\mathrm{DF}}$ for the Migdal (tungsten, gold, and lead)
and half for the BK (carbon, aluminum, and iron) approach,
respectively. Only iron has the same preference for the same theory at
both energies. It is therefore safe to conclude that the SLAC E-146
data do not allow to unambiguously single out one of the approaches as
the best.

\begin{figure}[t!!!!]
  \centering
  \begin{tabular}{c}
    \includegraphics[width=.425\textwidth]{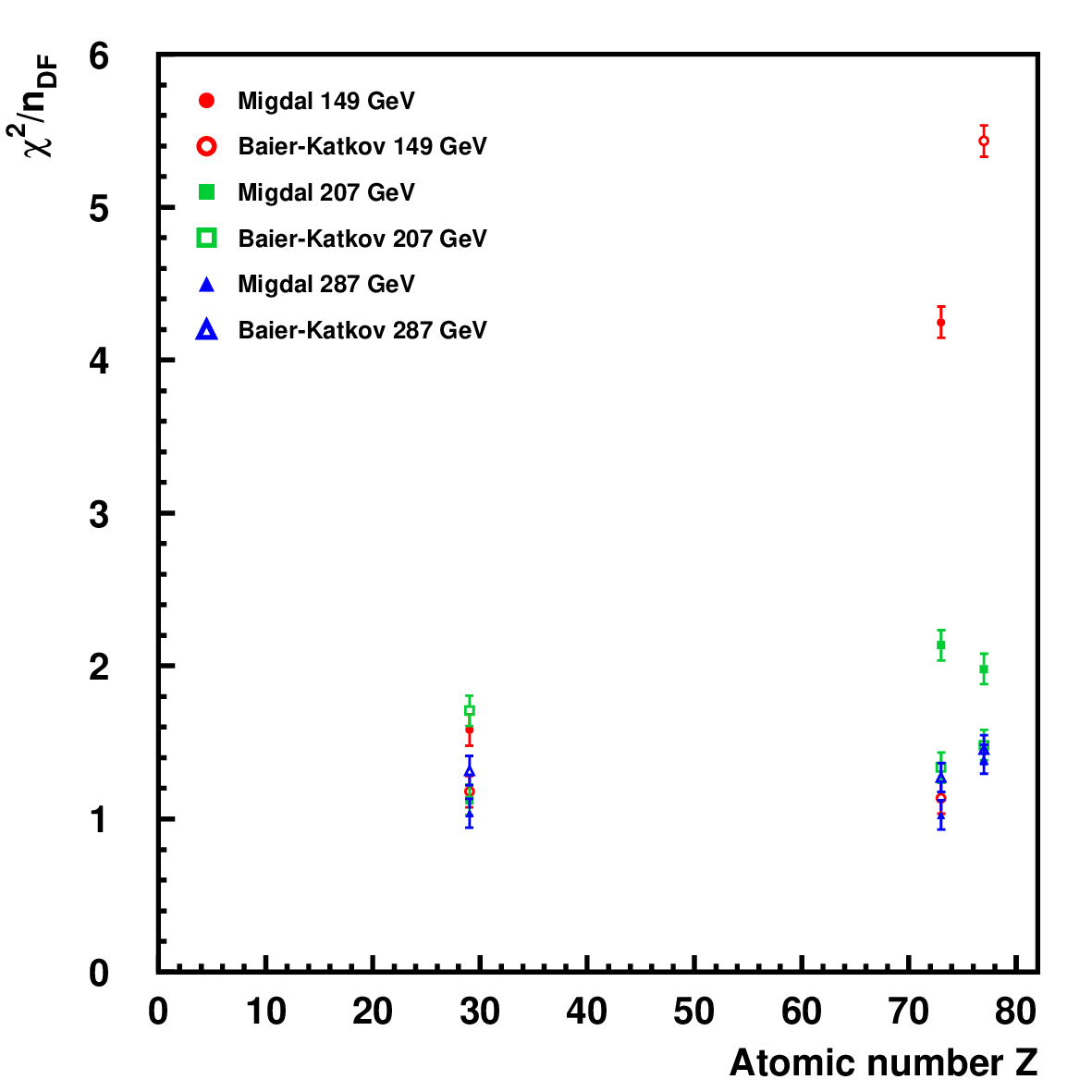}
  \end{tabular}
  \caption{\label{fig:cern_lpm_chi2}Values of
    $\chi^{2}/n_{\mathrm{DF}}$ for the comparison of the simulations
    based on the Migdal (solid symbols) and BK approaches (open
    symbols) with the CERN LPM data at $149$ (circles), $207$
    (squares), and $287$~GeV (triangles) as a function of the atomic
    number of the target $Z$. The bars have the same meaning as in
    Fig.~\ref{fig:slac_chi2}.}
\end{figure}

The comparison of the final full Monte Carlo simulations and the CERN
LPM data are displayed in Figs.~\ref{fig:cern_149GeV},
\ref{fig:cern_207GeV}, and~\ref{fig:cern_287GeV} for beam energies of
$E=149$, $207$, and $287$~GeV, respectively. All targets have been
considered (note that even though a carbon one was measured, it was
used to determine the efficiency of the calorimeter~\cite{Hansen:2004}
and therefore does not offer independent information). The background
subtraction procedure has been applied to the simulations to be
consistent with the data (see
Sec.~\ref{sec:multiph_bksub}). Especially for Ta and Ir, the Migdal
approach remains clearly higher than the BK one by $\approx 5$~\%
around the maximum of the power spectrum. Both approaches merge into
the BH limit at the SWL and get again rather close towards the
low-energy end: no discriminating power is expected in those
regions. Therefore, it is important to measure the part of the
spectrum around the maximum, which was done only by the CERN LPM
experiment.

The $\chi^{2}/n_{\mathrm{DF}}$ of the comparison between simulations
and data is plotted in Fig.~\ref{fig:cern_lpm_chi2} and the values are
given in Table~\ref{tab:cern_lpm_chi2} (again both have the same $25$
logarithmic bins per decade and no interpolation is necessary in
calculating the $\chi^{2}$). Despite the difference between the Migdal
and BK theories in the region of the maximum being apparent, as
mentioned above, it is unfortunately of the same order of the
statistical uncertainties of the measurements so that, overall, the
results of the comparison are not different from those of the SLAC
E-146 case: indeed, of the $9$ target--energy combinations, $5$ favor
the BK approach (clustered at $E=149$ and $207$~GeV). Note that one of
these cases is iridium at $E=149$~GeV, which shows the worst
$\chi^{2}/n_{\mathrm{DF}}$ of all (as a matter of fact the point
corresponding to the Migdal theory falls outside the scale of
Fig.~\ref{fig:cern_lpm_chi2}). Inspection of iridium at $E=149$~GeV in
Fig.~\ref{fig:cern_149GeV} actually reveals that indeed the
measurements remain well below the data even in the SWL limit, where
both cross section options merge into the BH value. In general, the
overall values of $\chi^{2}/n_{\mathrm{DF}}$ are all reasonable except
for the mentioned case of iridium at $149$~GeV, for both theories, and
tantalum at $E=149$~GeV, for the Migdal one. Unfortunately, from the
CERN LPM experiment, it is also not possible to conclude that one of
the approaches is to be preferred.

\begin{table}[b]
  \centering
  \begin{tabular}{||r|c|c|c||}\hline
&
$\chi^{2}/n_{\mathrm{DF}}$ &
$\chi^{2}/n_{\mathrm{DF}}$ &
$n_{\mathrm{DF}}$ \\
&
Migdal &
BK &
\\
&
&
&
\\\hline
\multicolumn{4}{||c||}{149~GeV}\\\hline
Cu 4 \% &
  $  1.6$ &
  $  1.2$ &
  $ 48$ \\
Ta 4 \% &
  $  4.2$ &
  $  1.1$ &
  $ 48$ \\
Ir 4 \% &
  $  9.6$ &
  $  5.4$ &
  $ 48$ \\
\hline
\multicolumn{4}{||c||}{207~GeV}\\\hline
Cu 4 \% &
  $  1.1$ &
  $  1.7$ &
  $ 51$ \\
Ta 4 \% &
  $  2.1$ &
  $  1.3$ &
  $ 51$ \\
Ir 4 \% &
  $  2.0$ &
  $  1.5$ &
  $ 51$ \\
\hline
\multicolumn{4}{||c||}{287~GeV}\\\hline
Cu 4 \% &
  $  1.0$ &
  $  1.3$ &
  $ 55$ \\
Ta 4 \% &
  $  1.0$ &
  $  1.3$ &
  $ 55$ \\
Ir 4 \% &
  $  1.4$ &
  $  1.5$ &
  $ 55$ \\
\hline
\end{tabular}

  \vspace{0.5cm}
  \caption{Values of $\chi^{2}/n_{\mathrm{DF}}$ for the comparison of
    the simulations with the CERN LPM data. All the measured energy
    range from $2$~GeV to the SWL has been used.}
  \label{tab:cern_lpm_chi2}
\end{table}

\begin{figure*}[hp!!!!]
  \centering
  \begin{tabular}{cc}
    \includegraphics[width=.425\textwidth]{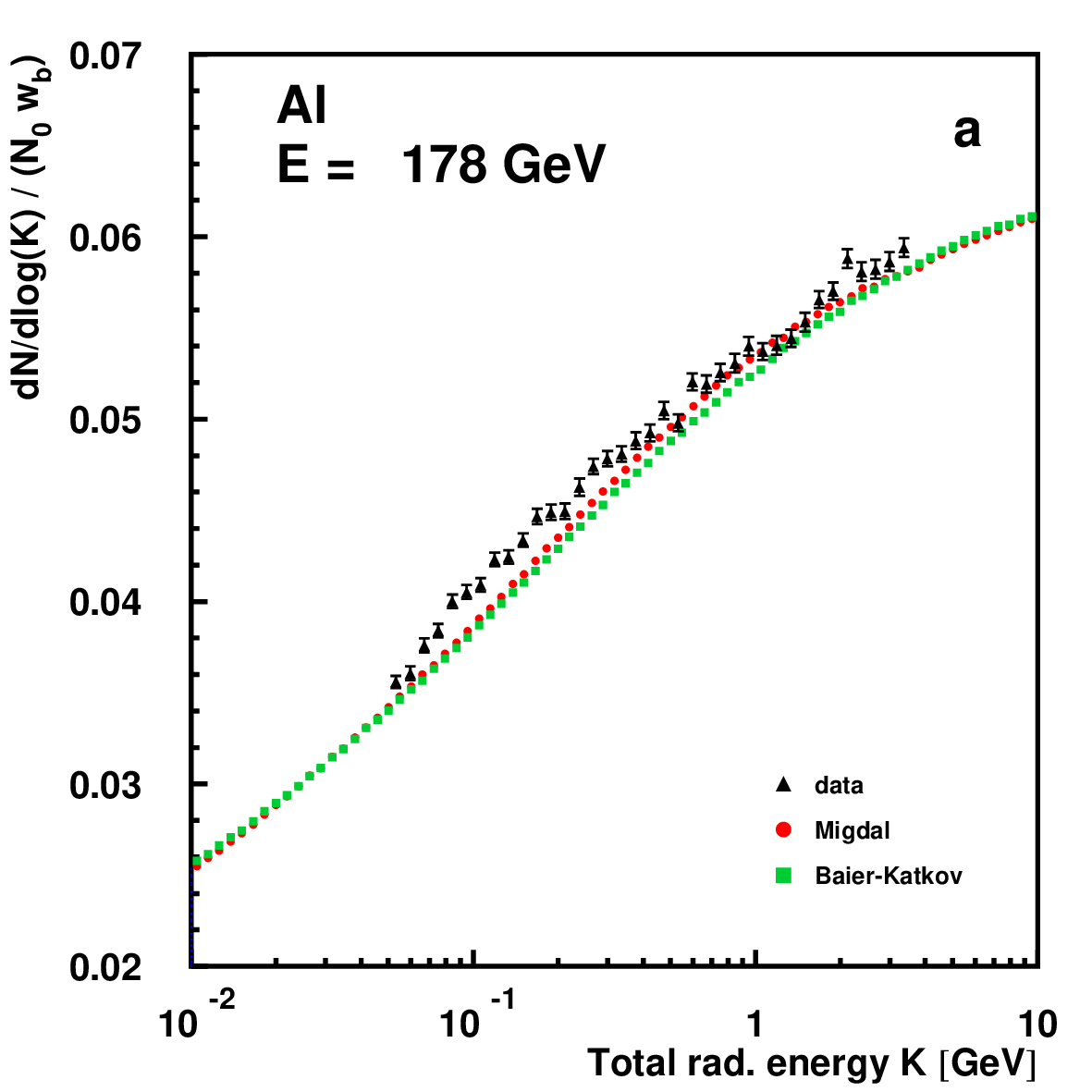} &
    \includegraphics[width=.425\textwidth]{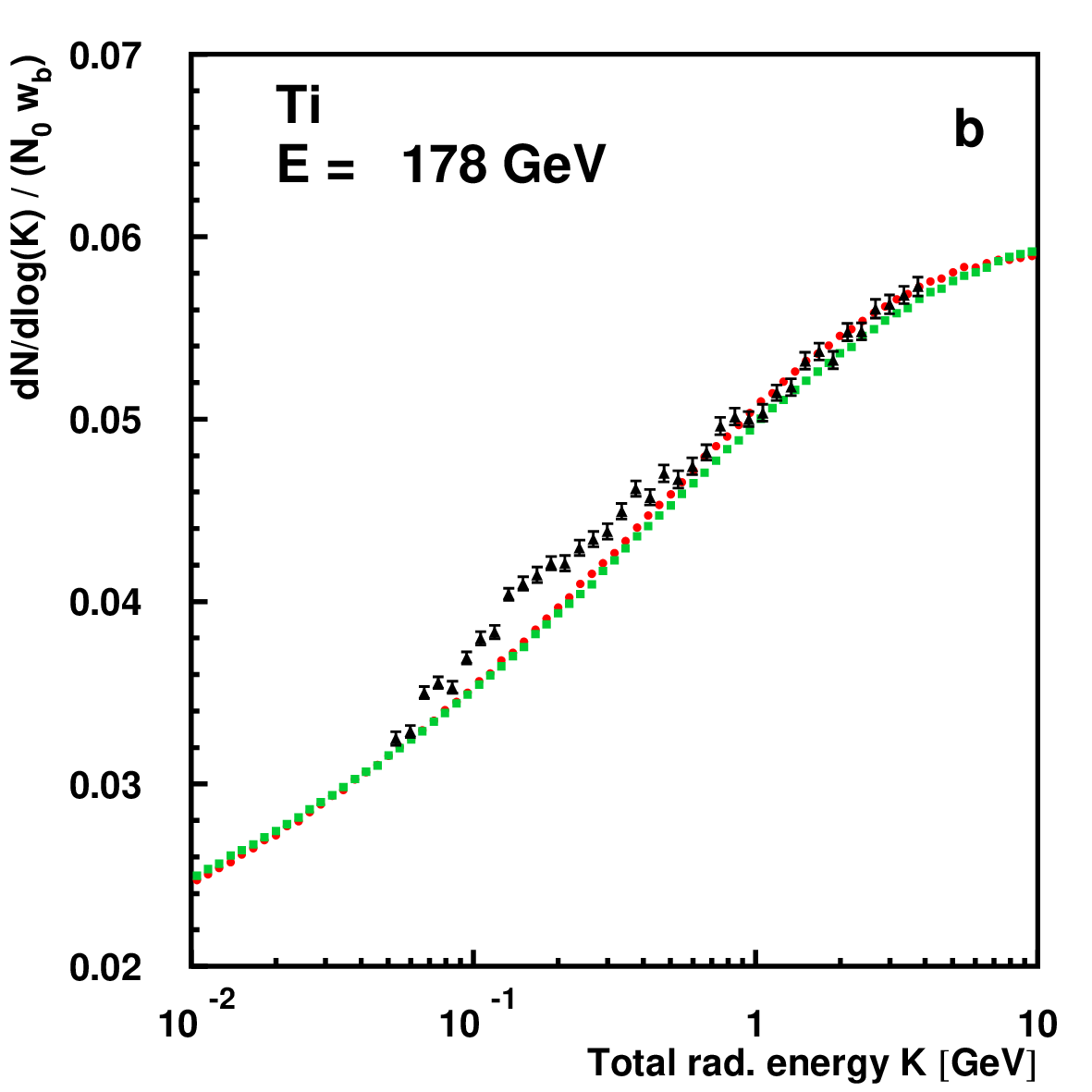} \\
    \includegraphics[width=.425\textwidth]{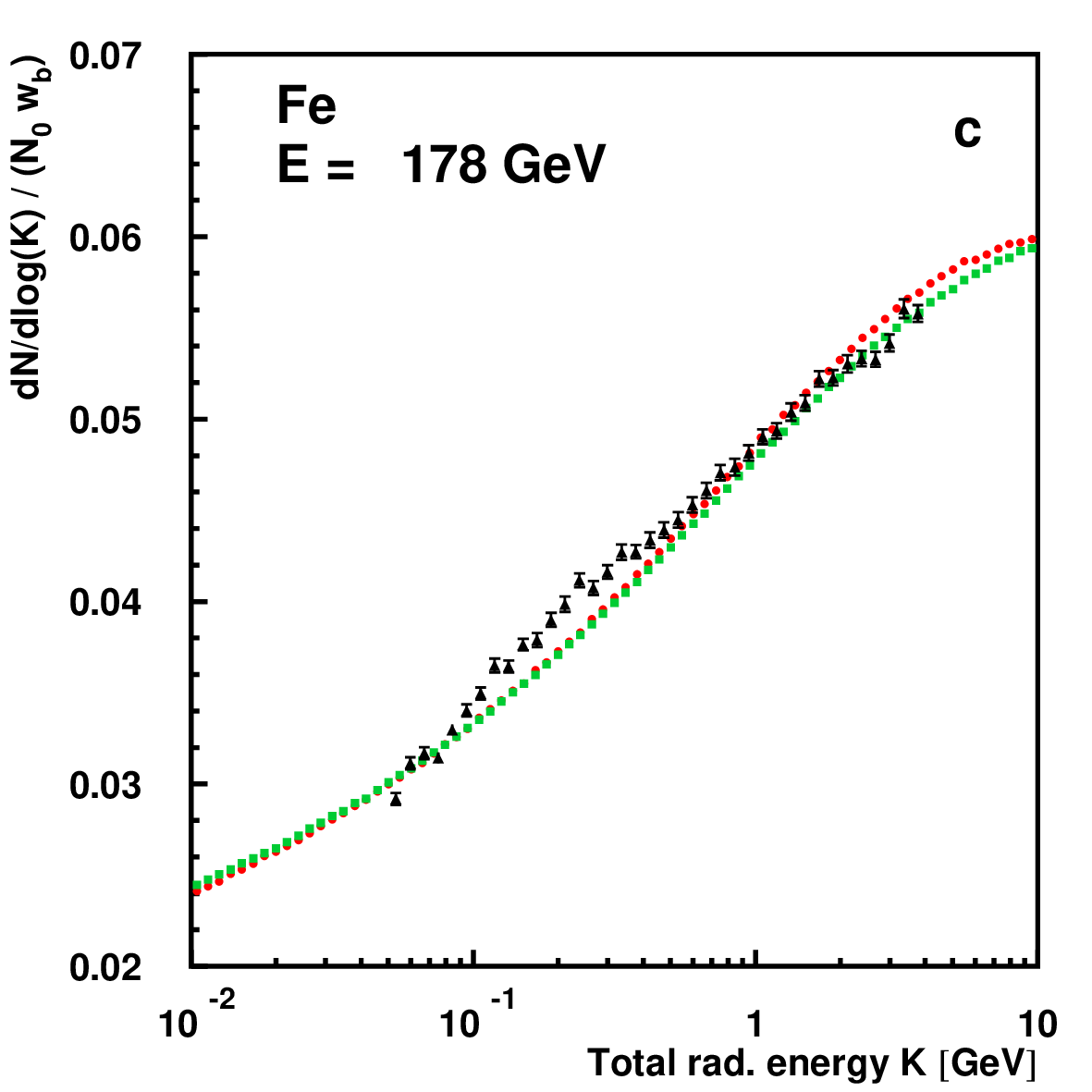} &
    \includegraphics[width=.425\textwidth]{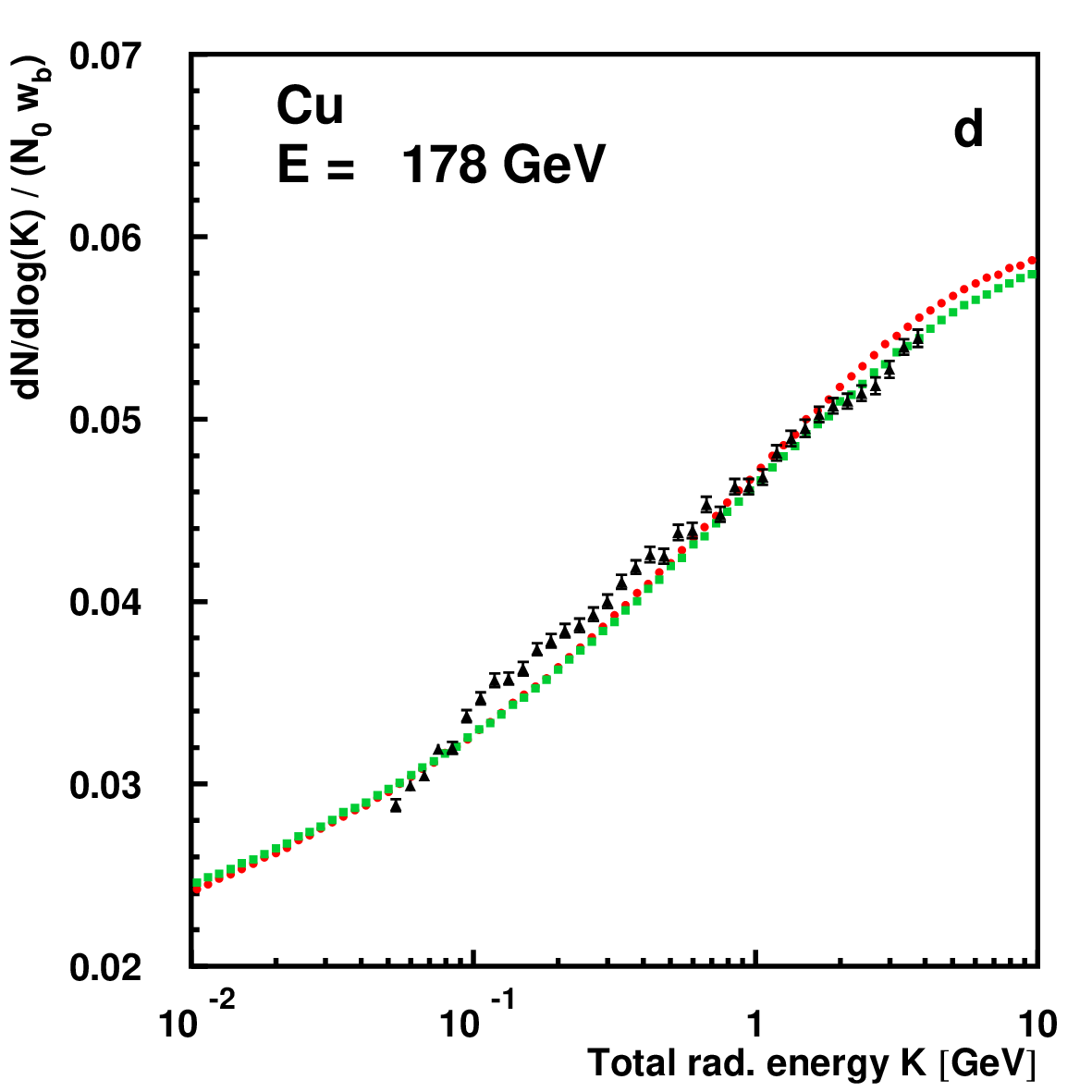} \\
    \includegraphics[width=.425\textwidth]{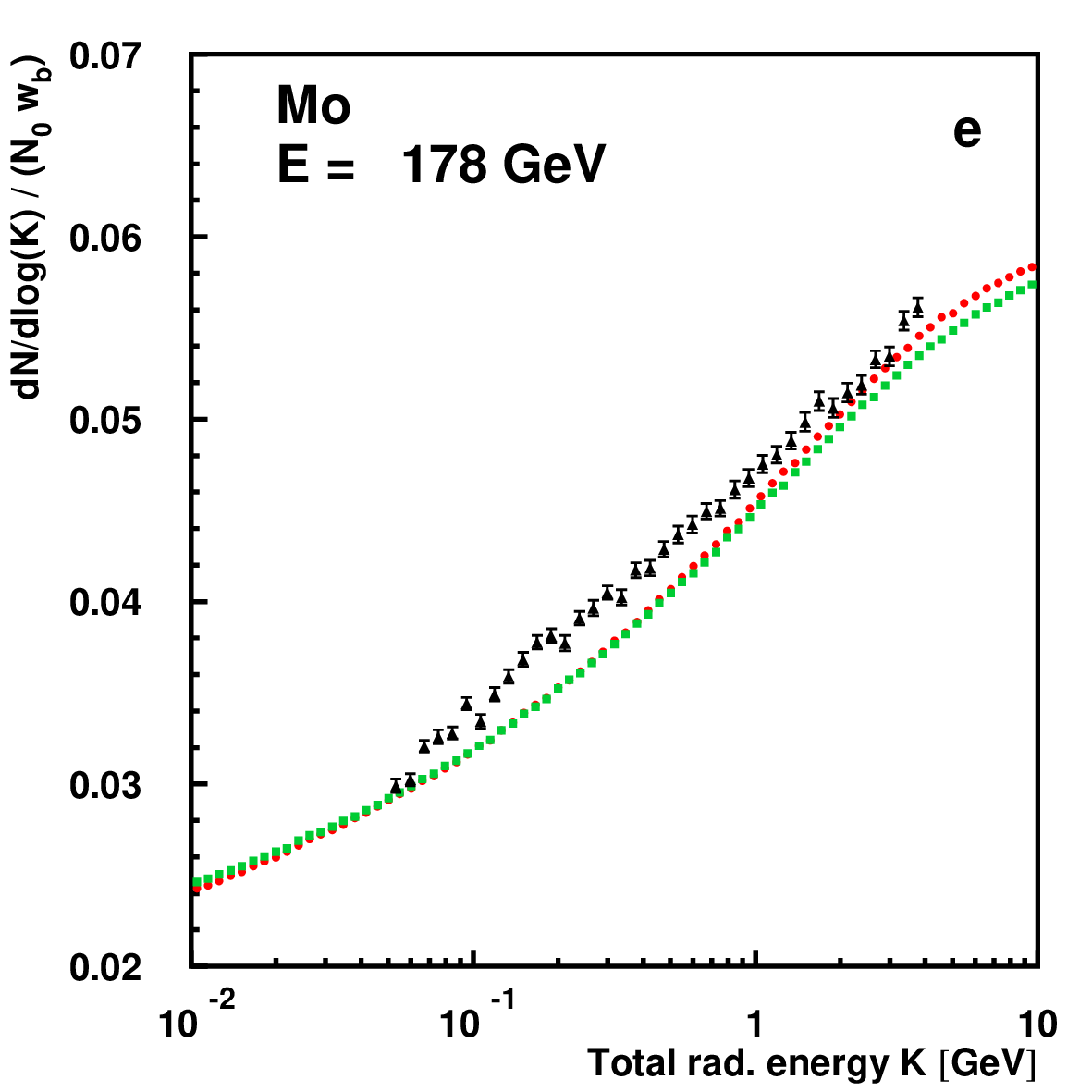} &
    \includegraphics[width=.425\textwidth]{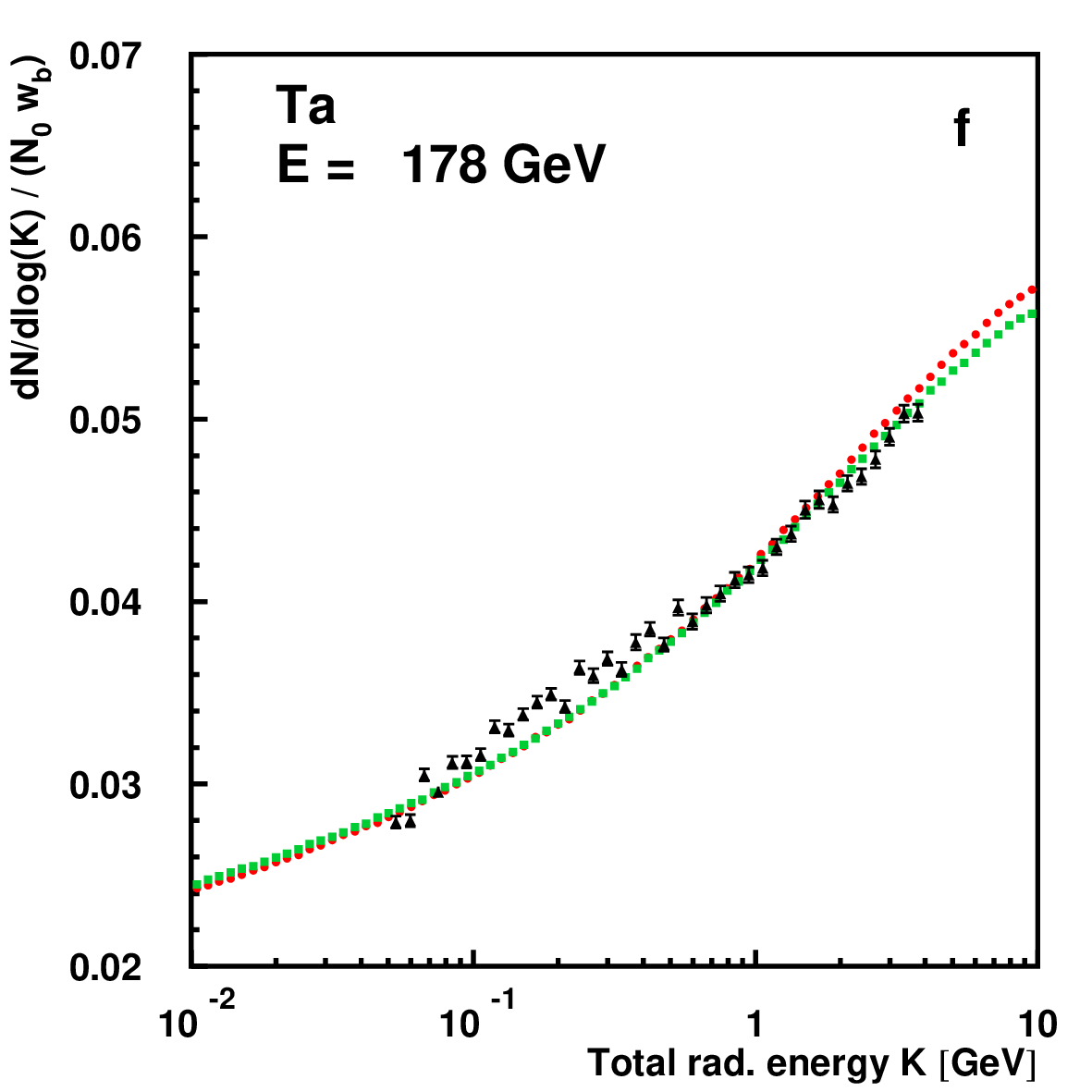}
  \end{tabular}
  \caption{\label{fig:cern_178GeV}Comparison of the simulations based
    on the Migdal (solid circles) and BK approaches (solid squares)
    with the CERN LOW-$Z$ data at $178$~GeV (solid triangles).}
\end{figure*}

The comparison of the final full Monte Carlo simulations and the CERN
LOW-$Z$ data for the beam energy of $E=178$~GeV is depicted in
Fig.~\ref{fig:cern_178GeV}. The aluminum reference target has not been
included because it consisted of $80$ foils with a thickness of
$25$~$\mu$m, to avoid any LPM suppression, and it was used in the
calibration of the calorimeter~\cite{Andersen:2013}. The LDPE one is
also not considered since our code cannot handle mixtures, but again
there is no loss of information since its shape is almost BH. The
background was not subtracted in the experimental analysis to avoid
distortions, as mentioned, and is included in the simulation (mind
that the scale does not start from zero in
Fig.~\ref{fig:cern_178GeV}). As in the SLAC E-146 case, the Migdal and
BK approaches give very close results because the covered energy range
is below the maximum of the power spectrum and it is, once more, not
possible to discriminate between them within the statistical and
systematic uncertainties of the measurements.

\begin{table}[t!!!!]
  \centering
  \begin{tabular}{||r|c|c|c||}\hline
&
$\chi^{2}/n_{\mathrm{DF}}$ &
$\chi^{2}/n_{\mathrm{DF}}$ &
$n_{\mathrm{DF}}$ \\
&
Migdal &
BK &
\\
&
&
&
\\\hline
Al 2.90 \% &
  $ 11.2$ &
  $ 18.9$ &
  $ 37$ \\
Ti 2.68 \% &
  $ 16.4$ &
  $ 21.8$ &
  $ 38$ \\
Fe 2.73 \% &
  $ 11.2$ &
  $ 13.1$ &
  $ 38$ \\
Cu 2.58 \% &
  $  9.5$ &
  $ 10.3$ &
  $ 38$ \\
Mo 2.55 \% &
  $ 27.2$ &
  $ 32.1$ &
  $ 38$ \\
Ta 2.56 \% &
  $ 10.1$ &
  $  8.8$ &
  $ 38$ \\
\hline
\end{tabular}

  \vspace{0.5cm}
  \caption{Values of $\chi^{2}/n_{\mathrm{DF}}$ for the comparison
    of the simulations with the CERN LOW-$Z$ data at $178$~GeV.
    All the measured energy range from $0.05$ to $3.8$~GeV has
    been used.}
  \label{tab:cern_lowz_chi2}
\end{table}

The $\chi^{2}/n_{\mathrm{DF}}$ of the comparison between simulations
and data is reported in Table~\ref{tab:cern_lowz_chi2} (here the
experiment does not follow the $25$ logarithmic bins per decade
adopted in the simulations and a linear interpolation has been
performed). There is a general preference for the Migdal approach with
only one target, Ta, showing a preference for the BK
one. Unfortunately, the larger values of $\chi^{2}/n_{\mathrm{DF}}$,
possibly indicating somewhat higher systematic uncertainties, do not
allow, once more, to reach a firm conclusion about which is the best
theory.

\section{\label{sec:con}Conclusions}

Quantum coherence effects, resulting in the LPM suppression, are an
essential ingredient to reproduce bremsstrahlung measurements for
high-energy electrons. Traditionally, they have been coupled to the
Coulomb correction rescaling the Migdal cross section, which does not
include such a feature, by a realistic radiation length. The
Baier-Katkov approach to quantum coherence effects embodies the
Coulomb correction in a consistent way. The first Monte Carlo
implementation of this latter theory has been presented. Enough
details have been provided to allow an independent development of a
new code. We have also shown how the Baier-Katkov formulae can be
rewritten to allow the reference to an accurate radiation length.

The Monte Carlo simulations have enabled the comparison of the Migdal
and Baier-Katkov approaches to all available data collected for
amorphous targets with accelerators under controlled conditions
including multiphoton emission, attenuation by pair production, and,
finally, pile-up with photons from the background. When all these
effects have been accounted for, the two theories end up being very
close and, unfortunately, currently available experiments cannot
discriminate between them within present statistical and experimental
inaccuracies. It would require a large effort to reduce those
uncertainties at the point of supporting one of the two versions. If
such an endeavor is undertaken in the future, the present work clearly
demonstrates the need to cover the photon energy range around the
maximum of the power spectrum, where the LPM suppression sets in. So
it appears that the best results could possibly be achieved by an
experiment carried out under conditions similar to those that produced
the CERN LPM data. However, it would be necessary to attain a
background suppression at a fraction of percent of $X_{0}$ and
combined statistical and systematic uncertainties better than $2$~\%
around the maximum of the power spectrum. All in all, there are no
compelling reasons to modify the Monte Carlo codes generally used to
simulate the response of calorimeters at the LHC or extensive air
showers, all based on the Migdal approach.

The LHCf collaboration reported inconsistencies in the initial profile
of the shower, where the LPM suppression is strong. If this evidence
should gain more support, the present work suggests that the reasons
should be looked for in additional limitations of current theories. In
particular, the general arguments about the formation length would
lead to predict a correlation between the magnitude of the scattering
angle and the amount of suppression. No experimental evidence was
found within present uncertainties~\cite{Hansen:2004}, but theoretical
predictions for the angular distribution under LPM suppression were
made~\cite{Varfolomeev:1974}. For the accurate simulation of extensive
air showers, the issue of the LPM suppression in electron-electron
bremsstrahlung, which is always assumed to be the same as in nuclear
bremsstrahlung, should also receive attention in the future. Finally,
it would be interesting to develop a treatment of the LPM suppression
beyond the full screening approximation, especially for the simulation
of showers, when it is necessary to follow the degradation of the
initial energy down to the detection threshold.

\section*{Acknowledgments}

AM acknowledges supported by Funda\c{c}\~{a}o de Amparo \`{a} Pesquisa
do Estado de S\~{a}o Paulo (FAPESP) under Contracts No.~2013/15634-5
and No.~2016/13116-5 and by Conselho Nacional de Desenvolvimento
Cient\'{\i}fico e Tecnol\'{o}gico (CNPq) under Contracts
No.~306331/2016-0 and No.~311915/2020-5.

\bibliographystyle{unsrt}
\bibliography{lpm}

\appendix

\section{\label{sec:imp_D1_D2}Approximation of the $D_{1}$ and $D_{2}$ functions}

\begin{figure}[t!!!!!]
  \centering
  \begin{tabular}{c}
    \includegraphics[width=.425\textwidth]{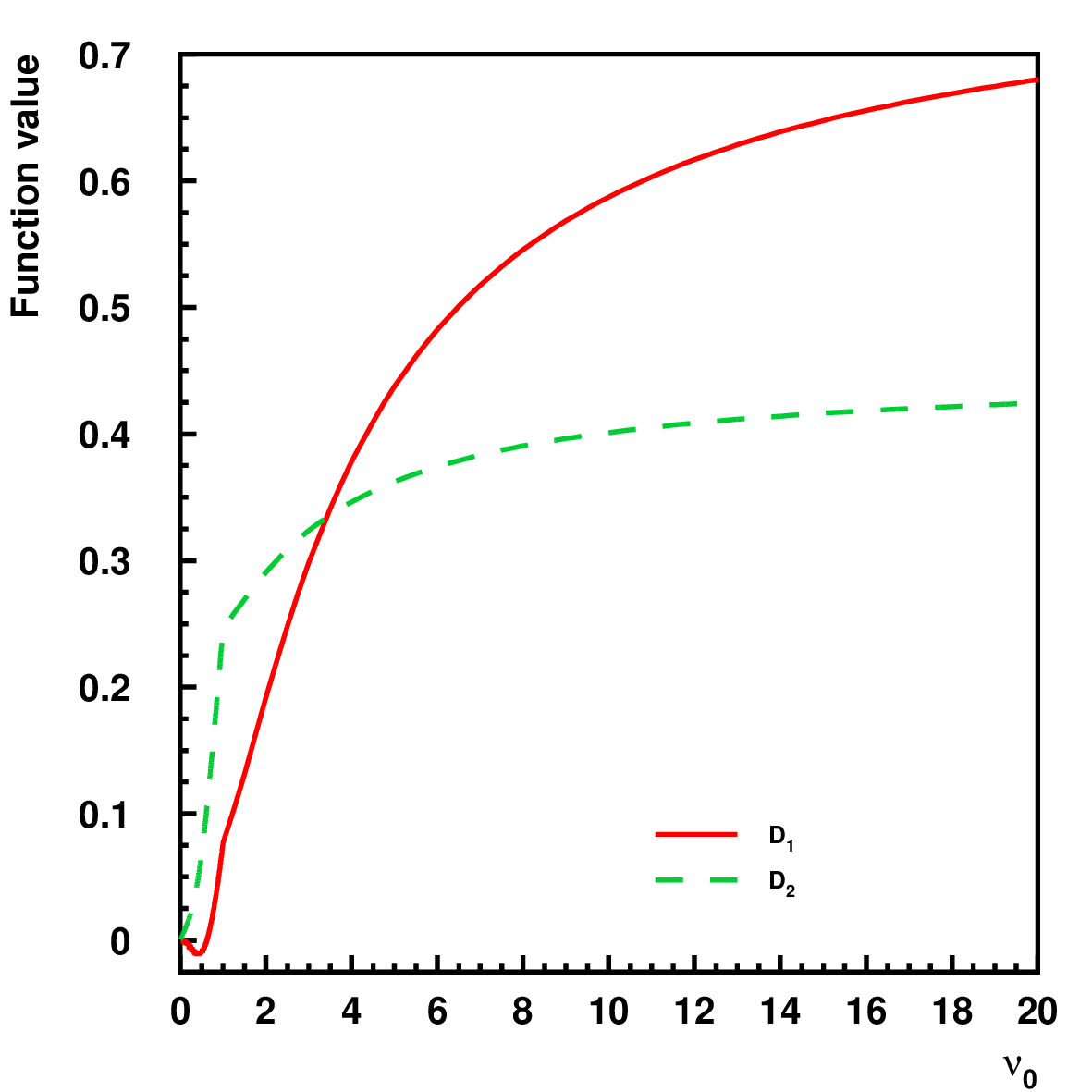}
  \end{tabular}
  \caption{\label{fig:d1_d2}Values of the $D_{1}(\nu_{0})$ and
    $D_{2}(\nu_{0})$ functions defined by Baier and Katkov obtained
    with the numerical procedure described in the text.}
\end{figure}

As anticipated in Sec.~\ref{sec:BK}, we discuss here the evaluation of
the functions $D_{1}(\nu_{0})$ and $D_{2}(\nu_{0})$ together with
their fit by means of rational functions of $\nu_{0}$. For the
numerical integration of the functions appearing in
Eqs.~(\ref{eq:BK_D1}) and~(\ref{eq:BK_D2}), the routines included in
the QUADPACK quadrature package appropriate for each case are
used~\cite{Piessens:1983}. For $D_{1}$, the integral is split in two
ranges, namely $0\leq\nu_{0}\leq 1$ and $1<\nu_{0}<\infty$ and no
problem arises in reaching a relative precision of $10^{-5}$ for both
ranges.  The integral appearing in the expression of $D_{2}$ is more
problematic as the functions $d(z)$ and $G(z)$ defined in
Sec.~\ref{sec:BK} both suffer from a logarithmic divergence which
cancels in the limit $z\rightarrow 0$.  Again, the integral has been
split in two ranges $0\leq\nu_{0}\leq z_{\mathrm{min}}$ and
$z_{\mathrm{min}}<\nu_{0}<\infty$, where the choice of
$z_{\mathrm{min}}$ will be explained in a while.  The numerical
integration on the second range is easily performed. Instead, to
execute the first integral, the integrand is developed in a McLaurin
series up to third order in $z$ and the integral is computed
analytically choosing $z_{\mathrm{min}}$ in such a way that the
percentage relative difference between the integral including the
third-order term and the one including only up to the second-order
term is smaller than $0.001$. The numerical values of $D_{1}$ and
$D_{2}$, obtained by the procedure described, are shown in
Fig.~\ref{fig:d1_d2} for $\nu_{0}$ from zero up to $20$. Note in
particular the discontinuity for $\nu_{0}=1$ and the negative minimum
of $D_{1}$ close to $\nu_{0}\approx 0.5$. It has been checked that the
values calculated by the numerical procedure described here agree with
those shown in Fig.~2 of Ref.~\cite{Baier:2005}.

\begin{table}[t!!!!!]
  \caption{Parameters of the fit to the $D_{1}$ and $D_{2}$
    functions by Baier and Katkov with Eqs.~(\ref{eq:fitD1})
    and~(\ref{eq:fitD2}).}
  \begin{tabular}{ll|ll}\hline
    Coef.  & Fit to $D_{1}$      &  Coef.  & Fit to $D_{2}$ \\\hline
    $a_{0}$ & $+0.20892$         & $e_{0}$ & $+0.25752$        \\
    $a_{1}$ & $-0.39046$         & $e_{1}$ & $+0.10318$        \\
    $a_{2}$ & $+0.35727$         & $e_{2}$ & $+0.28050$        \\
    $a_{3}$ & $-0.11675$E$-3$    & $e_{3}$ & $+0.26376$E$-4$   \\
    $a_{4}$ & $+0.115128$E$-5$   & $e_{4}$ & $-0.33742$E$-6$   \\
    $b_{1}$ & $+0.82588$         & $f_{1}$ & $+0.97830$        \\
    $b_{2}$ & $+0.45111$         & $f_{2}$ & $+0.62401$        \\
    $c_{0}$ & $-0.40185$E$-3$    & $g_{0}$ & $-0.94159$E$-3$   \\
    $c_{1}$ & $+0.80181$E$-2$    & $g_{1}$ & $+0.99852$E$-1$   \\
    $c_{2}$ & $-0.15769$         & $g_{2}$ & $-0.23823$        \\
    $c_{3}$ & $+0.15393$         & $g_{3}$ & $+0.67476$        \\
    $c_{4}$ & $+0.12994$         &        &                   \\
    $d_{1}$ & $-1.0680$          & $h_{1}$ & $-0.87428$        \\
    $d_{2}$ & $+1.8065$          & $h_{2}$ & $+2.0489$         \\
    $p_{1}$ & $-0.48175254$E$-4$ & $q_{1}$ & $+0.7862369$E$-1$ \\
    $p_{2}$ & $-0.11035494$      &        &                    \\
    \hline
  \end{tabular}
  \label{tab:fit_d1_d2}
\end{table}

\begin{figure}[t!!!!]
  \centering
  \begin{tabular}{c}
    \includegraphics[width=.425\textwidth]{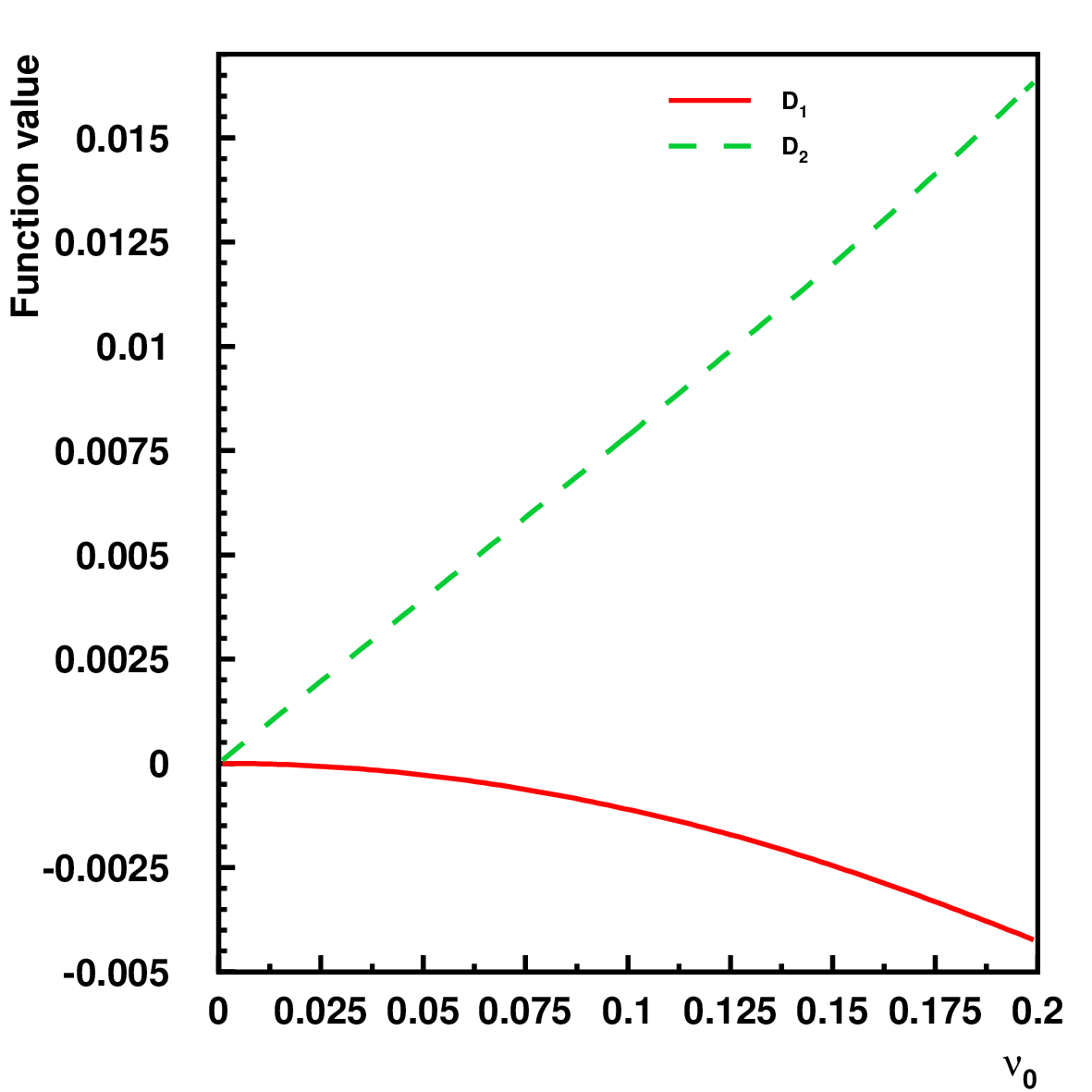}
  \end{tabular}
  \caption{\label{fig:d1_d2_fit}Plot of the rational function
    approximations given by Eqs.~(\ref{eq:fitD1}) and~(\ref{eq:fitD2})
    to $D_{1}$ and $D_{2}$ with the numerical values of the parameters
    reported in Table~\ref{tab:fit_d1_d2}. Note that two different
    parametrizations are used below and above $\nu_{0}=0.1$ and they
    join smoothly both in value and slope.}
\end{figure}

Since the functions $D_{1}$ and $D_{2}$ have to be used in lengthy
Monte Carlo calculations, a suitable approximation to the ``exact''
calculated values is in order. A fit by means of rational functions is
chosen for each of three ranges of $\nu_{0}$, namely
\begin{equation}
  D_{1}=
  \begin{cases}
  \frac{a_{0}+a_{1}\,\nu_{0}+a_{2}\,\nu_{0}^{2}+a_{3}\,\nu_{0}^{3}+a_{4}\,\nu_{0}^{4}}
  {1+b_{1}\,\nu_{0}+b_{2}\,\nu_{0}^{2}} &\text{if $\nu_{0}\geq 1$ }\\\\
  \frac{c_{0}+c_{1}\,\nu_{0}+c_{2}\,\nu_{0}^{2}+c_{3}\,\nu_{0}^{3}+c_{4}\,\nu_{0}^{4}}
  {1+d_{1}\,\nu_{0}+d_{2}\,\nu_{0}^{2}} &\text{if $0.1\leq\nu_{0}\leq 1$}\\\\
  p_{1}\,\nu_{0}+p_{2}\,\nu_{0}^{2} &\text{if $\nu_{0}\leq 0.1$ }
  \end{cases}
  \;\;,
  \label{eq:fitD1}
\end{equation}
and similarly
\begin{equation}
  D_{2}=
  \begin{cases}
  \frac{e_{0}+e_{1}\,\nu_{0}+e_{2}\,\nu_{0}^{2}+e_{3}\,\nu_{0}^{3}+e_{4}\,\nu_{0}^{4}}
  {1+f_{1}\,\nu_{0}+f_{2}\,\nu_{0}^{2}} &\text{if $\nu_{0}\geq 1$ }\\\\
  \frac{g_{0}+g_{1}\,\nu_{0}+g_{2}\,\nu_{0}^{2}+g_{3}\,\nu_{0}^{3}}
  {1+h_{1}\,\nu_{0}+h_{2}\,\nu_{0}^{2}} &\text{if $0.1\leq\nu_{0}\leq 1$} \\\\
  q_{1}\,\nu_{0} &\text{if $ \nu_{0}\leq 0.1$ }
  \end{cases}
  \;\;.
  \label{eq:fitD2}
\end{equation}
Actually, for the lowest range of $\nu_{0}$, the coefficients $p_{1}$
and $p_{2}$ have been determined by a fitting procedure and
subsequently rescaled to satisfy the requirement of the continuity of
$D_{1}$ for $\nu_{0}=0.1$ (as fixed by the fit in the range
$0.1\leq\nu_{0}\leq 1$). The coefficient $q_{1}$ has been determined
by requiring the continuity of $D_{2}$ in $\nu_{0}=0.1$. The numerical
values of the parameters appearing in Eqs.~(\ref{eq:fitD1})
and~(\ref{eq:fitD2}) are listed in Table~\ref{tab:fit_d1_d2}. In
principle, this adjustment procedure does not grant the continuity of
the first derivative in $\nu_{0}=0.1$. The fits to the functions
$D_{1}$ and $D_{2}$ are plotted in Fig.~\ref{fig:d1_d2_fit} for the
range $0\leq\nu_{0}\leq 0.2$. It is seen that, nevertheless, the slope
of the fit to $D_{1}$ and $D_{2}$ in the $0.1\leq \nu_{0}\leq 1$
region joins smoothly in $\nu_{0}=0.1$ to the slope for $\nu_{0}<0.1$.

\begin{figure}[b!!!!]
  \centering
  \begin{tabular}{c}
    \includegraphics[width=.425\textwidth]{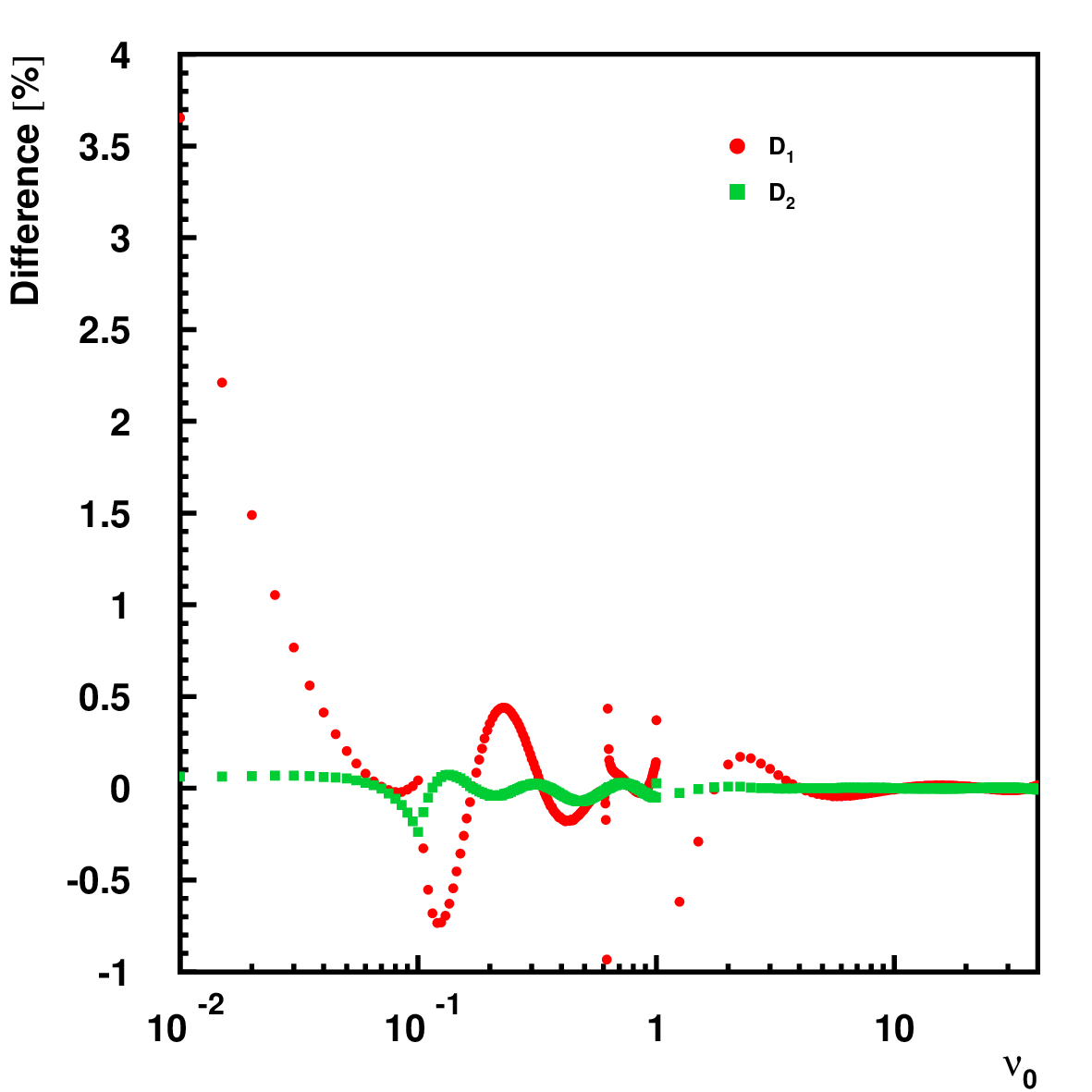}
  \end{tabular}
  \caption{\label{fig:d1_d2_dif}Plot of the relative deviation of the
    rational function approximations given by Eqs.~(\ref{eq:fitD1})
    and~(\ref{eq:fitD2}) to the true values of $D_{1}$ and $D_{2}$.}
\end{figure}

Finally, the percentage deviation of the fitted values from the exact
ones is reported for both $D_{1}$ and $D_{2}$ in
Fig.~\ref{fig:d1_d2_dif}. It is seen that for $D_{1}$ the maximum
deviation is below $0.3$~\% while for $D_{2}$ it reaches $3.5$~\% for
the lowest values of $\nu_{0}$. It should be noted that this deviation
refers to a first-order correction to the main term for the total
radiation intensity, i.e to a contribution that is, in the worst case,
no larger than $10$~\% of the main term (see Fig.~\ref{fig:comp_th_Ir}
in Sec.~\ref{sec:BK}). Moreover, the mentioned largest deviation is
reached at small values of $\nu_{0}$, where the correct behavior in
the limit $\nu_{0}\rightarrow 0$, i.e. quadratic for $D_{1}$ and
linear for $D_{2}$, is more important then the exact reproduction of
the numerical values of the functions to ensure, when inserted in
Eq.~(\ref{eq:BK_corr}), that the correction term to the cross section
vanishes in the short-wavelength limit at the tip of the spectrum (see
again Fig.~\ref{fig:comp_th_Ir} in Sec.~\ref{sec:BK}). For this
reason, it is not convenient to increase the order of the
parametrizations in the part $\nu_{0}\leq 0.1$.

\section{\label{sec:BK_ldisc}Location of the discontinuity}

\begin{figure}[t!!!!]
  \centering
  \begin{tabular}{c}
    \includegraphics[width=.425\textwidth]{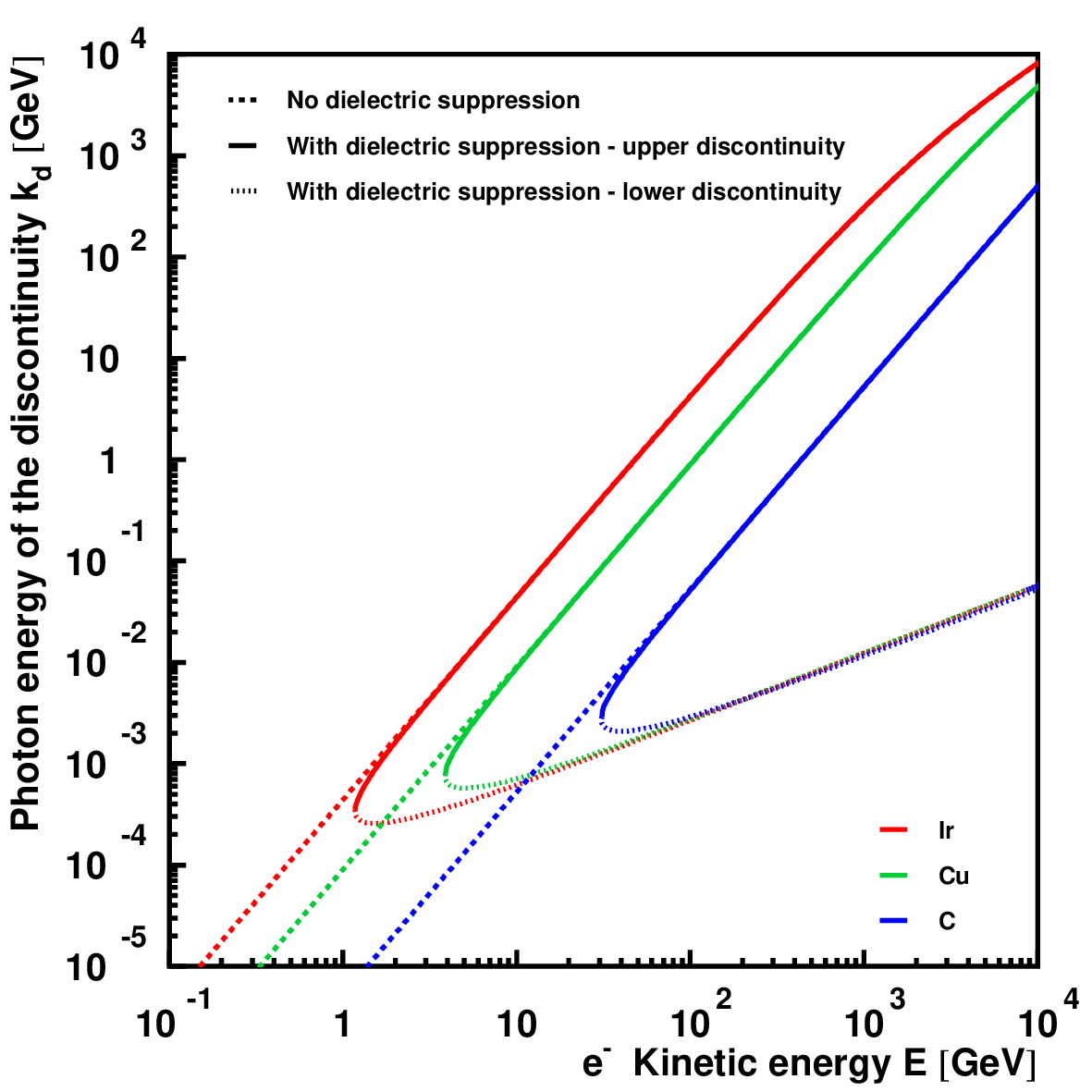}
  \end{tabular}
  \caption{\label{fig:kd_en}Value of the photon energy
    $k=k_{\mathrm{d}}$ at which $\rho_{\mathrm{c}}=1$ (no dielectric
    suppression) and $\tilde{\rho}_{\mathrm{c}}=1$ (dielectric
    suppression) as a function of the impinging electron energy $E$
    for three materials. Without dielectric suppression, there is
    always a value of $k=k_{\mathrm{d}}$ for every value of $E$
    (dashed line). With dielectric suppression, there is no value of
    $k=k_{\mathrm{d}}$ for $E$ below a minimum value
    $E_{\mathrm{d}}$. For $E$ above $E_{\mathrm{d}}$, there are two
    values of $k=k_{\mathrm{d}}^{\mathrm{l}}$ and
    $k=k_{\mathrm{d}}^{\mathrm{u}}$ for one value of $E$: therefore a
    lower (dotted line) and an upper (solid line) branch are
    present. The two branches join for $E=E_{\mathrm{d}}$.}
\end{figure}

For simplicity, in what follows, we refer to $k=k_{\mathrm{d}}$ as
discontinuity, being understood that we mean the one in the first
derivative.  The behavior of the discontinuity is markedly different
according to whether dielectric suppression is taken into account or
not. The two cases are correspondingly handled differently in the
program. When dielectric suppression is not enabled, for a given
material and a given impinging electron energy $E$, there is always
one unique minimum value of the photon energy $k=k_{\mathrm{d}}$
resulting in $\rho_{\mathrm{c}}=1$ (see Sec.~\ref{sec:BK_nodiel}). It
can actually be directly determined analytically (this is in contrast
to the Migdal approach, where $k_{\mathrm{d}}$ has to be found
numerically, even when no dielectric suppression is
present~\cite{Mangiarotti:2011}) and it is shown in
Fig.~\ref{fig:kd_en} with a dashed line. It clearly increases with $E$
and, for a fixed $E$, it decreases for lower atomic numbers $Z$. To
quote a value, for $E=300$~GeV electrons impinging on iridium,
$k_{\mathrm{d}}\approx 35$~ GeV, well in the region covered by the
CERN LPM data, and for $E=8$~GeV, $k_{\mathrm{d}}\approx 29$~MeV,
again well in the region covered by the SLAC E-146 data.

When dielectric suppression is present, three cases are possible. For
a certain unique value of $E=E_{\mathrm{d}}$, there is again one
unique minimum value of the photon energy $k=k_{\mathrm{d}}$ resulting
in $\tilde{\rho}_{\mathrm{c}}=1$. However, for $E<E_{\mathrm{d}}$, it
is always $\tilde{\rho}_{\mathrm{c}}<1$, while for $E>E_{\mathrm{d}}$,
there is an interval of values
$k_{\mathrm{d}}^{\mathrm{l}}<k<k_{\mathrm{d}}^{\mathrm{u}}$ for which
$\tilde{\rho}_{\mathrm{c}}=1$ resulting in the presence of two
discontinuities: one for $k=k_{\mathrm{d}}^{\mathrm{l}}$ and another
for $k=k_{\mathrm{d}}^{\mathrm{u}}$ (see Sec.~\ref{sec:BK_diel}). The
values of $k_{\mathrm{d}}^{\mathrm{l}}$ and
$k_{\mathrm{d}}^{\mathrm{u}}$ are plotted in Fig.~\ref{fig:kd_en} with
dotted and continuous lines, respectively: they have to be determined
numerically. The simple bisection method is used because of its
guaranteed convergence; here again typically $\approx 20$ steps are
required to achieve a relative precision better than $10^{-8}$. This
part of the code works in double-precision (i.e.~64-bits) arithmetics
to further improve the reliability. For $E$ above few times
$E_{\mathrm{d}}$, the values of $k_{\mathrm{d}}^{\mathrm{u}}$ are very
close to the corresponding ones of $k_{\mathrm{d}}$ for no dielectric
suppression. Around $E_{\mathrm{d}}$, $k_{\mathrm{d}}^{\mathrm{u}}$
gets reduced since it has to merge with $k_{\mathrm{d}}^{\mathrm{l}}$:
this behavior is actually used in the code to tune the bracketing
strategy of the bisection method. An interesting feature of
$k_{\mathrm{d}}^{\mathrm{l}}$ apparent in Fig.~\ref{fig:kd_en} is that
it rises much less slowly with energy than
$k_{\mathrm{d}}^{\mathrm{u}}$ and it is also much less sensitive to
the atomic number $Z$ of the target material. To quote a value, for an
$E=300$~GeV electron impinging on iridium,
$k_{\mathrm{d}}^{\mathrm{l}}\approx 6$~MeV. This is a kind of ``worst
case'' scenario for the simulations of the CERN LPM data (not very
sensitive to $Z$), where a low-energy cut of $\mathrm{TCUT}=50$~MeV is
applied for photon propagation. For $E=25$~GeV,
$k_{\mathrm{d}}^{\mathrm{l}}\approx 1$~MeV, which is not below the
value of $\mathrm{TCUT}=10$~keV used in the simulations for the SLAC
E-146 data.  This situation is very similar to what was found for
Migdal approach in Ref.~\cite{Mangiarotti:2011}.

\begin{figure}[t!!!!]
  \centering
  \begin{tabular}{c}
    \includegraphics[width=.425\textwidth]{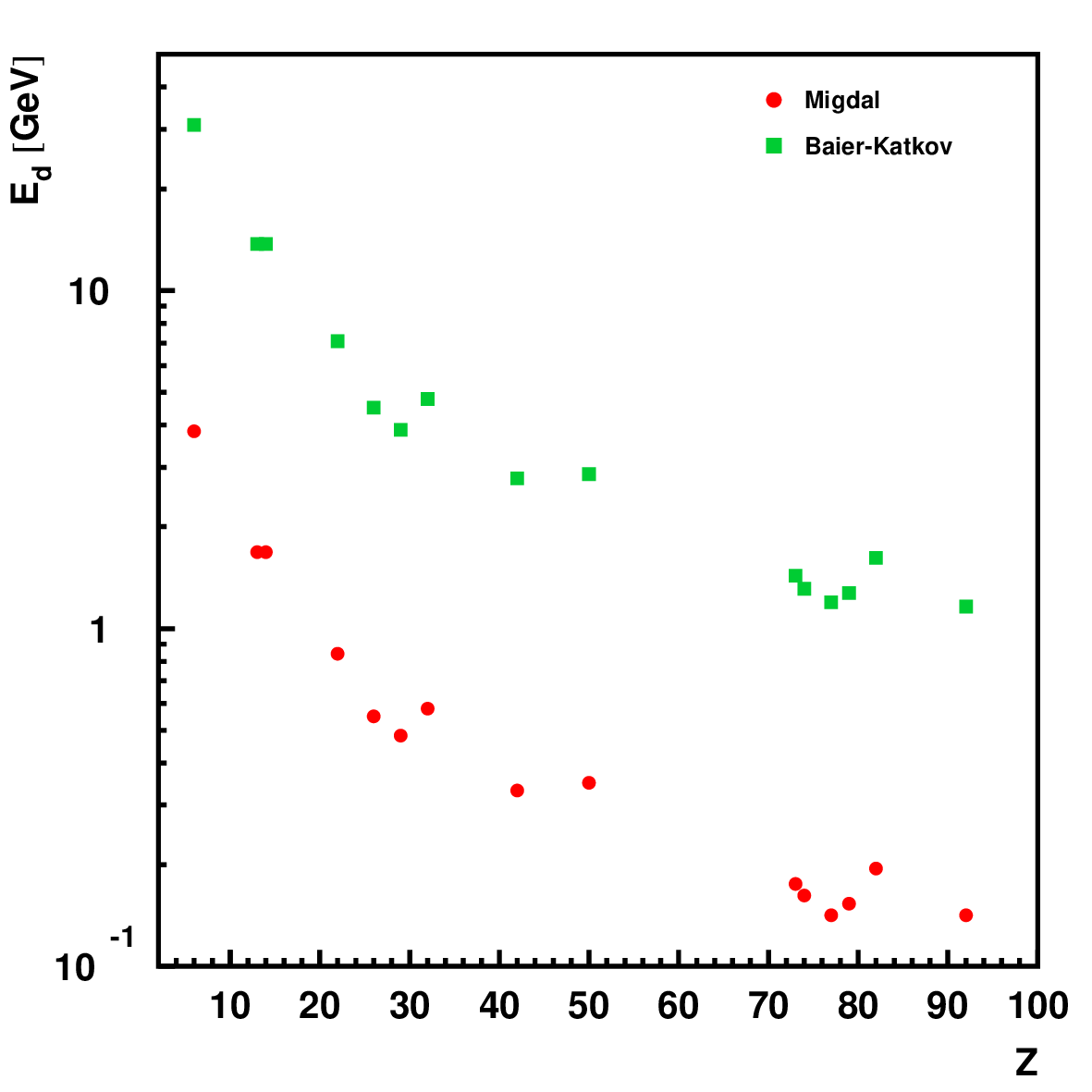}
  \end{tabular}
  \caption{\label{fig:ed_z}Minimum value of the impinging electron
    energy $E=E_{\mathrm{d}}$ for which the discontinuity in the first
    derivative of the cross section is present, when dielectric
    suppression is taken into account, as a function of the atomic
    number of the target element. Both the Migdal (solid circles) and
    BK approaches (solid squares) are considered.}
\end{figure}

Finally, the values of $E_{\mathrm{d}}$ for both the Migdal and BK
approaches are compared in Fig.~\ref{fig:ed_z}. The general behavior
is quite similar, but the values for the former are approximately one
order of magnitude larger than for the latter.

\section{\label{sec:BK_sigmatot}Total cross section}

The probability for an impinging electron with energy $E$ to emit a
photon of energy $k$ is controlled by the total cross section, which
has to be supplied to the Monte Carlo code to fill the internal
tables. It is given basically by the sum of Eq.~(\ref{eq:BK_main})
and~(\ref{eq:BK_corr})) [multiplied by the factor $1/(n\,E)$ to
convert radiation probability per unit length to differential cross
section]:
\begin{equation}
\begin{split}
  \frac{d\sigma}{dk}=&
  \frac{\alpha}{12\pi\,\hbar c\,n}\\
  &\bigg\{\frac{\nu_{0}^{2}}{\gamma^{2}}
  \left[R_{1}\,G\left(\frac{s_{\mathrm{BK}}}{2}\right)+
  2\,R_{2}\Phi\left(\frac{s_{\mathrm{BK}}}{2}\right)\right]\\
  &+\frac{3}{\gamma^{2}\,L_{\mathrm{c}}}
  \big[D_{1}(\nu_{0})\,R_{1}+D_{2}(\nu_{0})\,
  R_{2}\,\sqrt{2}\,\nu_{0}\big]
  \bigg\}
  \;\;,
\end{split}
\label{eq:BK_dsdk}
\end{equation}
where $R_{1}$ and $R_{2}$ have been defined in
Eqs.~(\ref{eq:BK_R1R2}), the functions $G$ and $\Phi$ in
Eqs.~(\ref{eq:M_PhiG}), and the functions $D_{1}$ and $D_{2}$ in
Eqs.~(\ref{eq:BK_D1D2}). The dielectric suppression is taken into
account with the substitutions presented in Sec.~\ref{sec:BK_diel}. A
low-energy cut $\mathrm{TCUT}$ is applied for photon propagation in
the Monte Carlo code, so that Eq.~(\ref{eq:BK_dsdk}) must be
integrated numerically over the range $\mathrm{TCUT}$ to
$E$. Actually, the maximum allowed value of $k$ is given by the
kinetic energy of the electron $T_{\mathrm{E}}$ corresponding to $E$,
but this is a small difference.  In the present case, all formulae are
valid in the ultrarelativistic limit, but the same correction
described in Appendix~\ref{sec:BK_samp} is applied to enforce energy
conservation under all conditions during the simulations.  The
adaptive Gaussian quadrature routine \cal{DADAPT} from the CERN
mathlib library~\cite{Cernlib:1996p92} is used and then a change of
integration variable from $k$ to $\ln(k)$ is performed to regularize
the $1/k$ BH divergence of Eq.~(\ref{eq:BK_dsdk}), contained in
$\nu_{0}^{2}$. To improve the stability, if the discontinuity is
located in the region of integration, i.e. if
$k_{\mathrm{d}}>\mathrm{TCUT}$, the integral is split into two parts:
one from $\mathrm{TCUT}$ to $k_{\mathrm{d}}-\delta k_{\mathrm{d}}$ and
another from $k_{\mathrm{d}}+\delta k_{\mathrm{d}}$ to $E$, where
$\delta k_{\mathrm{d}}$ is the estimated uncertainty in the numerical
localization of $k_{\mathrm{d}}$. The contribution from
$k_{\mathrm{d}}-\delta k_{\mathrm{d}}$ to $k_{\mathrm{d}}+\delta
k_{\mathrm{d}}$ is simply handled by one application of the
trapezoidal rule.  If dielectric suppression is active, only the upper
discontinuity is handled explicitly following
Ref.~\cite{Mangiarotti:2011} and in such a case
$k_{\mathrm{d}}=k_{\mathrm{d}}^{\mathrm{u}}$. Explicit handling of
$k_{\mathrm{d}}^{\mathrm{l}}$, which falls within the interval of
integration only for the simulations to be compared with the SLAC
E-146 data (see Appendix~\ref{sec:BK_ldisc}), would increase too much
the complexity of the code. No problems have been detected during the
tests described in Sec.~\ref{sec:tests}. All the calculations are
again made with double-precision (i.e.~64-bits) arithmetics to further
improve the reliability with the absolute tolerance in \cal{DADAPT}
set to zero and the relative tolerance to $10^{-6}$. A typical value
for $\delta k_{\mathrm{d}}/k_{\mathrm{d}}$ is $10^{-8}$, as
mentioned. The minimum number of initial subdivisions performed by
\cal{DADAPT} is set to $25$.

\begin{figure}[t!!!!]
  \centering
  \begin{tabular}{c}
    \includegraphics[width=.425\textwidth]{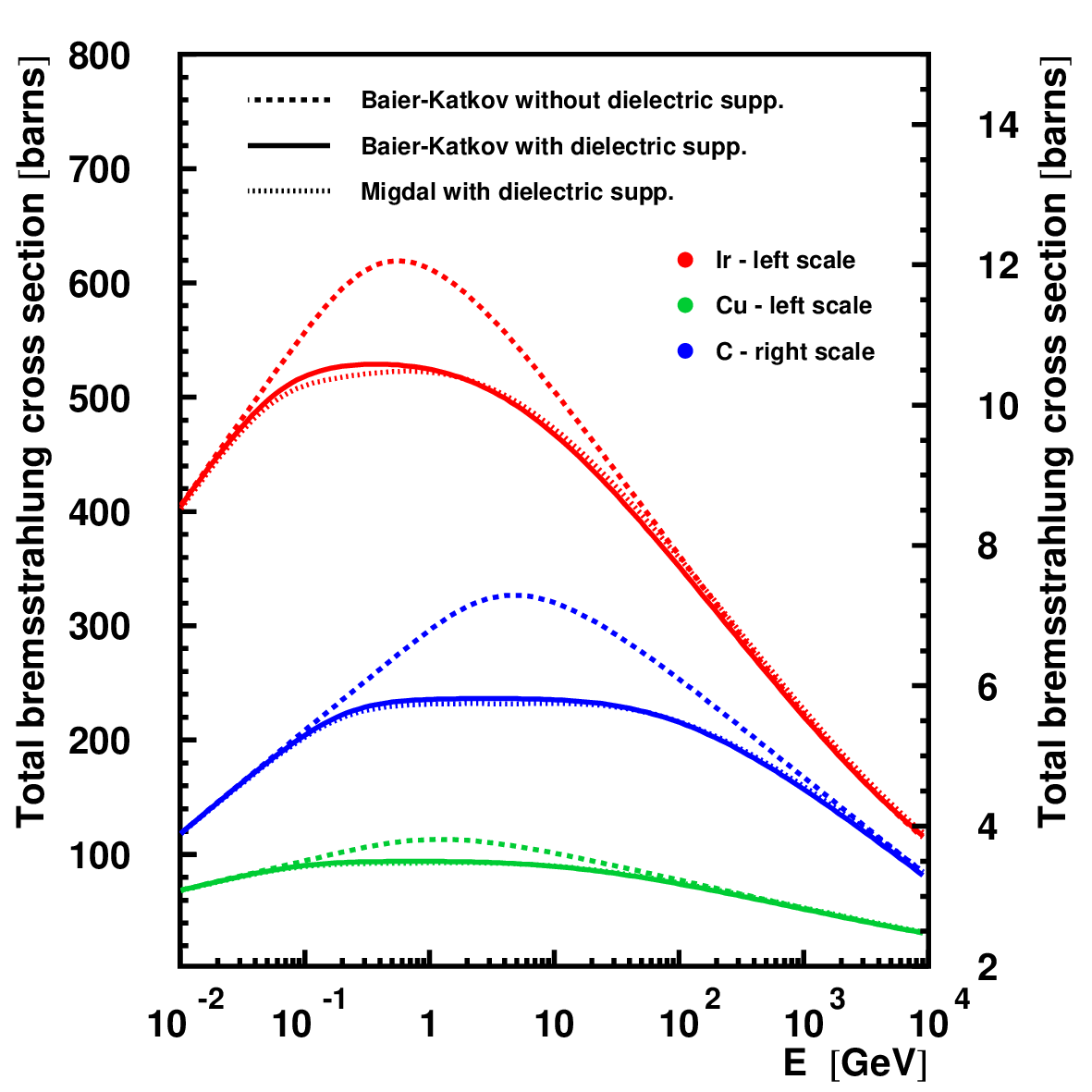}
  \end{tabular}
  \caption{\label{fig:sigmat}Total bremsstrahlung cross section for
    the radiation of a photon with energy $k>\mathrm{TCUT}=10$~keV by
    an electron with energy $E$ impinging on the indicated
    material. The vertical scale on the left is used for copper and
    iridium and the one on the right for carbon. Three approaches are
    considered: the BK one without dielectric suppression (dashed
    lines), the BK one with dielectric suppression (solid lines), and,
    finally, the one by Migdal with dielectric suppression (dotted
    lines).}
\end{figure}

Typical results are shown in Fig.~\ref{fig:sigmat}. It is apparent how
the LPM effect brings the initial growth of the total cross section to
saturation for increasing $E$ and then steadily reduces it. The
inclusion of the dielectric suppression even enhances this
trend. Finally, a comparison of the Migdal and BK approaches reveals
that they are quite close, which is to be expected from the discussion
in Sec.~\ref{sec:th_comp}.

\section{\label{sec:BK_samp}Sampling of the differential cross section}

Once a bremsstrahlung emission has been selected to happen by the
code, it is necessary to sample the photon energy $k$ according to the
differential cross section given by Eq.~(\ref{eq:BK_dsdk}). Of course,
it is not possible to find analytically the primitive of
Eq.~(\ref{eq:BK_dsdk}) and use directly the transformation method. The
straightforward application of the acceptance-rejection method would
be quite inefficient because of the $1/k$ BH divergence in
Eq.~(\ref{eq:BK_dsdk}). This problem was tackled by Butcher and Messel
to perform the first Monte Carlo simulations of electromagnetic
showers in the fifties: they introduced the ``composition and
rejection'' method~\cite{Butcher:1958}. The natural variable for the
sampling is the ratio $y=k/T_{\mathrm{E}}$ which spans the interval
$\mathrm{TCUT}/T_{\mathrm{E}}\leq y\leq 1$, where $\mathrm{TCUT}$ is
the low-energy cut imposed on photon propagation in the Monte Carlo
code. Then $y$ can be sampled from the function
$1/(y\,\ln(T_{\mathrm{E}}/\mathrm{TCUT}))$ by the transformation
\begin{equation}
  y=\exp(r_{1}\,\ln(\mathrm{TCUT}/T_{\mathrm{E}}))
  \;\;,
  \label{eq:y_tr}
\end{equation}
where the random number $r_{1}$ is uniformly distributed over the
$0$--$1$ range. Afterwards, the acceptance-rejection method is applied
to the rejection function
\begin{equation}
  r(y)=\frac{q(y)}{q_{\mathrm{max}}}
  \label{eq:rejf}
\end{equation}
by drawing a second random number $r_{2}$, also uniformly distributed
between $0$ and $1$, and accepting it if $r_{2}\leq r(y)$. If this is
not the case, the whole procedure is repeated from the sampling of
$r_{1}$. Since the $1/k$ BH divergence of Eq.~(\ref{eq:BK_dsdk}) is
already taken into account in Eq.~(\ref{eq:y_tr}), the corresponding
expression for $q(x)$ reads
\begin{equation}
\begin{split}
  q(x)=&k\,\bigg\{
  \frac{\nu_{0}^{2}}{\gamma^{2}}
  \left[R_{1}\,G\left(\frac{s_{\mathrm{BK}}}{2}\right)+
  2\,R_{2}\Phi\left(\frac{s_{\mathrm{BK}}}{2}\right)\right]\\
  &+\frac{3}{\gamma^{2}\,L_{\mathrm{c}}}
  \big[D_{1}(\nu_{0})\,R_{1}+D_{2}(\nu_{0})\,
  R_{2}\,\sqrt{2}\,\nu_{0}\big]
  \bigg\}
  \;\;.
\end{split}
\label{eq:BK_q}
\end{equation}
The dielectric suppression is included by means of the substitutions
given in Sec.~\ref{sec:BK_diel}. Although Eqs.~(\ref{eq:BK_dsdk})
and~(\ref{eq:BK_q}) are only valid in the ultrarelativistic limit, it
is necessary to enforce energy conservation exactly in the Monte Carlo
code; thus, it is assumed $x=y\,T_{\mathrm{E}}/E$. This distinction is
always negligible except very close to the lower cut
$T_{\mathrm{min}}=50$~MeV, imposed for electron propagation. Indeed,
the details of the cross section close to this threshold do not affect
appreciably the final result of the simulation.

\begin{figure}[t!!!!]
  \centering
  \begin{tabular}{c}
    \includegraphics[width=.425\textwidth]{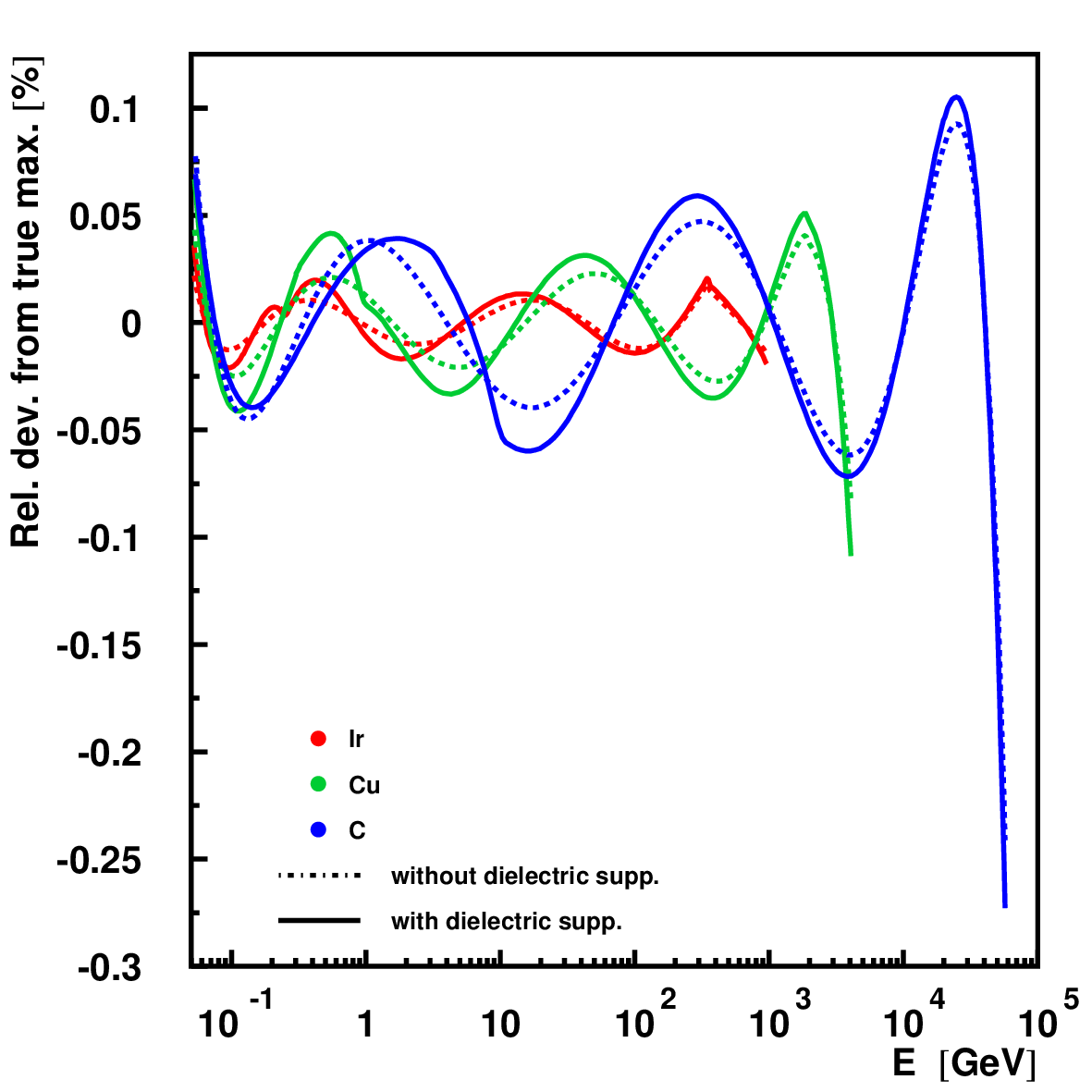}
  \end{tabular}
  \caption{\label{fig:fmin_dev}Difference between the exact value of
    the maximum $q_{\mathrm{max}}(E)$ of the function $q$ [see
    Eq.~(\ref{eq:BK_q})] and the result obtained by the sixth-order
    polynomials in $E$ fitted to a set of precalculated values. Both
    the case without (dashed lines) and with (solid lines) dielectric
    suppression are considered.}
\end{figure}

The value of $q_{\mathrm{max}}$ in Eq.~(\ref{eq:rejf}) has to be
determined by finding the maximum of Eq.~(\ref{eq:BK_q}). Of course,
$q_{\mathrm{max}}$ depends on the target material and the electron
energy $E$. The task is performed by using the subroutine \cal{DMINFC}
from the CERN mathlib library~\cite{Cernlib:1996p131}. All the
calculations are again made with double-precision (i.e.~64-bits)
arithmetics to further improve the reliability, the relative accuracy
parameter is set to $10^{-8}$, and the tolerance for boundary checking
is $10$ times larger, following the recommendations given in
Ref.~\cite{Cernlib:1996p131}. Unpredictable behavior of \cal{DMINFC}
with jumps towards the limits of the search intervals have been
observed due to the discontinuity at $k_{\mathrm{d}}$. Therefore, it
is mandatory to search for the maximum separately in two regions: the
first below $\ln(k_{\mathrm{d}}/T_{\mathrm{E}})$ and the second
between $\ln(k_{\mathrm{d}}/T_{\mathrm{E}})$ and $0$. It has been
found advantageous to search for the minimum using the variable
$\ln(k/T_{\mathrm{E}})$, since $k/T_{\mathrm{E}}\rightarrow 0$ for
decreasing $Z$. The maximum is found to move from the branch above
$k_{\mathrm{d}}$ to the branch below $k_{\mathrm{d}}$ when the energy
$E$ increases. This is true both when dielectric suppression is
enabled or disabled, contrary to the case of Migdal approach, where,
without dielectric suppression, the maximum is always found below
$k_{\mathrm{d}}$~\cite{Mangiarotti:2011}. In the case of dielectric
suppression and for the energy range of interest here, where
$T_{\mathrm{min}}=50$~MeV, it is sufficient to consider
$k_{\mathrm{d}}=k_{\mathrm{d}}^{\mathrm{u}}$ disregarding the presence
of $k_{\mathrm{d}}^{\mathrm{l}}$. To avoid searching the maximum
during each step of the Monte Carlo, this is done once for every
specific material at initialization time for $1750$ energy values
between $T_{\mathrm{min}}=50$~MeV and $10$~TeV. Then a fit is
performed with a sixth degree polynomial to represent the evolution of
$q_{\mathrm{max}}$ with $E$. To give an idea of the inaccuracies
involved, the deviation of the value calculated from the polynomial
with respect to the true one is shown, over the full $E$ range
considered for the fit, for the usual representative target elements
in Fig.~\ref{fig:fmin_dev}. Even in the worst case, the deviation does
not exceed $0.3$~\%. Due to the smoother behavior of the BK approach,
it has not been found useful to use two separate sixth degree
polynomials to represent the evolution of $q_{\mathrm{max}}$ below and
above $k_{\mathrm{d}}$, as it is necessary for Migdal
approach~\cite{Mangiarotti:2011}. To save memory, the $1750$ values
for the fit are stored in a temporary ZEBRA bank, which is dropped
afterwards, and only the coefficients of the polynomial are stored in
the \cal{JMATE} data bank of GEANT for each specific material. Then,
these parameters can be easily retrieved at tracking time and employed
to evaluate Eq.~(\ref{eq:rejf}).

\end{document}